%% file: main.tex
\documentclass[9pt,twocolumn,twoside,lineno]{pnas-new}
\templatetype{pnasresearcharticle} 
\usepackage{caption}
\usepackage{color,soul}
\usepackage[T1]{fontenc}
\usepackage{geometry}
\usepackage{pdflscape}
\usepackage{framed}
\usepackage{outlines}
\usepackage{subcaption}
\usepackage{subfiles} 
\usepackage{textcomp}
\usepackage[mathletters]{ucs}

\newcommand*{\rig}[1]{rig#1}


\title{The Facebook Algorithm's Active Role in Climate Advertisement Delivery}
\author[a,*]{Aruna Sankaranarayanan}
\author[a]{Erik Hemberg}
\author[a]{Una-May O'Reilly}
\affil[a]{Computer Science and Artificial Intelligence Laboratory, MIT}
\leadauthor{Sankaranarayanan} 
\authorcontributions{A.S conceived the question, A.S., E.H. and U.O designed the experiment, A.S. developed the experimental stimuli and conducted the experiments, A.S. analyzed the results, A.S., E.H and U.M wrote the manuscript.}
\correspondingauthor{\textsuperscript{*}To whom correspondence should be addressed. E-mail: arunas@mit.edu}
\include{abstract}
\keywords{advertising $|$ climate change $|$ social media $|$ polarization}

\begin{document}
\maketitle
\ifthenelse{\boolean{shortarticle}}{\ifthenelse{\boolean{singlecolumn}}{\abscontentformatted}{\abscontent}}{}

\input{introduction}
\input{part1-1-main}
\input{part2-1}

\input{discussion}
\input{consent}
\input{acknowledgement}
\bibliography{main}

\subfile{appendix}

\end{document}

%% file: abstract.tex
\begin{abstract}Communication strongly influences attitudes on climate change. Within sponsored communication, high spend and high reach advertising dominates. In the advertising ecosystem we can distinguish actors with adversarial stances: organizations with contrarian or advocacy communication goals, who direct the advertisement delivery algorithm to launch ads in different destinations by specifying targets and campaign objectives. We present an observational (N=275,632) and a controlled (N=650) study which collectively indicate that the advertising delivery algorithm could itself be an actor, asserting statistically significant influence over advertisement destinations, characterized by U.S. state, gender type, or age range. This algorithmic behaviour may not entirely be understood by the advertising platform (and its creators). These findings have implications for climate communications and misinformation research, revealing that targeting intentions are not always fulfilled as requested and that delivery itself could be manipulated. \end{abstract}

%% file: introduction.tex
\noindent In 2022, the Intergovernmental Panel on Climate Change (IPCC) identified ``rhetoric, misinformation, and politicization of science'' as key barriers to climate action.  The report, accepted by all members of the IPCC, stated explicitly that ``vested economic and political interests have organized and financed  misinformation and contrarian climate change communications''. 
Scholars across disciplines have documented the deceptive nature of contrarian climate communications~\cite{supran2017assessing, supran2021rhetoric, supran2023assessing, cook2022understanding, farrell2019evidence, lewandowsky2021climate, maertens2020combatting,franta2021litigation, franta2021early, franta2022weaponizing, grasso2019oily, kaupa2021smoke, mccright2000challenging, mulvey2015climate, ucs2007smoke, supran2017assessing, supran2021rhetoric, supran2023assessing}. 
Many of these studies (and the IPCC report) focus on communication through traditional media - print and broadcast ~\cite{bolsen2018us, zhou2016boomerangs,mccright2011politicization, merkley2018party, boykoff2004balance, antilla2005climate,koehler2016can, bruggemann2017beyond}, while climate discourse also occurs on social media platforms such as Twitter and Facebook.  
Like all discourse on social media platforms, climate discourse occurs ``organically'' through user content posts, up-voting, and sharing. It also occurs through advertising.  

Studies of climate discourse on social media platforms, have the unique opportunity to examine and report about platform users and advertisers in greater detail than in print or broadcast media. 
They are able to characterize the demographics of both users and advertisers, answering questions such as \textit{who is paying for content?} \textit{how much are they spending?} \textit{where are they targeting campaigns?}, and \textit{where are ads delivered and to whom in terms of age and gender?}. 
This more granular demographic knowledge can potentially improve the effectiveness of online climate action campaigns, support litigation\cite{wentz2023research}, and help inform effective inoculation and communication strategies against climate disinformation  \cite{bruine2007individual,treen2020online,wentz2023research, rolfe2011republicans,goldberg2021shifting}, \cite{weaver2022sponsored, aisenpreis2023us, scheufele2019science}.

This contribution examines climate related advertising activity(more succinctly, climate ads) on Facebook.  
Noting the adversarial nature of climate ad activity between actors who can be considered contrarians or advocates, our investigation starts with observational data -- the  data provided by Facebook of its historical advertising delivery activity. 
We are able to divide the data by adversary -- ads sponsored by contrarians and ads sponsored by advocates.  To consider the impacts of engagement, we subdivide the data for each adversary by the magnitude of impressions. This axis, related to delivery volume, linearly relates to advertising spend.  

Along the impression axis, beyond asking \textit{``who?''}, \textit{``where?''}, and indirectly \textit{``how much?''}, we also investigate the possible presence of algorithmic bias.  
Current studies have restricted the notion of climate actors  to underlying trade, organizational, and financial organizations influencing climate discourse. 
However, the algorithms and recommendation systems on social media may also deserve recognition as actors influencing climate discourse. 
Is it possible that Facebook's advertising algorithmic decision system (more succinctly, ADS or algorithm)  is itself a climate discourse actor of significance? 
Like most digital advertising platforms, Facebook's ADS is designed to provide maximum engagement for the cheapest cost.  Micro-targeting features that deliver ads to users most likely to engage with an advertiser's content enable advertisers - both contrarians and advocates, to act with detailed intentions.
Facebook's ad  algorithm has previously been shown to exhibit gender, racial, and political bias \cite{ali2019discrimination, ali2021ad,venkatadri2019investigating,sapiezynski2022algorithms,ali2022all}. 
For example, when the algorithm is tasked with delivering political ads for Democratic and Republican candidates, it tends to deliver them in larger quantities to Democratic and Republican voters respectively \cite{ali2021ad}, even when no targeting parameters are specified. 
Similarly, advertising algorithms have been shown to ``see colour'', propagating communications featuring individuals of a particular race to Facebook audiences of the same race\cite{sapiezynski2022algorithms}; ``see gender'', propagating communications featuring objects of stereotypical interest to males and females to Facebook audiences of the same gender\cite{ali2019discrimination}. These algorithms have also been shown to influence labour\cite{imana2021auditing}.
Because algorithmic bias can accelerate the spread of disinformation\cite{vosoughi2018spread}, in this contribution we first examine algorithmic bias while comparing advocacy ad and contrarian ad activity in observational data.

Detecting algorithmic bias requires complete transparency of target intent and ad delivery. Effectively, for 2 types of content if the same targeting parameters result in different delivery patterns, this constitutes algorithmic bias, interference, or skew. Our observational study turns out to be partially adequate to detect algorithmic skew. 
The data provided by Facebook only documents delivery statistics, and does not provide the corresponding campaign information entered by its advertisers. This obscures their intentions around targeting. Advertisers may choose specific or general targeting parameters to distribute an ad. In some circumstances, such as when we observe ads going only to one state or to users of a single gender or age segment, we can surmise an intent to target and the target itself.
When ads are delivered to all possible locations or user segments, we can surmise there has been no intent to target. In other circumstances, advertiser intentions remain obscure. Within these constraints and with these surmissions, we find apparent algorithmic skew. 

Therefore, to more fully investigate skew and assess whether it influences ad delivery, we conduct an experiment where we assume the role of (under controlled and ethical conditions) an advertiser, thus making the intent of targeting totally transparent. With details found in Section~\ref{section:experiment}, to uncover the foundations of bias,
we design the simple and informative experiments. This points to ads completely consisting of images, given that past experiments demonstrate that ad delivery can be significantly affected by the image alone. Choosing just two objects for ad images, one featuring an oil rig, i.e highly negatively aligned with climate action, and the other featuring a solar cell, i.e highly positively aligned with climate action, drawn from contemporary ads on Facebook, we completely ablate specific state, gender or age targeting parameters. This leaves the fate of these two ``experimental probes''' delivery entirely to Facebook's algorithmic decision system. Subsequently checking for differences in the delivery of ads featuring these two object in U.S. state, gender, and age based ad destinations, teases out apparent algorithmic bias. We go on to analyze whether affixing the logos of contrarian or advocacy actors on an ad image impacts delivery in ad destinations, and if the delivery we observe in an ad destination is proportional to Facebook's estimates of audiences in these destinations.
Finally, we discuss the implications of such algorithmic decision making on climate communication, and the climate discourse. \\

This work is divided into two sections. Section 1 presents an observational study of climate ads on Facebook between May 2018 - May 2023, paying close attention to delivery patterns. The analysis of the observational data is done with respect to different advertisement destinations (age, gender, location) and for different ad targeting strategies (targeted and non-targeted) as surmised from the data. Section 2 presents an experimental study to isolate the influence of the algorithm on delivery. We launch climate ads yielding full control of the delivery to the Facebook advertising algorithm and report results. The results show preferential delivery emerging solely from the type of ad content featured, in ad destinations characterized again by age, gender, and location. We discuss the implications of the algorithm's active role in the last section.

%% file: part1-1-main.tex
\section{An Observational Study of Past Climate Ads}
We conduct an observational study of 274K climate ads delivered on Facebook in the U.S. We first define terms pertaining to digital advertising. We go on to identify contrarian and advocacy actors, and assemble a dataset of past ads from these actors. We finally analyze the differences between the delivery patterns of advocacy and contrarian ads.\\
Digital advertising uses specific vocabulary. An advertiser publishes an ad along with `targeting' parameters that describe the intended audience. These can be the gender, age, location, interests, or even personally identifiable information of an audience group. Advertisers also specify `optimization criteria' to maximize the returns from an advertisement. For example, an advertiser could request the platform to maximize the ad view/ad click count, or increase traffic towards a website or a store. Lastly, advertisements contain `delivery' information. It is similar to the targeting information, but is inserted after the Facebook platform has delivered the ad. It describes the gender, age, and location compositions of audiences who were shown the ad. Delivery information also includes ad impressions, a value which describe the number of times the ad was shown on screen. The dataset we analyze does not contain targeting information or optimization criteria associated with an ad. It does contain partial delivery information: the location, gender, and age composition of an ad audience and the range of impressions received and expenditure made on the ad. We analyze this delivery information later in this section.
\input{part1-3-dataset}
\input{part1-4-methods}
\input{part1-5-1-results}

%% file: part1-3-dataset.tex
\subsection{Dataset}
To assemble our dataset, we begin by creating a list of contrarian and advocacy actors identified by peer-reviewed research\cite{farrell2016corporate, desmog, brulle2021networks}. Contrarians are restricted to fossil fuel corporations and groups who advertise on their behalf, and advocates are restricted to environmental groups and renewable energy providers. Any ad published by these groups is considered a `climate' ad. We extract 81,248 ads published by 260 contrarian actors and 171,877 ads published by 482 advocacy actors between May 2018 - May 2023. Collectively, the ads in our dataset are viewed for 1.36M days, shown between 5.4B - 6.4B times on screen, and involve an expenditure of \$79M - \$133M. We remove duplicate ads and ads delivered to locations outside the U.S. to yield 63,542 contrarian  and 139,012 advocacy ads. We sub-divide and aggregate these ads by impression counts (See Appendix \ref{appendix:impClassesAll} for a full list of 39 impression classes) into 5 groups: ads receiving $< 1K$, $1K-10K$, $10K-100K$, $100K-1M$, $1M+$ impressions; See \ref{appendix:dataset} for a link to the dataset.

\begin{table*}[]
    \centering
    \begin{tabular}{c|c|c}
         \hline
         & Contrarians & Advocates \\
         \hline
         \hline
         Expenditure (\$) &  \$34M  - \$47M & \$45M - \$67M\\
         Impressions &  1.9B - 2.2B & 4.5B - 6.7B  \\
         Impressions/\$ & 40 - 64 impressions/\$ & 52- 91 impressions/\$ \\
         Top 5 targeted states (In order) & Texas, Michigan, New Mexico, Pennsylvania, Colorado & Michigan, California, Pennsylvania, North Carolina, Colorado \\
         Top 5 non-targeted states (In order) &  Texas, Florida, Ohio, Pennsylvania, North Carolina & California, New York, Florida, Texas, Pennsylvania\\
         Top 5 states overall (In order) & Texas, Florida, Pennsylvania, Ohio, California & California, New York, Florida, Texas, Pennsylvania\\
         \hline
    \end{tabular}
    \newline
    \newline
    \caption{Comparisons of ad impressions and ad spend for contrarians and advocates}
    \label{tab:desc_stats}
\end{table*}

%% file: part1-4-methods.tex
\noindent After assembling the dataset, we analyze it. Ignoring the content of an ad, we focus our attention on the delivery information. We now describe possible ad destinations and corresponding delivery information, and proxies to surmise targeting intent from delivery.
\paragraph{Ad Destinations and Delivery Information} Each ad in the dataset contains attributes describing delivery information for three ad destinations: U.S. states, gender, and age. 
\begin{itemize}
    \item \textit{U.S. States} - The $\texttt{delivery\_by\_region}$ attribute of each ad contains delivery information for U.S. state destinations. This attribute is a list of $N$ tuples, $N \in [1, 52]$. The $i^{th}$  tuple is $(\texttt{region}_i, \texttt{delivery\_percentage}_i)$, where $\texttt{region}_i$ can be one of 52 locations, comprising 50 U.S. states, Washington D.C and an `Unknown' category, and $\texttt{delivery\_percentage}_i$ is a fraction such that $\texttt{delivery\_percentage}_i \in [0,1]$ and $\Sigma_{i=1}^{52}\texttt{delivery\_percentage}_i = 1$.
    \item \textit{Age and gender destinations} - The $\texttt{demographic\_distribution}$ attribute of each ad contains delivery information for age and gender destinations.  This attribute is a list of $N$ tuples, $N \in [1, 24]$. The $i^{th}$ tuple is $(\texttt{gender\_i}, \texttt{age\_i}, \texttt{delivery\_percentage}_i)$, where $\texttt{age\_i} \in [\text{18-24, } \text{25-34, } \text{35-44, } \text{45-54, } \text{55-64, }\text{65+}]$ and $\texttt{gender\_i} \in [\text{male, } \text{female, } \text{unknown}]$. $\texttt{delivery\_percentage}_i$ is a fraction such that $\texttt{delivery\_percentage}_i \in [0,1]$ and $\Sigma_{i=1}^{52}\texttt{delivery\_percentage}_i = 1$.
\end{itemize}
The delivery information reveals the audiences reached by the contrarians and advocates, their locations, and indicates the role of the algorithm in ad delivery. We would ideally use both targeting and delivery information associated with an ad, to uncover these insights. However, while the dataset contains delivery information along three ad destinations - U.S. states, age, and gender - it contains no targeting information. We therefore use proxies in the delivery data to deduce targeting intent.
\paragraph{Targeting Proxies} We consider two proxies to surmise the presence or absence of targeting intent. Ads delivered to only one U.S. state, gender or age destination are assumed to be targeted by the advertiser, and called `Targeted' ads. 
Ads delivered to audiences that are diverse in composition, i.e ads delivered to at least 48 U.S. states\footnote{Facebook delivers ads to 52 locations -- 50 states, Washington D.C. and an `Unknown' category. In ~700 campaigns targeted at audiences across U.S. states, genders and ages, we found that while the ad was always delivered to audiences belonging to all gender and age categories, sometimes ads were only delivered to 48 locations. Changing this to other values between 48 and 52 had little effect on our findings.} and to all genders and age groups are called `Non-targeted' ads. We hypothesize that the role of algorithmic decision making is visible in the delivery patterns of ads reaching audiences of diverse compositions based on findings in past research\cite{ali2021ad,ali2019discrimination,imana2021auditing}\footnote{The role of the algorithm is visible even when advertisements are targeted using advertiser-controlled targeting features such as personally identifiable information based targeting or custom audience list based targeting.}.

%% file: part1-5-1-results.tex
\subsection{ANALYSIS}
 We use the delivery information data to analyze and compare the advertising behaviours of contrarians and advocates in the overall dataset across state, age, and gender based ad destinations. We also analyze delivery patterns from the two groups for different targeting strategies as defined by our targeting proxies. The methods used for the analysis are outlined in \ref{appendix:part1-methods}.
\input{part1-5-2-all-ads}
\input{part1-5-3-targeted-ads}
\input{part1-5-4-non-targeted-ads}

%% file: part1-5-2-all-ads.tex
\subsubsection{ANALYSIS OF ALL ADS}
\begin{figure*}[t!]
    \centering
    \hfill
    \begin{subfigure}[b]{\textwidth}
        \begin{subfigure}[b]{0.75\textwidth}
            \centering
            \includegraphics[width=\textwidth]{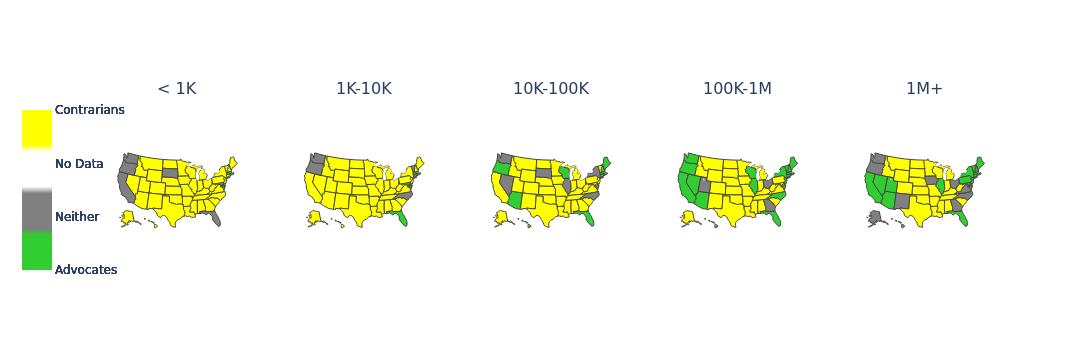}
            \caption{Comparisons of delivery percentage samples for contrarian and advocacy ads across different impression sub-divisions. States in yellow had higher average delivery percentages for, and thus were more effectively prioritized by, contrarian ads, while states in green had higher average delivery percentages for, and were more effectively prioritized by, advocacy ads. When comparing delivery percentages of low-impression ads ($<$ 10K impressions), contrarians more effectively prioritize 80\% of states while advertising. Only Delaware, Maryland, Massachusetts and Rhode Island are more likely to be frequented by low-impression advocacy ads, with Florida receiving higher percentages of advocacy ads in the $1K - 10K$ category. For ads receiving a moderate number of impressions, $10K-100K$, 65\% of states see higher average delivery percentages of contrarian ads as compared to advocacy ads. Notably, contrarians are more effective in California with low to moderate impression ads, and advocates are more effective in Florida with moderate to high impression ads.}
            \label{fig:all-ads-state-map}
        \end{subfigure}
        \hfill
        \begin{subfigure}[b]{0.23\textwidth}
            \centering
            \includegraphics[width=\textwidth, angle=-90]{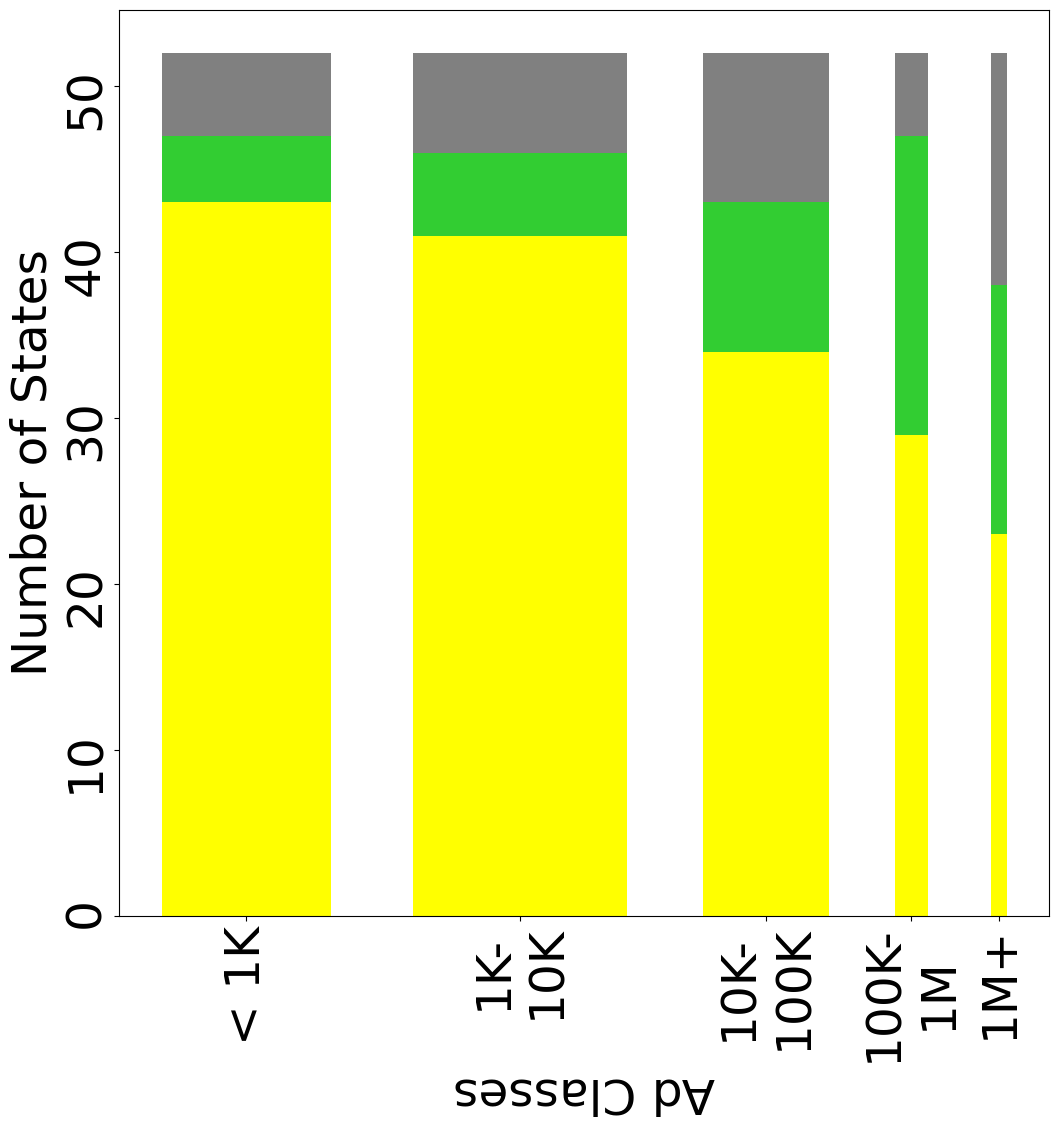}
            \caption{Number of states where contrarian or advocacy ads are delivered more frequently, on average, for each impression sub-division. In a majority of states, contrarian ads are delivered to more individuals across all impression sub-divisions.}
            \label{fig:all-ads-state-counts}
        \end{subfigure}
        \caption*{\textbf{U.S. State Destinations} - Comparing delivery percentages of advocacy and contrarian ads in each U.S. state.}
        \label{fig:all-ads-state}
    \end{subfigure}
    \hfill
    \begin{subfigure}[b]{\textwidth}
        \begin{subfigure}[b]{0.48\textwidth}
            \centering
            \includegraphics[width=\textwidth]{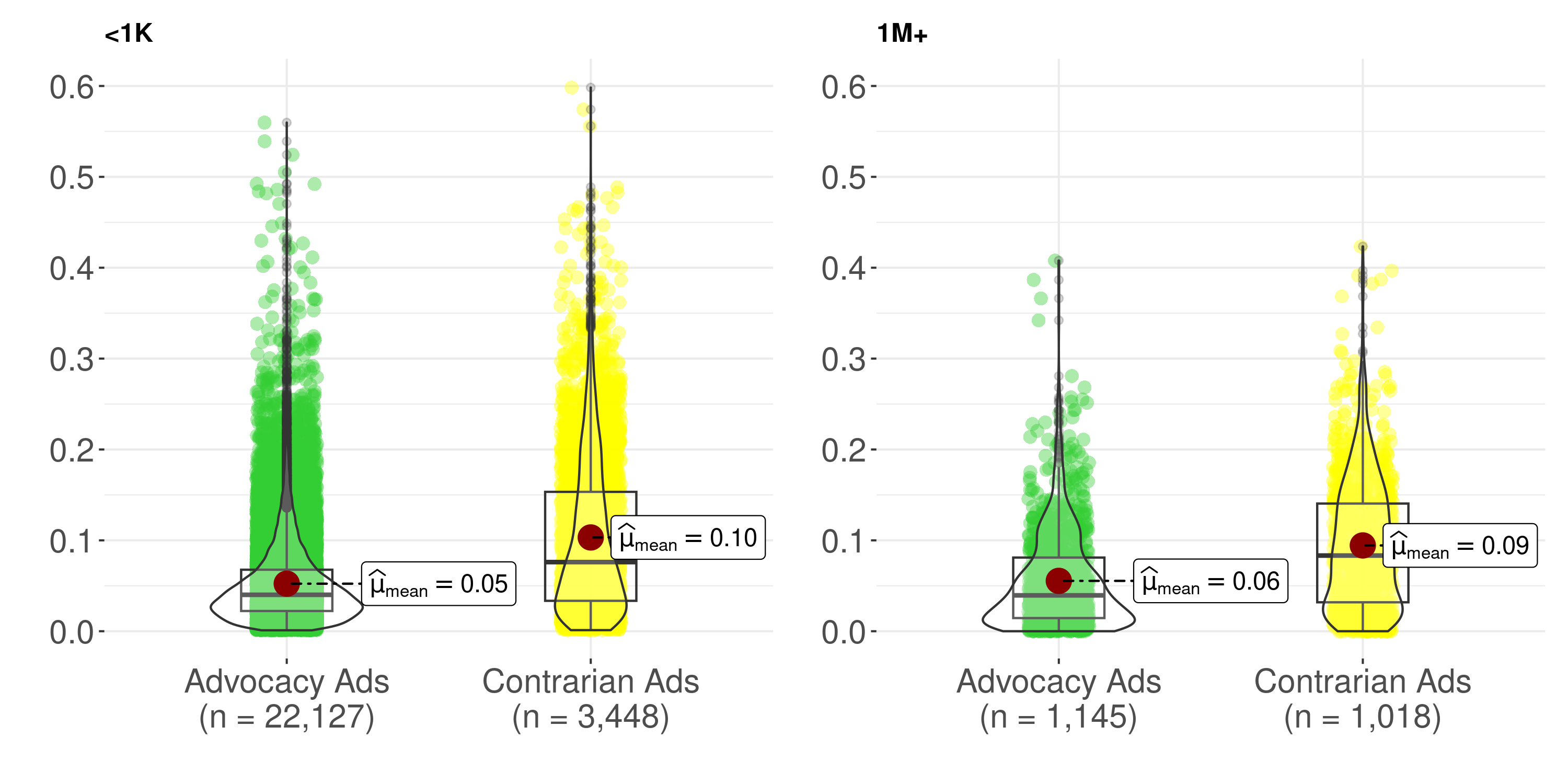}
            \hfill
            \caption{Males receive higher delivery percentages of contrarian ads on average across all impression classes.}
            \label{fig:all-ads-males}
        \end{subfigure}
        \begin{subfigure}[b]{0.48\textwidth}
            \centering
            \includegraphics[width=\textwidth]{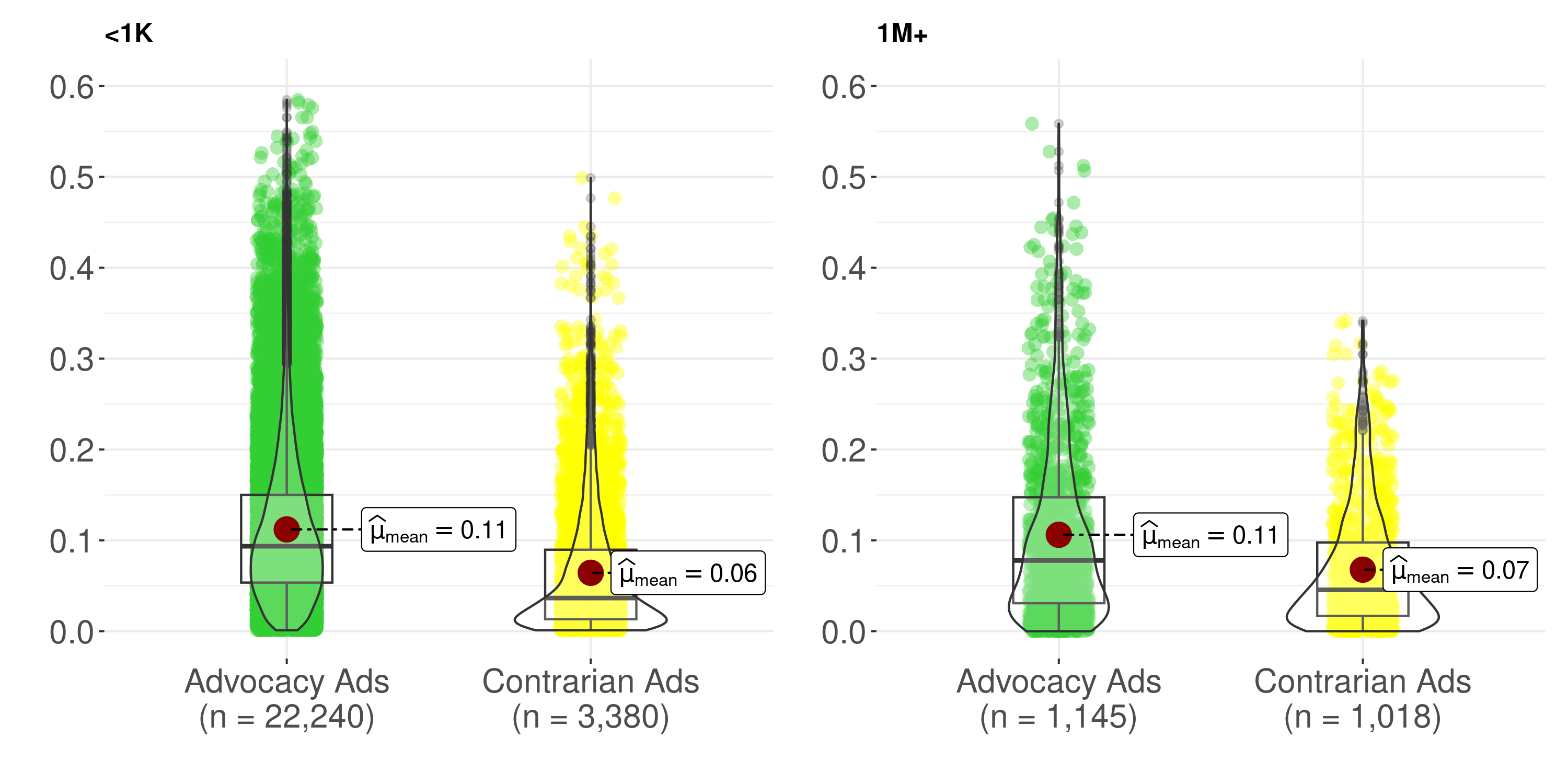}
            \hfill
            \caption{Females receive higher delivery percentages of advocacy ads on average across all impression classes.}
            \label{fig:all-ads-females}
        \end{subfigure}
        \caption*{\textbf{Gender Destinations} - Comparing delivery percentages of advocacy and contrarian ads for gender-based ad destinations. Only the lowest and highest impression class are shown here. 
        }
        \label{fig:all-ads-gender}
    \end{subfigure}
    \begin{subfigure}[b]{\textwidth}
        \begin{subfigure}[b]{0.48\textwidth}
            \centering
            \includegraphics[width=\textwidth]{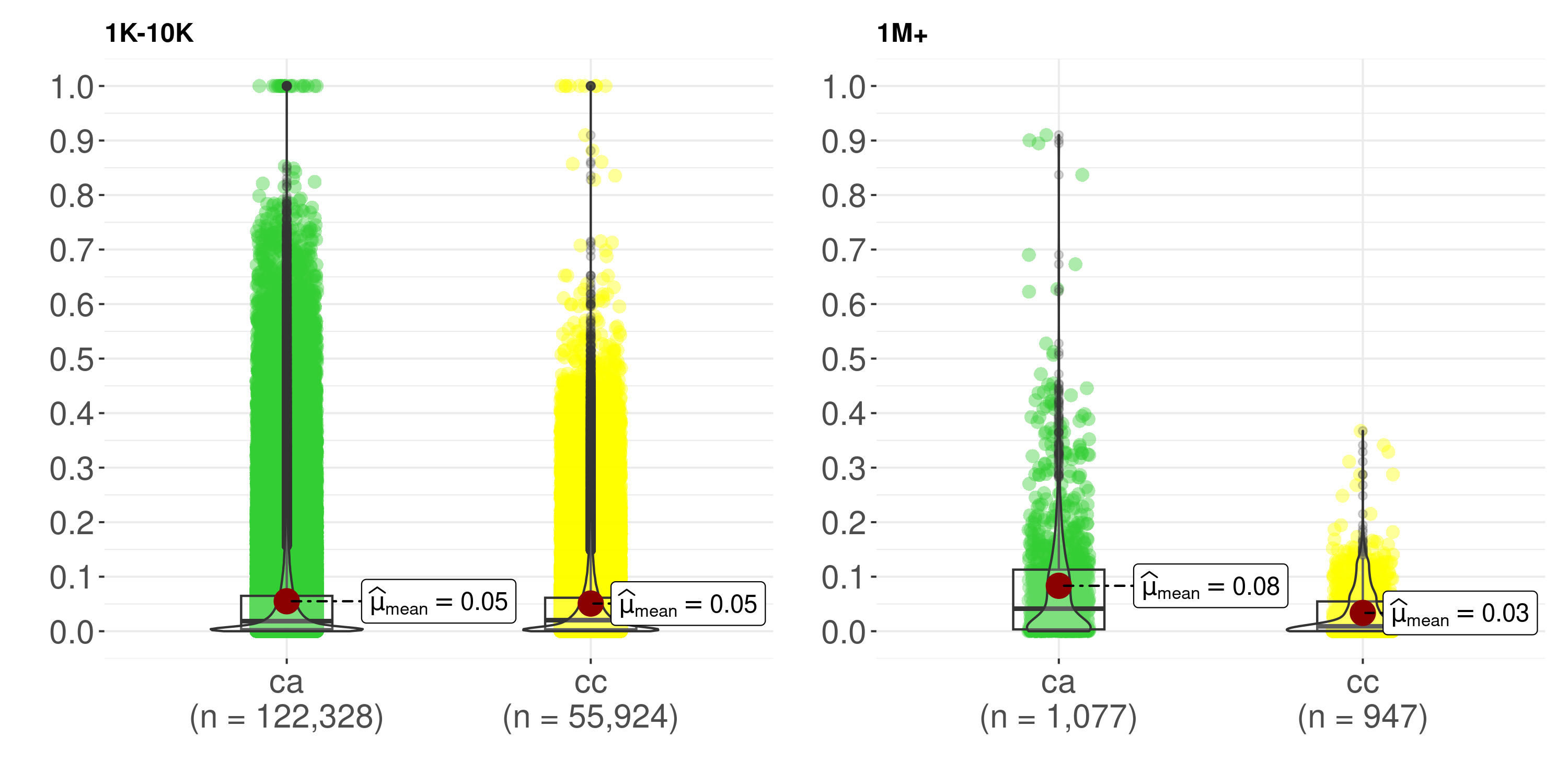}
            \hfill
            \caption{Younger audiences receive higher delivery percentages of advocacy ads on average across all impression sub-divisions.}
            \label{fig:all-ads-young}
        \end{subfigure}
        \begin{subfigure}[b]{0.48\textwidth}
            \centering
            \includegraphics[width=\textwidth]{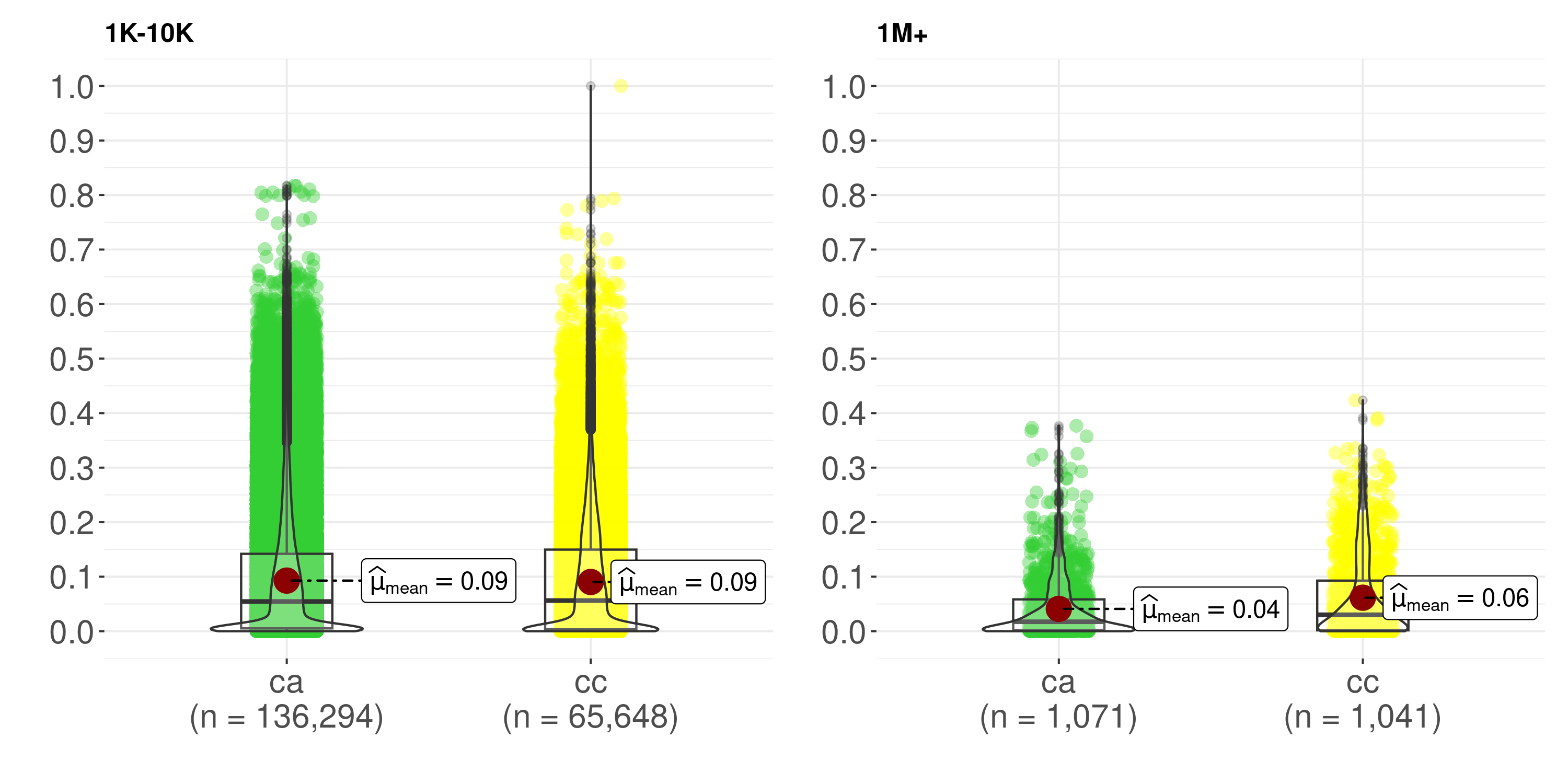}
            \hfill
            \caption{Older audiences receive higher delivery percentages of contrarian ads on average across all impression sub-divisions.}
            \label{fig:all-ads-old}
        \end{subfigure}
        \caption*{\textbf{Age Destinations} - Comparing delivery percentages of advocacy and contrarian ads for age-based ad destinations. Only the lowest and highest impression class are shown here. Delivery samples of advocacy and contrarian ads are significantly different across all impression sub-divisions, but the effect size (Cohen's d) is small for low-impression ads. 
        }
    \label{fig:all-ads-age}
    \end{subfigure}
    \caption{\textbf{ALL ADVERTISEMENTS} Comparing delivery percentages of advocacy and contrarian ads across different ad destinations for all ads in the dataset.}
    \label{fig:all-ads}
\end{figure*}

There are a total of 63,542 climate contrarian ads and roughly twice as many climate advocacy ads (139,012) in the ads dataset. We investigate the delivery percentage samples of advocacy and contrarian ads for each U.S. state, gender, and age based ad destination (See Fig \ref{fig:all-ads}). The delivery percentage samples tell us how contrarian and advocate ads are delivered on priority to different state, age, and gender based destinations - the higher the delivery percentage the more highly a destination was prioritized during delivery. Table \ref{tab:desc_stats} contains a list of states ranked by their prioritization across various targeting parameters in the dataset.
\begin{enumerate}
    \item U.S. State Destinations -  On average, delivery percentages of contrarian ads are higher than delivery percentages of advocacy ads. This indicates that contrarians run ads are delivered on priority to a single U.S. state or a few U.S. states as compared to advocates who run ads delivered to a larger cluster of states, across all impression categories. Advocates change strategy to prioritize delivery to a single state when running high-impression ads ($>100K$ impressions), but still lag behind contrarians in effectiveness (See Fig \ref{fig:all-ads-state-map} and \ref{fig:all-ads-state-counts}). See tables \ref{tab:obs_table_region_1K_all}, \ref{tab:obs_table_region_10K_all}, \ref{tab:obs_table_region_100K_all}, \ref{tab:obs_table_region_1M_all}, and \ref{tab:obs_table_region_1M+_all} for the values from our statistical analyses.
    \item Gender Destinations - On average, contrarians prioritize delivering their ads to males, while advocates prioritize delivering their ads to females across all impression classes (See Fig. \ref{fig:all-ads-males} and \ref{fig:all-ads-females}). See table \ref{tab:obs_table_gender_all} for values from our statistical analyses.
    \item Age Destinations - On average, contrarians prioritize delivering their ads to older audiences while advocates prioritize younger audiences. Facebook audience estimates suggest that the largest group of users on the platform are in the ages of 25-34, with users over the age of 65 being the smallest group, suggesting that a larger fraction of older individuals are receiving contrarian advertisements (See Fig. \ref{fig:all-ads-young} and \ref{fig:all-ads-old}). See table \ref{tab:obs_table_age_all} for values from our statistical analysis.
\end{enumerate}

%% file: part1-5-3-targeted-ads.tex
\subsubsection{ANALYSIS OF TARGETED ADS}
\begin{figure*}[t!]
    \centering
    \begin{subfigure}[b]{0.75\textwidth}
        \centering
            \includegraphics[width=\textwidth]{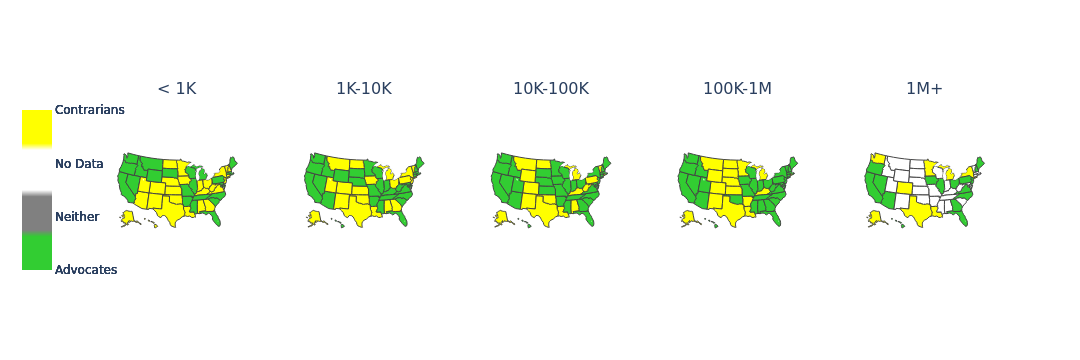}            \caption{\textbf{STATE-TARGETED ADVERTISEMENTS} - Comparisons of the fraction of targeted contrarian and advocacy ads. States in green show states where $\frac{\texttt{Advocacy Ads targeted at the state}}{\texttt{Count of targeted Advocacy Ads}} > \frac{\texttt{Contrarian Ads targeted at the state}}{\texttt{Count of targeted Contrarian Ads}}$, and states in yellow show states where $\frac{\texttt{Advocacy Ads targeted at the state}}{\texttt{Count of targeted Advocacy Ads}} < \frac{\texttt{Contrarian Ads targeted at the state}}{\texttt{Count of targeted Contrarian Ads}}$. In these maps, we see states that are important to the two groups. While advocates more frequently target advertisements at the two coasts, contrarians focus on the interior regions, and this pattern becomes more evident in higher impression classes. In ads receiving between 10K-100K impressions, Michigan, Vermont, Pennsylvania, West Virginia, Kentucky, and Alabama receive higher percentages of contrarian ads, while in the 100K-1M category, this is reduced to Kentucky and Michigan. In the category of ads receiving 1M+ impressions, Washington, Texas, Alaska, Louisiana, Minnesota, Michigan, Colorado, New York are dominated by contrarian ads. Alaska, one of the smallest states by Facebook user counts is a contrarian stronghold through and through. Similarly, Florida seems to be consistently dominated by advocacy advertisements.}
            \label{fig:targeted-states-map}
    \end{subfigure}
    \hfill
    \begin{subfigure}[b]{0.23\textwidth}
        \centering
        \includegraphics[width=\textwidth, angle=-90]{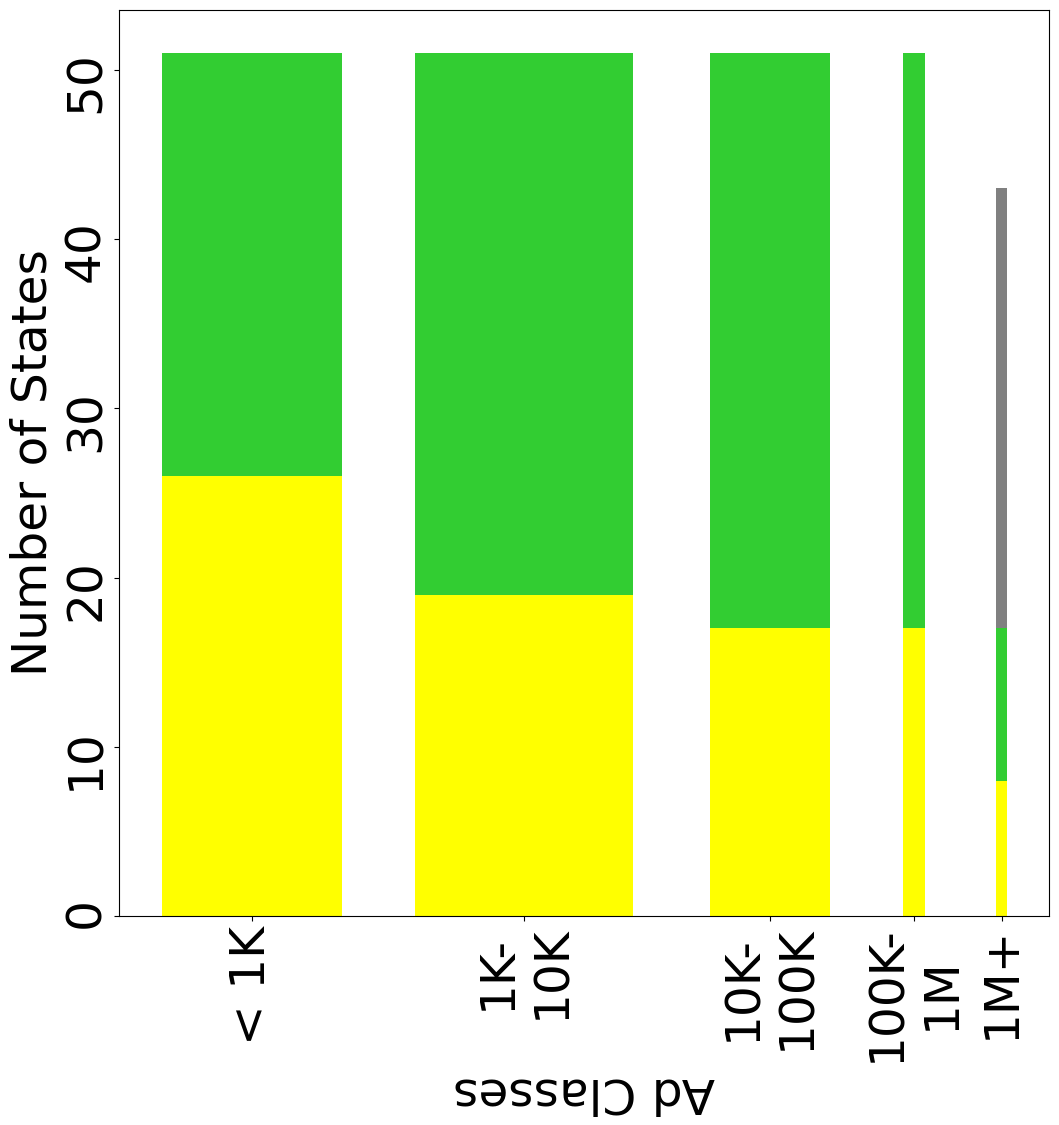}
        \caption{Count comparisons for states where contrarians and advocates target ads more frequently, across different impression classes.}
        \label{fig:targeted-states-counts}
    \end{subfigure}
    \caption{\textbf{TARGETED ADVERTISEMENTS} Comparisons of targeted contrarian and advocacy ads}
    \label{fig:targeted-states}
\end{figure*}

We investigate the delivery percentage samples of advocacy and contrarian ads that were delivered to at most 1 U.S. state, gender, or age destination (See Fig \ref{fig:targeted-states-map} and \ref{fig:targeted-states-counts})
\begin{enumerate}
    \item U.S State Destinations - 36K (56\%) contrarian and 46K (33\%) advocacy ads are targeted at a single U.S. state (Fig. \ref{fig:targeted-states-map}). 
    Particularly, Texas, New Mexico, Alaska, Louisiana, Colorado, Iowa, Montana, New York, and Utah receive a higher number of targeted contrarian ads in multiple impression sub-divisions, in spite of our dataset having twice as many advocacy ads as contrarian ads. The fraction of targeted ads being delivered to a state, advocates dominate in most states. See tables \ref{tab:obs_table_region_1K_t}, \ref{tab:obs_table_region_10K_t}, \ref{tab:obs_table_region_100K_t}, \ref{tab:obs_table_region_1M_t}, and \ref{tab:obs_table_region_1M+_t} for the exact values used in our analyses.
    \item Gender Destinations - 2,591 (3.1\%) contrarian ads and no advocacy ads, roughly 60\% of these were delivered to males, and 40\% were delivered to females. On average, contrarian ads receiving $< 1K$ impressions were delivered in higher percentages to females while in all other impression categories, they were delivered in higher percentages to males. The ads targeted at females went to all states, but predominantly to Alaska, Arizona, California, Texas, and Florida
    The ads targeted at males went to all states, but predominantly to Alaska, California, Hawaii, Kentucky, and Louisiana
    The contrarian ads targeted at males were more frequently delivered to older audiences while those targeted at females were more frequently delivered to younger audiences. See table \ref{tab:obs_table_gender_targeted} for the exact vaues from our analyses.
    \item Age Destinations - 248 (0.3\%) contrarian ads and 0 advocacy ads employ age group based targeting strategies.
\end{enumerate}

%% file: part1-5-4-non-targeted-ads.tex
\subsubsection{ANALYSIS OF NON-TARGETED ADS}
\begin{figure*}[t!]
    \centering
    \begin{subfigure}[b]{\textwidth}
        \centering
        \begin{subfigure}[b]{0.75\textwidth}
            \centering
            \includegraphics[width=\textwidth]{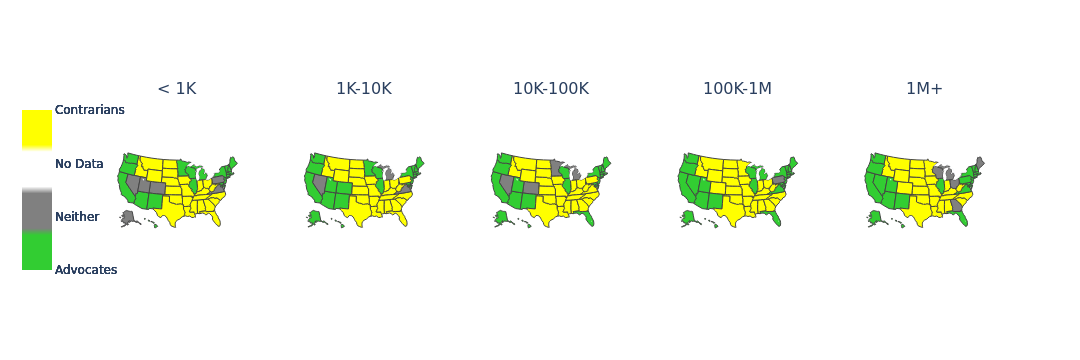}
            \caption{States in yellow had higher average delivery percentages for, and thus were more effectively prioritized by, contrarian ads, while states in green had higher average delivery percentages for, and were more effectively prioritized by, advocacy ads. Advocacy ads have higher average delivery percentages in states along the coasts, while contrarian ads have higher delivery percentages in the interior regions.}
            \label{fig:non-targeted-states-map}    
        \end{subfigure}
        \hfill
        \begin{subfigure}[b]{0.23\textwidth}
            \centering
            \includegraphics[width=\textwidth, angle=-90]{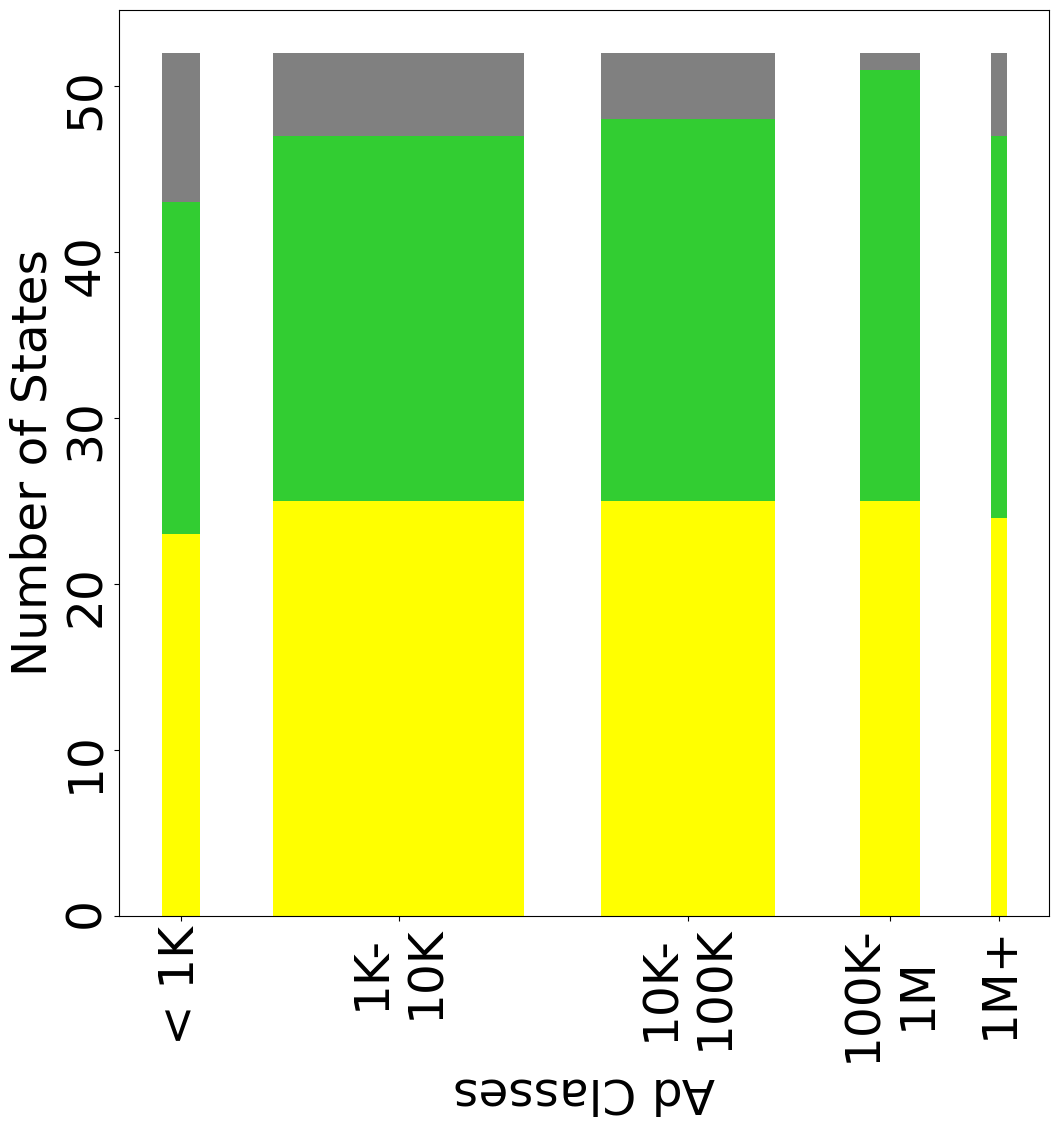}
            \caption{Number of states where contrarian or advocacy ads are delivered more frequently on average for each impression sub-division. Contrarian ads are more frequent in a slightly larger number of states in all impression sub-divisions except the 100K-1M category.}
            \label{fig:non-targeted-states-counts}
        \end{subfigure}
        \caption*{\textbf{U.S. State Destinations} - Comparing delivery percentages of non-targeted advocacy and contrarian ads in each U.S. state destination.}
        \label{fig:non-targeted-states}
    \end{subfigure}
    \begin{subfigure}[b]{\textwidth}
        \begin{subfigure}[b]{0.48\textwidth}
            \centering
            \includegraphics[width=\textwidth]{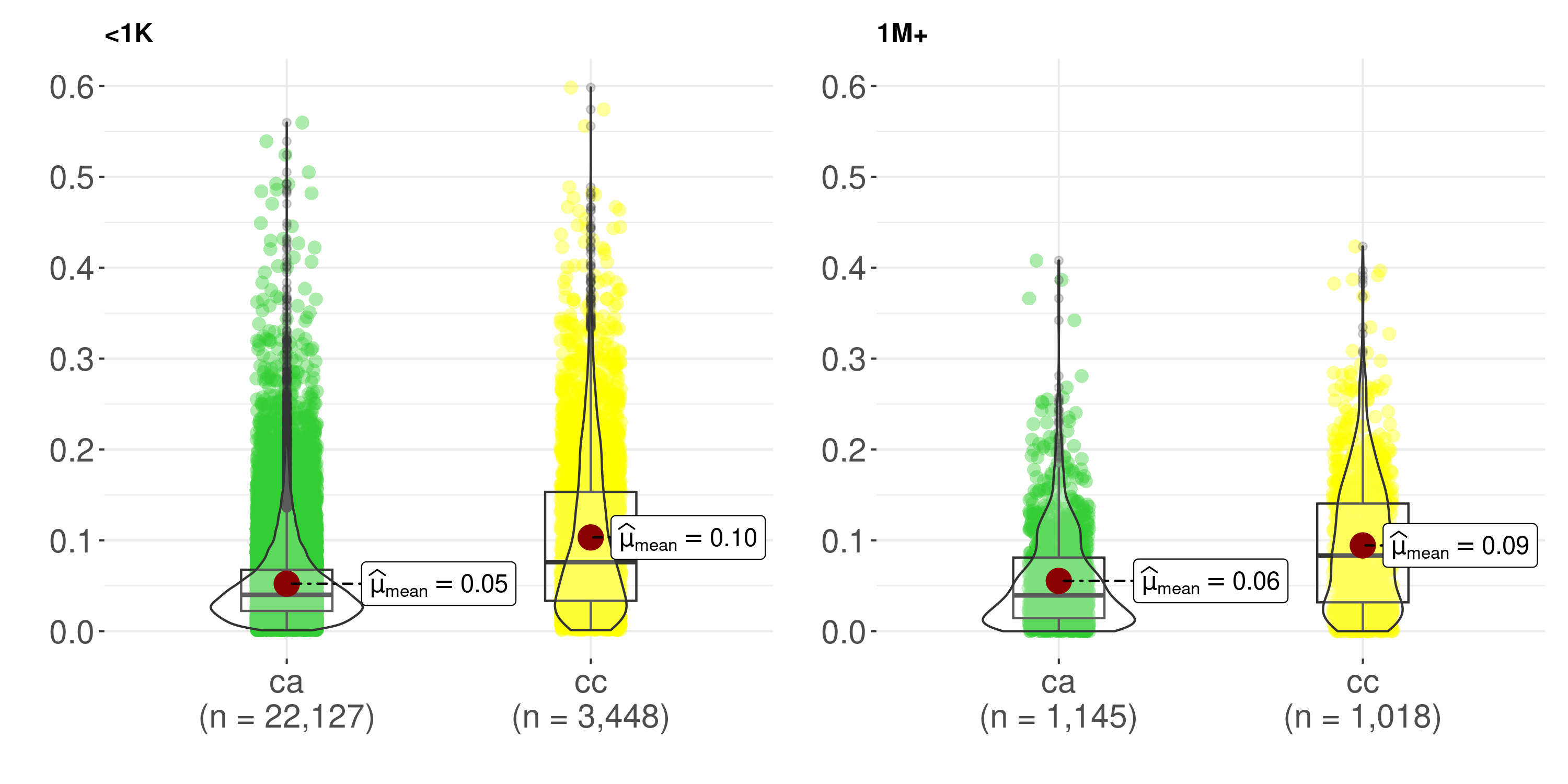}
            \hfill
            \caption{Males receive higher delivery percentages of, and are more effectively prioritized by, climate contrarian ads on average}
            \label{fig:male-nt-del-vol-comp}
        \end{subfigure}
        \begin{subfigure}[b]{0.48\textwidth}
            \centering
            \includegraphics[width=\textwidth]{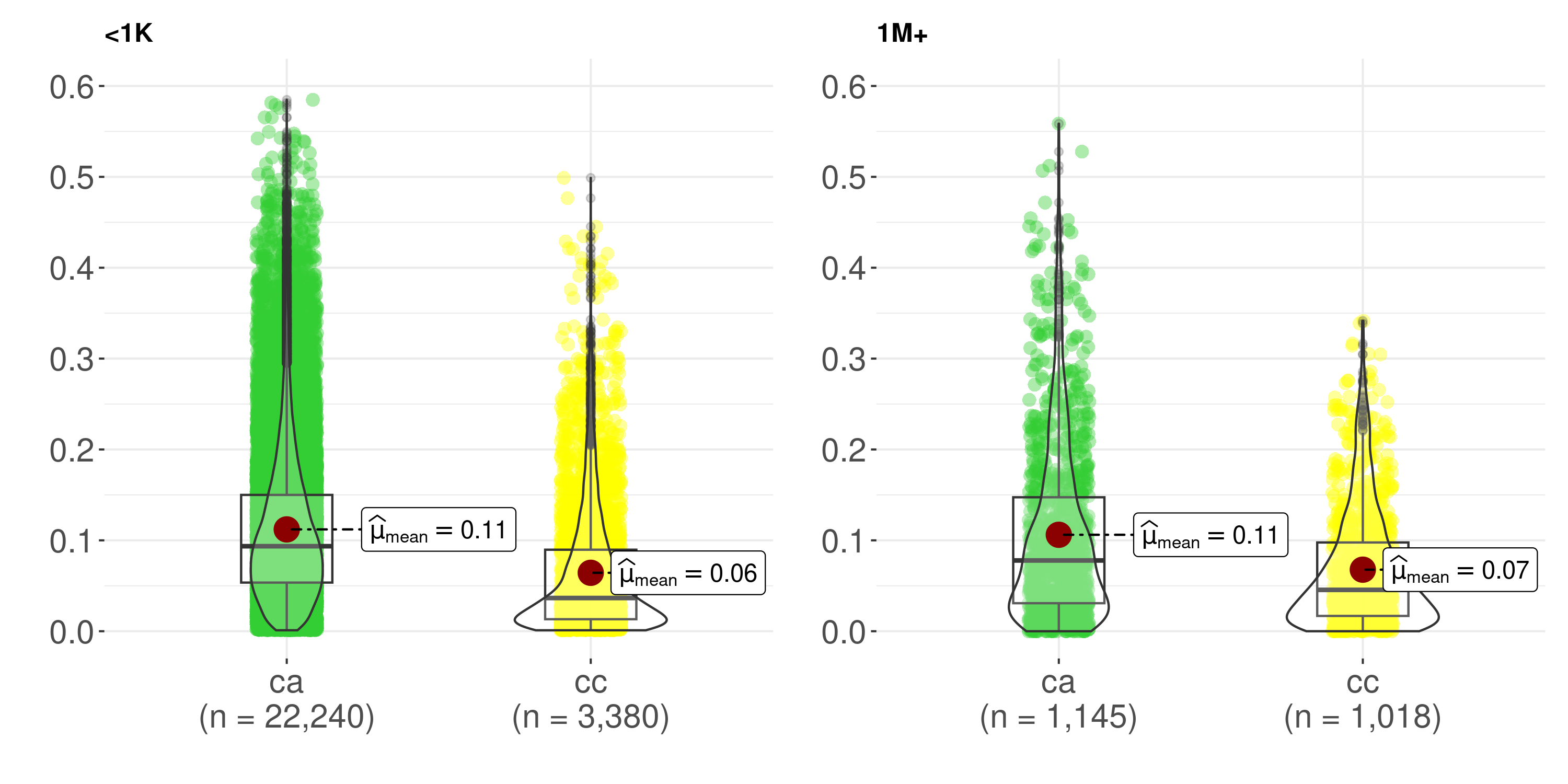}
            \hfill
            \caption{Females receive higher delivery percentages of, and are more effectively prioritized by, climate advocacy ads on average}
            \label{fig:female-nt-del-vol-comp}
        \end{subfigure}
        \caption*{\textbf{Gender Destinations} - Comparing delivery percentages of advocacy and contrarian ads for gender-based ad destinations. Only the lowest and highest impression classes are shown here.}
    \end{subfigure}
    \begin{subfigure}[b]{\textwidth}
        \begin{subfigure}[b]{0.48\textwidth}
            \centering
            \includegraphics[width=\textwidth]{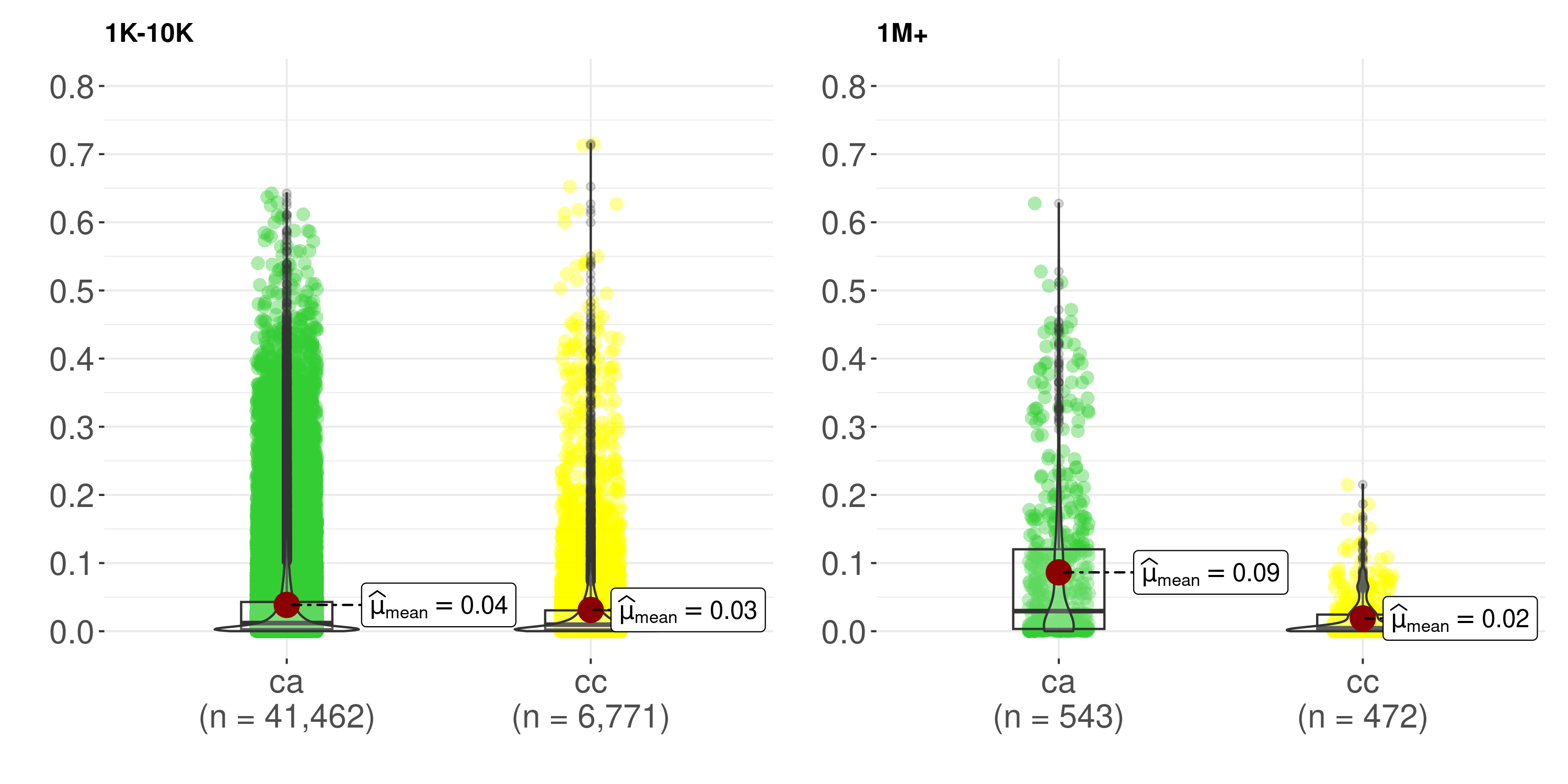}
            \hfill
            \caption{Younger Audiences}
            \label{fig:younger-nt-del-vol-comp}
        \end{subfigure}
        \begin{subfigure}[b]{0.48\textwidth}
            \centering
            \includegraphics[width=\textwidth]{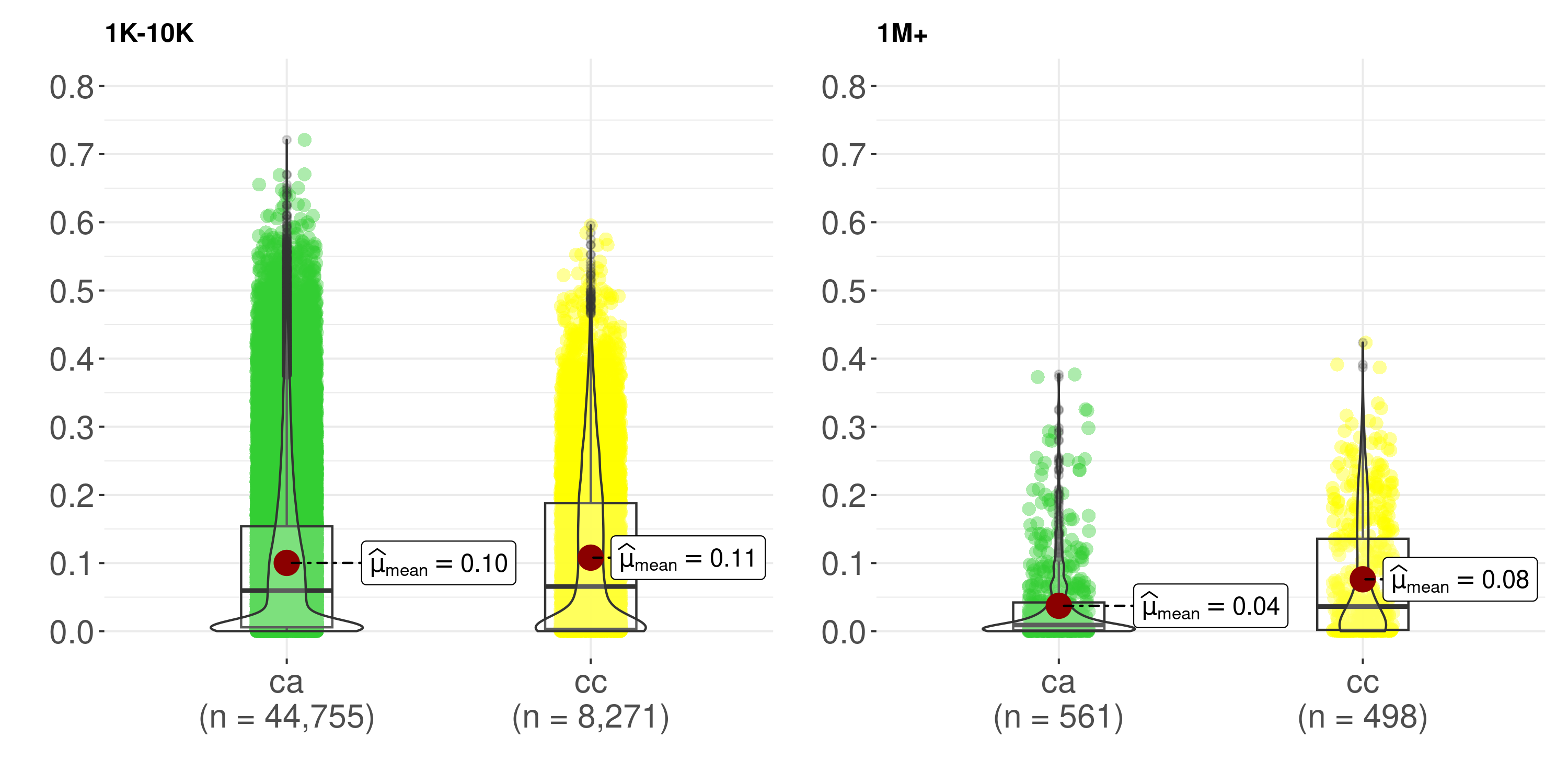}
            \hfill
            \caption{Older Audiences}
            \label{fig:older-nt-del-vol-comp}
        \end{subfigure}
        \caption*{\textbf{Age Destinations} - Comparing delivery percentages of advocacy and contrarian ads for age-based ad destinations. Younger audiences receive higher delivery percentages of, and are more effectively prioritized by, advocacy ads and older audiences receive higher delivery percentages of,  and are more effectively prioritized by, contrarian ads. However the effect size (Cohen's d) is small in the lower impression sub-divisions.}
    \end{subfigure}
    \caption{\textbf{NON-TARGETED ADVERTISEMENTS} Comparing delivery percentages of advocacy and contrarian ads across different ad destinations for non-targeted ads in the dataset.}
    \label{fig:on-targeted}
\end{figure*}

We filter a set of 7.5K contrarian ads and 44K advocacy ads that are not-targeted, i.e ads that reached $> 48$ U.S. states, all 3 genders, and 6 age groups (See Fig.\ref{fig:non-targeted-states-map} and \ref{fig:non-targeted-states-counts}). We hypothesize that the delivery of this set of ads is influenced by algorithmic decision making\footnote{While Facebook does allow for personally identifying information (PII) based targeting and targeting using custom audiences which may well have influenced this category of ads, past research has shown that algorithms exert a statistically significant influence even when there is PII or custom-audience based targeting\cite{ali2019discrimination, ali2021ad,venkatadri2019investigating,sapiezynski2022algorithms,ali2022all} especially when the number of ad destinations are large in number}. We compare delivery percentages in each location, gender, and age based ad destination.
\begin{enumerate}
    \item U.S. State Destinations -- Delivery percentages of contrarian and advocacy ads in this category are evenly distributed across all states in the U.S., with contrarians having a slight edge over advocates in a large number of states and impression sub-divisions (See Fig. \ref{fig:non-targeted-states-map}). States receiving higher percentages of contrarian or advocacy ads are strongly correlated to a state's likelihood to vote Republican or Democrat respectively based on voting patterns in the last 4 elections (Cramer's V correlation, $\phi_c$ = 0.73). See tables \ref{tab:obs_table_region_1K_nt}, \ref{tab:obs_table_region_10K_nt}, \ref{tab:obs_table_region_100K_nt}, \ref{tab:obs_table_region_1M_nt}, and \ref{tab:obs_table_region_1M+_nt} for the exact values from our analyses.
    \item Gender Destinations -- There is a significant difference between the delivery percentages of contrarian and advocacy ads among males, females and audiences whose gender are unknown to Facebook. Among females, and those of unknown gender, advocacy ads are delivered in higher percentages than contrarian ads and among males, the converse is true. See table \ref{tab:obs_table_gender_non_targeted} for the exact values from our analyses.
    \item Age Destinations -- There is a significant difference between the delivery percentages of contrarian and advocacy ads among audiences of all age groups. We observe that advocacy ads are delivered in higher percentages to audiences of younger age groups while contrarian ads are delivered in higher percentages to audiences of older age groups. See table \ref{tab:obs_table_age_non_targeted} for the exact values from our analyses.
\end{enumerate}
\subsection{Discussion}

\noindent We analyze the distribution of climate advertisements on Facebook in the last 5 years, and show that in spite of advocacy ads being higher in number, and costing lesser to advertise overall (See Table \ref{tab:desc_stats}), contrarian ads seem to be employing more effective location, gender, and age-group based targeting. Further, upon investigating targeted contrarian ads, we reveal that contrarians and advocates prioritize a small set of states. Texas, Florida, Pennsylvania and California are states prioritized by both contrarians and advocates(See Table \ref{tab:desc_stats}).\\
Secondly, we find a linear relationship in the dataset between the spend made on an ad and the impressions received, suggesting that our findings from the observational study would hold true, even if we sub-divided ads based on the spend instead of the impressions.\\
Lastly, we find evidence indicating, but not concluding, that Facebook's advertising algorithm enables this difference in delivery patterns across ad destinations. The dataset from the study only provides the range, and not the absolute values, of ad impressions received and ad spends made on each ad. To clarify the role of the algorithm in determining delivery and spend, we design and launch ads that are controlled for both targeting and spend parameters. We analyze the delivery ensuing from these campaigns, and report cases of preferential delivery enabled by the algorithm.

%% file: part2-1.tex
\section{An Experimental Study of Non-Targeted Ads}\label{section:experiment}
In this section, we conduct an experimental study to investigate the role of algorithmic decision making in the delivery of climate ads, without directed targets for age, gender, or location based ad destinations. Algorithmic decision making plays an invisible but important role in the delivery of digital information. Past research shows that ads related to elections, labour, and social issues are delivered preferentially to certain audience groups\cite{ali2019discrimination, ali2021ad, imana2021auditing, sapiezynski2022algorithms}. We are interested in investigating the existence of bias, and quantifying it, in the delivery of climate ads. We now describe our experimental design and probes, investigative methods, research questions, the main findings, and a detailed analysis of the results.

\input{part2-2-experiment-design}
\input{part2-3-1-results}

%% file: part2-2-experiment-design.tex
\begin{figure}[ht!]
    \begin{subfigure}[b]{0.15\textwidth}
        \centering
        \includegraphics[width=\textwidth]{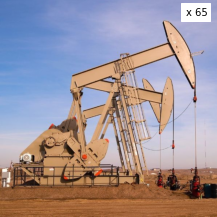}
        \caption{Oil \rig{s}}
    \end{subfigure}
    \begin{subfigure}[b]{0.15\textwidth}
        \centering
        \includegraphics[width=\textwidth]{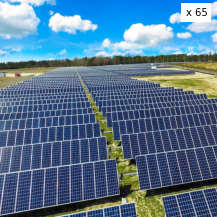}
        \caption{Solar cells}
    \end{subfigure}
    \begin{subfigure}[b]{0.15\textwidth}
        \centering
        \includegraphics[width=\textwidth]{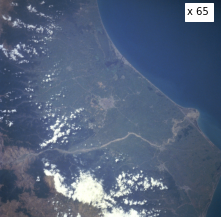}
        \caption{Controls}
    \end{subfigure}
     \caption{Images without logos}
     \label{fig:exp-design-images}
\end{figure}
\begin{figure}[ht!]
    \begin{subfigure}[b]{0.15\textwidth}
        \centering
        \includegraphics[width=\textwidth]{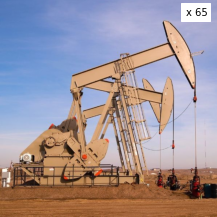}
        \caption{Oil \rig{s}}
    \end{subfigure}
    \begin{subfigure}[b]{0.15\textwidth}
        \centering
        \includegraphics[width=\textwidth]{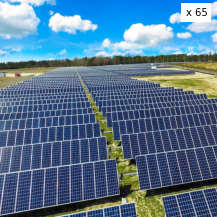}
        \caption{Solar cells}
    \end{subfigure}
    \begin{subfigure}[b]{0.15\textwidth}
        \centering
        \includegraphics[width=\textwidth]{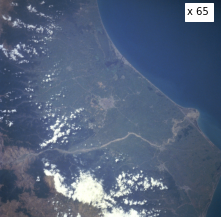}
        \caption{Controls}
    \end{subfigure}
     \caption{Duplicate of images without logos for a consistency check}
     \label{fig:exp-design-images-consistency}
\end{figure}
\begin{figure}[ht!]
    \begin{subfigure}[b]{0.11\textwidth}
        \centering
        \includegraphics[width=\textwidth]{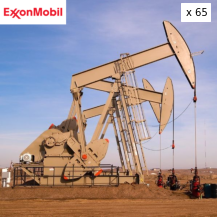}
        \caption{Oil rig + Contrarian Logo}
    \end{subfigure}
    \begin{subfigure}[b]{0.11\textwidth}
        \centering
        \includegraphics[width=\textwidth]{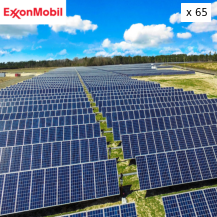}
        \caption{Solar cell + Contrarian Logo}
    \end{subfigure}
    \begin{subfigure}[b]{0.11\textwidth}
        \centering
        \includegraphics[width=\textwidth]{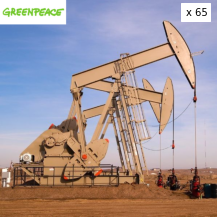}
        \caption{Oil rig + Advocacy Logo}
    \end{subfigure}
    \begin{subfigure}[b]{0.11\textwidth}
        \centering
        \includegraphics[width=\textwidth]{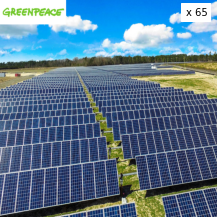}
        \caption{Solar cell + Advocacy Logo}
    \end{subfigure}
    \caption{Images with contrarian and advocacy logos}
    \label{fig:exp-design-images+logos}
\end{figure}
\begin{figure*}[ht!]
    \begin{subfigure}[b]{0.48\textwidth}
        \centering
        \includegraphics[width=\textwidth]{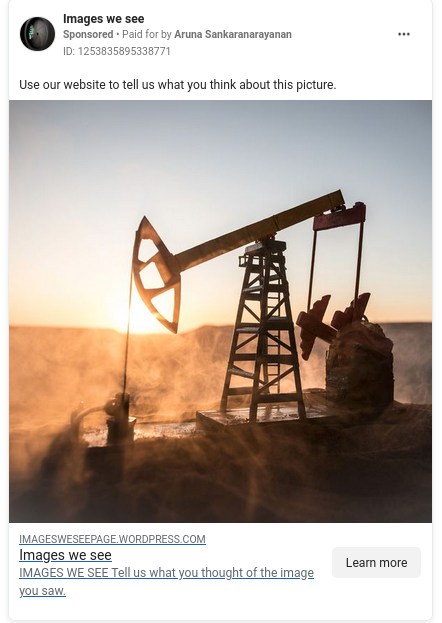}
        \caption{Oil \rig{s}}
    \end{subfigure}
    \begin{subfigure}[b]{0.48\textwidth}
        \centering
        \includegraphics[width=\textwidth]{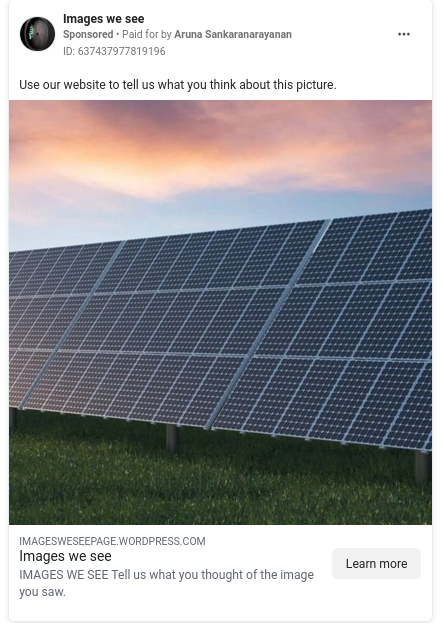}
        \caption{Solar cells}
    \end{subfigure}
     \caption{Example ads featuring an oil \rig{} and a solar cell.}
     \label{fig:example-ads}
\end{figure*}

\section{Experiment} We create 650 ads containing images, but no text (See fig. \ref{fig:example-ads}). This sample size is chosen based on an \textit{a priori} power analysis, which suggests selecting 65 images per group to uncover a small to moderate effect size of 0.25, with power=0.8. Ad images are sourced from the three experimental groups described below. We do not use any special ad targeting features provided by Facebook, i.e we request that the ads be delivered impartially to audiences in all U.S. states, and of all genders and ages. We launch the ads for a period of 24 hours, and  ask Facebook to optimize delivery to reach audiences likely to click on them. This is similar to prior research \cite{ali2019discrimination}. When a Facebook user clicks on the ad, they are taken to a \href{https://imagesweseepage.wordpress.com}{website}. At the end of the 24h experiment, ad metadata is collected from Facebook and comparisons are made between the delivery information of contrarian vs advocacy ads. Given that we run ads yielding full control over the ad targeting to Facebook, we measure the Delivery Ratio, $D^R$. $D^R$ includes both the observed delivery during the experiment and the expected delivery (which is derived from Facebook's ad audience estimates). To calculate $D^R$, first, the `Reach' information is collected for all launched ads. This contains the count of unique Facebook accounts that were shown one of our launched ads. This value is collected for each ad destination (U.S. state, gender, and age). Second, Facebook's self-reported population estimates for various ad destinations are collected. These provide a measure of the expected delivery count that is proportional to the audience size matching an ad destination. Facebook also advertises these population estimates as being the population sizes from which an ad audience sample will be drawn.
The `Delivery Ratio' ($D^R$), is given by $D_{ci}^R = \frac{O_{ci}}{E_{ci}}$. Here, $D^R \in \mathbb{R}^+$, and $O_{ci}, E_{ci} \in \mathbb{N}$. $O_{ci}$ is the unique number of times an ad $i$ was shown in an ad destination(U.S. state, gender, or age) $c$, and $E_{ci}$ is Facebook's estimated reach of the ad for the same category.

\subsection{Experimental Groups}
We design 3 experimental groups to investigate differences in delivery.
\begin{enumerate}
    \item \textbf{Images} - Solar cells (65 images) and oil \rig{s} (65 images) without any additional modifications. Additionally, two other sub-groups we consider in the \textit{Images} category are the following:
        \begin{enumerate}
            \item \textbf{Controls} - Control images (65 images)
            \item \textbf{Duplicates} - Duplicate ads using images from the  \textit{Images} and \textit{Controls} group to check that  delivery is consistently caused by ad content (65 images x 3).
        \end{enumerate}
    \item \textbf{Images + Contrarian Logo} - Solar cells and oil \rig{s} with the logo of a contrarian organization on the top left (65 images x 2).
    \item \textbf{Images + Advocacy Logo} - Solar cells and oil \rig{s} with the logo of an advocacy organization on the top left (65 images x 2).
\end{enumerate}

\noindent We use images of oil \rig{s} and solar cells in the experimental ads because these objects are found across the U.S., and featured in both contrarian and advocacy ads. While contrarians advertise oil \rig{s} to highlight engineering capabilities and economic advantages, advocates use them to campaign against drilling. Similarly, contrarian ads use solar cells to highlight their contributions to climate action, and advocates use them to promote the use of renewable energy. 65 images featuring each of these objects are selected. A state-of-the-art image classifier\cite{ridnik2021imagenet21k} is able to distinguish our probes with high accuracy. We also sample 65 controls from the ImageNet-21K dataset, using a sampling process that  excludes overlapping categories in the dataset. This ensures that controls are also able to be distinguished by a machine classifier. The algorithm to sample controls is provided in Appendix \ref{appendix:sampling-control}.

\section{Research Questions}
We pose the following research questions:
\begin{outline}[enumerate]
    \1 Does ad delivery ratio, $D^R$, differ based on the content of an ad image?
    \1 Does ad delivery ratio, $D^R$, differ when logos are present on an ad image? Is the effect similar for ads with solar cells and oil \rig{} images?
    \1 Can observed ad delivery be consistently attributed to the ad image?
    \1 Is observed ad delivery proportional to Facebook's population estimates in all ad destinations?
\end{outline}

%% file: part2-3-1-results.tex
\section{Results}
\input{part2-3-2-main-takeaways}
\input{part2-3-4-RQ1}
\input{part2-3-5-RQ2}
\input{part2-3-6-RQ3}
\input{part2-3-7-RQ4}

%% file: part2-3-2-main-takeaways.tex
\subsection{Main takeaways}
\begin{enumerate}
    \item Image matters: $D^R$ is significantly different based on the content of an ad image, in 46\% of U.S. states, and in all gender and age destinations.
    \item Logo influences are modest: $D^R$ is not significantly different for images featuring contrarian or advocacy logos. $D^R$ is also not influenced by the presence or absence of a logo in nearly any ad destination. Audiences in certain age destinations -- 18-24, 45-54, and 65+ -- are sensitive to the type of logo used on a solar cell image; $D^R$ is significantly different based on logo used in solar cell ads for these groups.
    \item Ad delivery is consistent with ad image: The audience sizes of ads featuring the same image is consistently similar in 90\% of the ads for U.S. state based ad destinations. In gender and age destinations, audience sizes are consistently similar for 100\% and 99\% of the ads we run respectively. 
    \item Observed audience sizes are not always proportional to Facebook's user populations: In U.S. state based ad destinations, audience sizes for \textit{Controls} are more likely (64\%) to be proportional to Facebook's population estimates than for images of solar cells and oil rigs (42.5\%). In gender based destinations, audience sizes for Solar cells and oil rigs are more likely (67\%) than controls (22\%) to be proportional to Facebook's population estimates. In age based destinations, neither images of of solar cell and oil rig images (0\%), nor controls (0\%), are proportional to Facebook's population estimates.
\end{enumerate}

\subsection{Methods}
Most of the samples we collect from Facebook are not normal, but are largely homoscedastic. To answer RQ1 and RQ2, we therefore use the One-Way Kruskall-Wallis non-parametric test to look for significant differences in the $D^R$ samples from different groups. Pairwise differences are further investigated using Dunn's Test with p-values adjusted using a Benjamini-Hochberg correction. To answer RQ3, we use Fisher's Test to investigate if the audience sizes in an ad destination is consistent when the same image is delivered twice. Lastly, to answer RQ4, we investigate if observed audience sizes are proportional to Facebook's self reported estimates using the Chi-Square Goodness of Fit Test with a Bonferroni correction.

%% file: part2-3-4-RQ1.tex
\subsection{Detailed Analysis}
\subsubsection{RQ1: Does ad delivery ratio, $D^R$, change based on the content of an ad image?} 
\begin{figure}[t!]
    \begin{framed}
    \begin{subfigure}[b]{\textwidth}
        \centering
        \includegraphics[width=\textwidth]{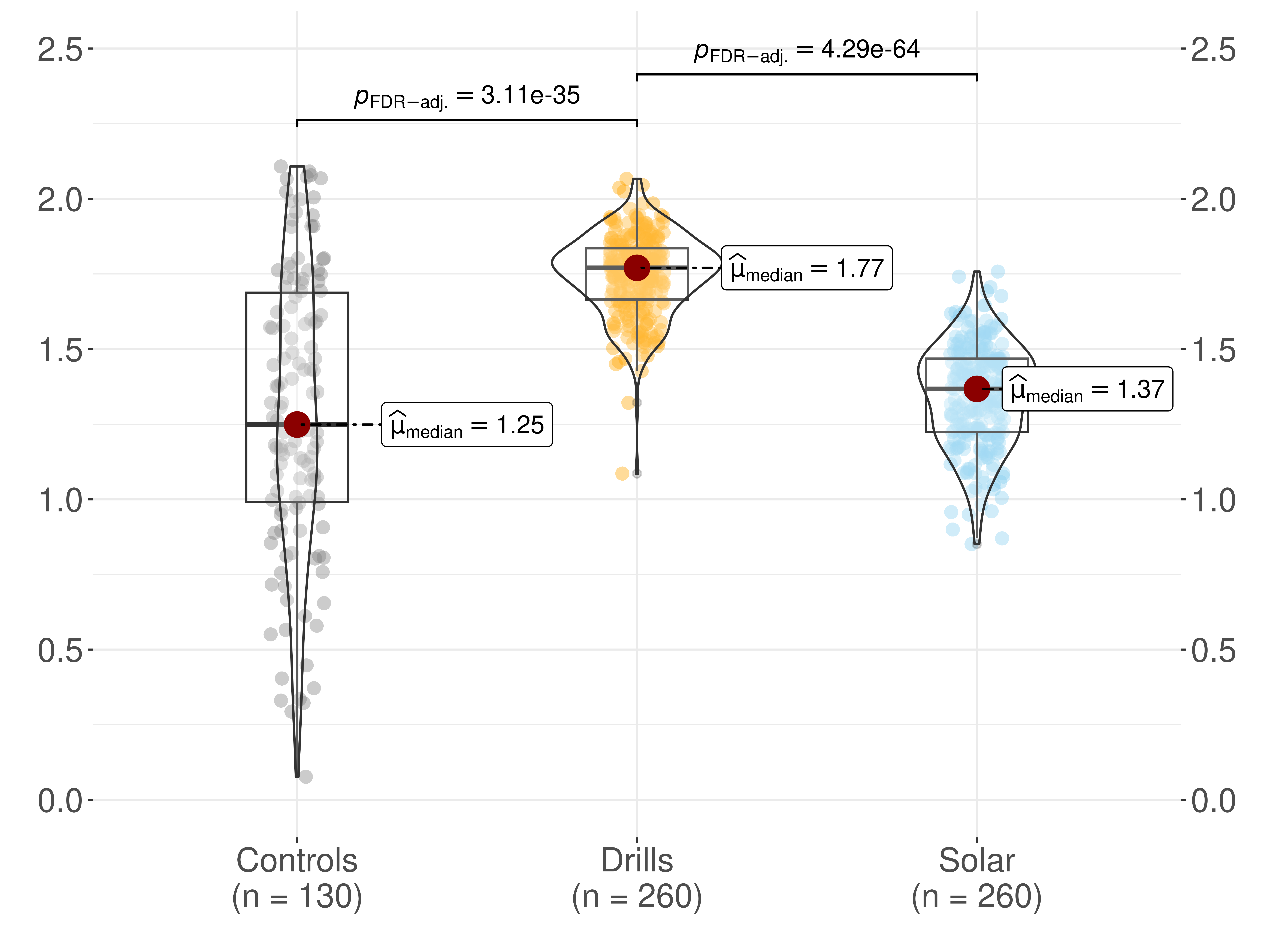}
        \hfill
        \caption{Males}
    \end{subfigure}
    \hfill
    \begin{subfigure}[b]{\textwidth}
        \centering
        \includegraphics[width=\textwidth]{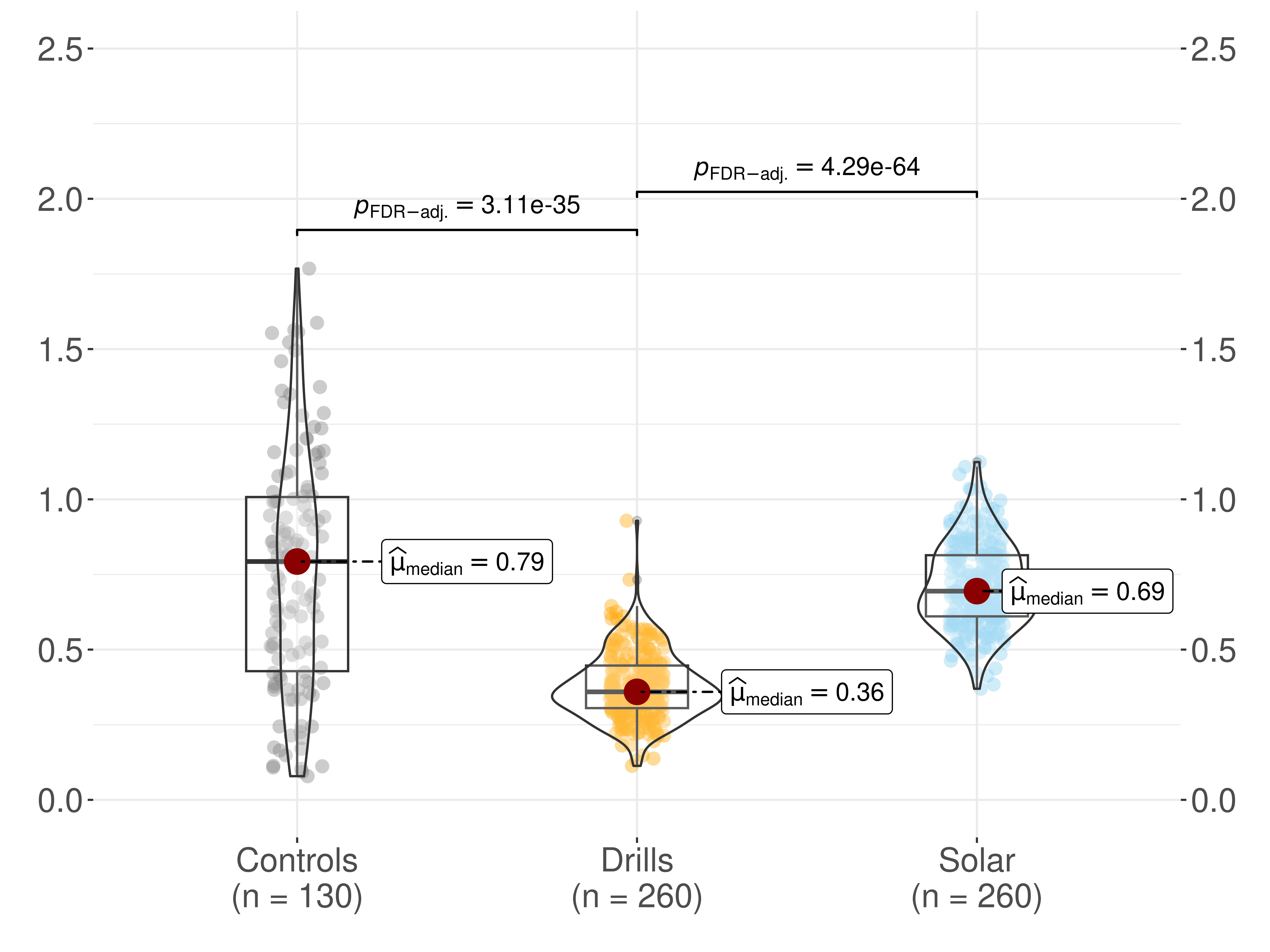}
        \hfill
        \caption{Females}
    \end{subfigure}
    \caption{$D^R$ samples of ads featuring oil \rig{s}, solar cells, and controls among male and female audiences. Note that images of oil rigs are delivered preferentially to males as compared to females. Similarly, images of solar cells are delivered preferentially to females as compared to males. The $D^R$ samples further reveal that male audiences are over-represented ($D^R > 1$), and female populations are under-represented ($D^R < 1$) compared to their composition in the Facebook population.}
    \label{fig:males-females-exp-comp}
    \end{framed}
\end{figure}

\begin{figure}[t!]
    \begin{framed}
    \begin{subfigure}[b]{\textwidth}
        \centering
        \includegraphics[width=\textwidth]{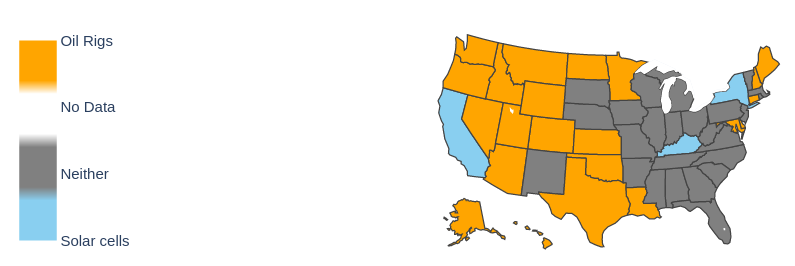}
        \hfill
        \caption{Regions where the $D^R$ of ads featuring solar cells is significantly different from the $D^R$ of ads featuring oil rigs. Regions in orange are regions where $D_{\text{oil rigs}}^R > D_{\text{solar cells}}^R$, and regions in blue are regions where $D_{\text{solar cells}}^R > D_{\text{oil rigs}}^R$. There are 25 states where there's a significant difference between the $D^R$ of ads featuring oil rigs and solar cells. In 88\% of these states, the $D_{\text{oil rigs}}^R > D_{\text{solar cells}}^R$}
    \end{subfigure}
    \hfill
    \begin{subfigure}[b]{\textwidth}
        \centering
        \includegraphics[width=\textwidth]{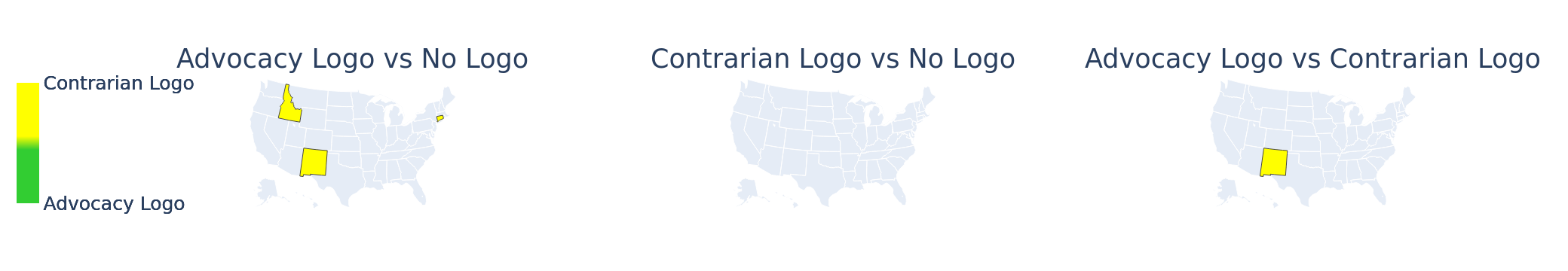}
        \hfill
        \caption{Regions where the $D^R$ of ads featuring oil \rig{s} is different based on the logo in the ad. Regions in yellow are regions where ads featuring contrarian logos had a higher median $D^R$ (Idaho, New Mexico, Connecticut) and regions in green are regions where ads featuring  advocacy logos had a higher median $D^R$.}
    \end{subfigure}
    \begin{subfigure}[b]{\textwidth}
        \centering
        \includegraphics[width=\textwidth]{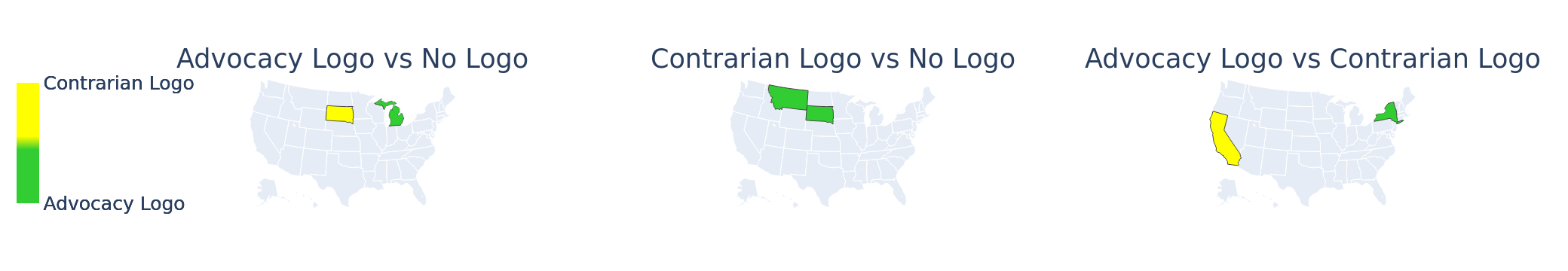}
        \hfill
        \caption{Regions where the $D^R$ of ads featuring solar cells is different based on the logo in the ad. Regions in yellow are regions where ads featuring contrarian logos had a higher median $D^R$ (South Dakota, California) and regions in green are regions where ads featuring  advocacy logos had a higher median $D^R$ (South Dakota, Montana, Michigan, New York).}
    \end{subfigure}
    \caption{}
    \label{fig:locations-rq1}
    \end{framed}
\end{figure}
We compare and investigate differences in the $D^R$ samples of ads featuring solar cells, oil \rig{s} and controls.
\paragraph{U.S. states} In 38 states, the $D^R$ sample of at least one of the three groups (Solar cells, oil \rig{s} and controls) is significantly different from the others (N=650, p < 0.05, k=3). Upon investigating the pairwise differences, we find that, in 25 states, there's a significant difference (N = 520, p < 0.05) between the $D^R$ samples of solar cells and oil \rig{s}. In 30 states, there's a significant difference (N=390, p < 0.05) between the $D^R$ samples of solar cells and controls and in 18 states there's a significant difference (N=390, p < 0.05) between the $D^R$ samples of oil \rig{s} and controls. See \ref{fig:locations-rq1} for a map visualizing these states, and table \ref{tab:rq1-region} for the Kruskall-Wallis H Statistic and p values.
\paragraph{Gender} In both male and female audiences, the $D^R$ samples of the three groups (solar cells, controls, or oil \rig{s}) are significantly different(N=650, p < 0.05, k=2) as seen in Fig. \ref{fig:males-females-exp-comp}. Upon investigating the pairwise differences, we find that in both male and female audiences, the $D^R$ samples of oil \rig{s} and solar cells are significantly different (N=520, p < 0.05, k=2), and of oil \rig{s} and controls (N=390, p < 0.05, k=2) are significantly different. Further, ads featuring oil \rig{s} are preferentially delivered to males while ads featuring solar cells are preferentially delivered to females. The $D^R$ samples further reveal that male populations are over represented ($D^R > 1$) while female populations are under represented in the ad audiences selected by Facebook ($D^R < 1$). See table \ref{tab:rq1-gender} for the Kruskall-Wallis H Statistic and p values.
\begin{figure}[t!] 
    \begin{framed}
    \begin{subfigure}[b]{\textwidth}
        \centering
        \includegraphics[width=\textwidth]{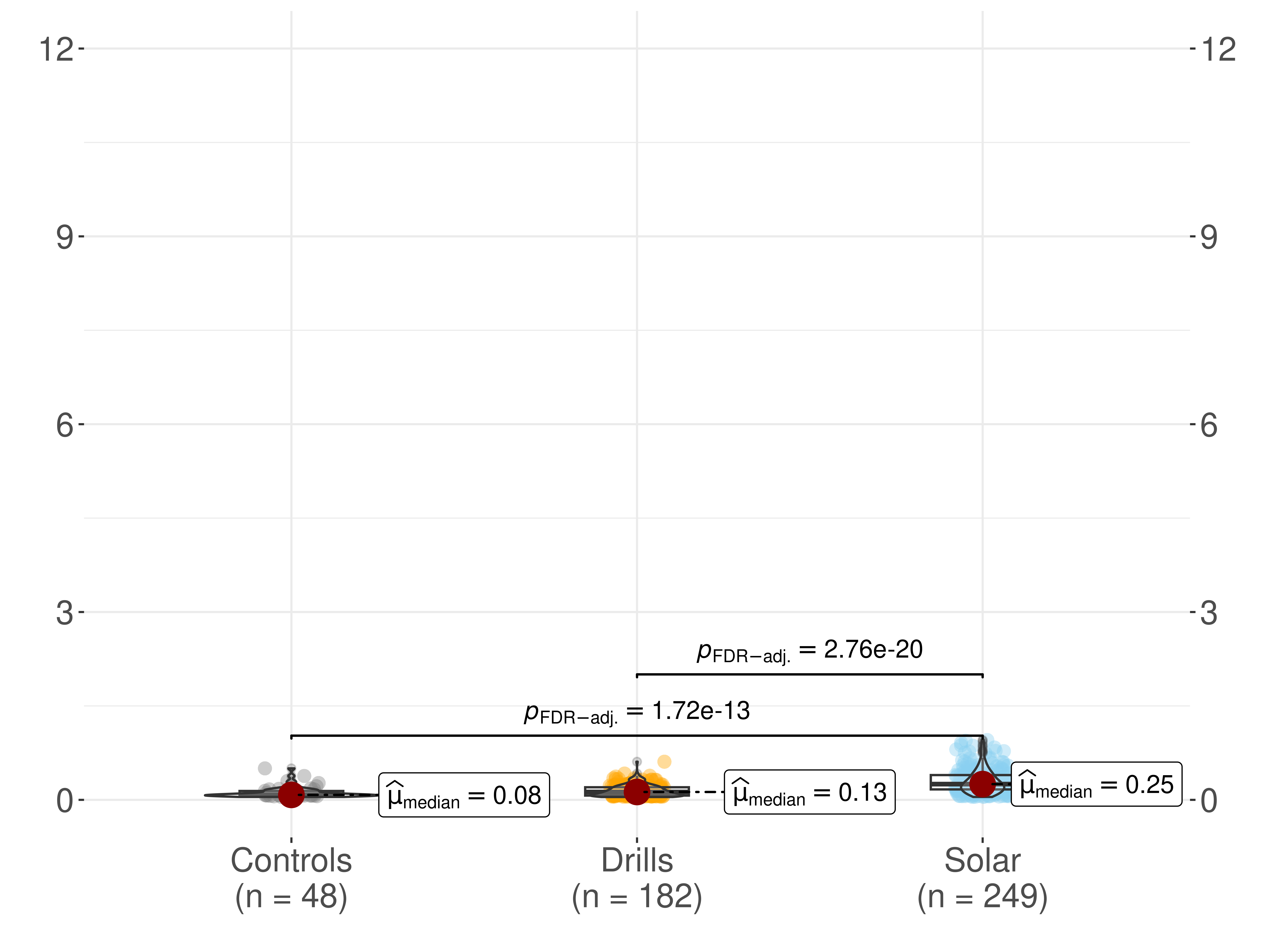}
        \hfill
        \caption{Audiences in the 18-24 age group}
    \end{subfigure}
    \hfill
    \begin{subfigure}[b]{\textwidth}
        \centering
        \includegraphics[width=\textwidth]{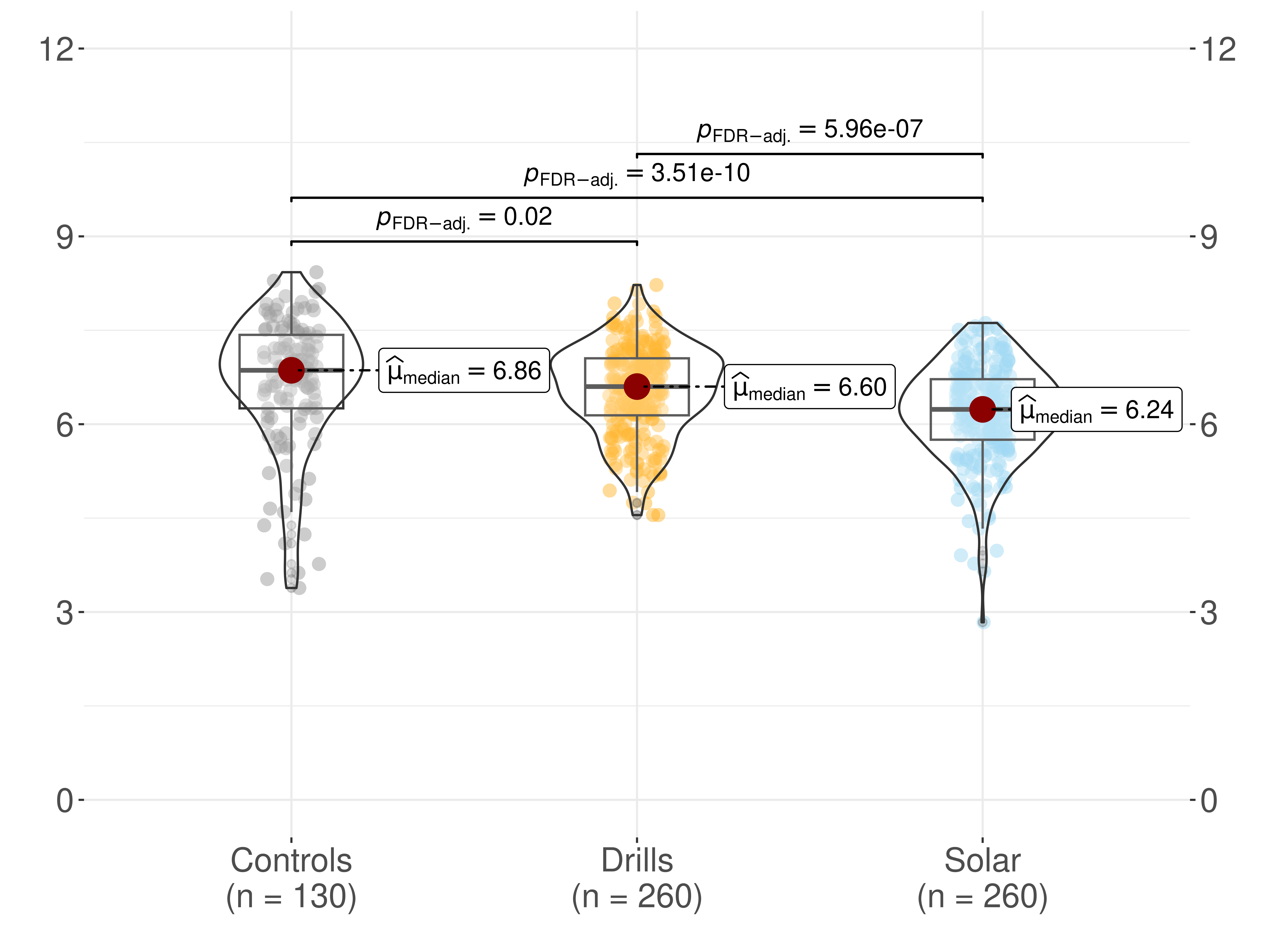}
        \hfill
        \caption{Audiences in the 65+ age group}
    \end{subfigure}
    \caption{$D^R$ samples of ads featuring oil \rig{s}, solar cells, and controls in younger and older audiences. Note that images of oil rigs are delivered preferentially to older audiences as compared to younger audiences. Similarly, images of solar cells are delivered preferentially to younger audiences as compared to older audiences. The mean $D^R$ further reveals that older audiences are over-represented, and younger populations are under-represented, as compared to their composition in the Facebook population.}
    \label{fig:young-old-exp-comp}
    \end{framed}
\end{figure}
\paragraph{Age} The $D^R$ samples of ads featuring solar cells, controls, or oil \rig{s} (Fig \ref{fig:young-old-exp-comp}) are significantly different (p < 0.05, N=650, k=3) in audiences belonging to all age groups (18-24, 25-34, 35-44, 45-54, 55-64, 65+). Upon investigating the pairwise differences, we find that except for audiences in the ages of 45-54, the $D^R$ samples of solar cells and oil \rig{s} are significantly different in all age groups (p < 0.05, N=520, k=2). Ads featuring oil \rig{s} are preferentially delivered to older audiences while ads featuring solar cells are preferentially delivered to younger audiences. See table \ref{tab:rq1-age} for the Kruskall-Wallis H Statistic and p values. \noindent Additional sub-analyses can be found in Appendix \ref{appendix:additionalAnalysisRQ1Region}

%% file: part2-3-5-RQ2.tex
\subsubsection{RQ2: Does ad delivery ratio, $D^R$, differ when logos are present on an ad image?} 
We compare $D^R$ samples of ads featuring images with a contrarian, advocacy, and no logos.
\subsubsubsection{RQ2a: Does $D^R$ differ based on logos present in an oil \rig{} images?} 
\paragraph{U.S. states} In 3 states (\textit{Connecticut},\textit{ New Mexico} and \textit{Idaho}), we find a significant difference (N=260, p < 0.05) in the $D^R$ samples of oil \rig{s} containing different types of logos (contrarian, advocacy, and none). Upon investigating the pairwise differences, we find that:
\begin{itemize}
    \item In \textit{New Mexico}, $D^R$ samples of oil \rig{s} with a contrarian logo are significantly different from those with an advocacy logo (N=130, p < 0.05).
    \item In \textit{New Mexico}, \textit{Connecticut}, and \textit{Idaho}, $D^R$ samples of oil \rig{s} with a contrarian logo are significantly different from those with no logo (N=195, p < 0.05).
    \item In \textit{Connecticut}, \textit{New Mexico} and \textit{Idaho}, $D^R$ samples of oil \rig{s} with an advocacy logo are significantly different from those with no logo (N=195, p < 0.05).
\end{itemize}
Values from our analyses are provided in table \ref{tab:rq2-region}.

\paragraph{Gender} $D^R$ samples of oil \rig{s} with different types of logos (contrarian, advocacy, none) are not significantly different in audiences of different genders. Values from our analyses are provided in table \ref{tab:rq2-gender}.

\paragraph{Age} $D^R$ samples of oil \rig{s} with different types of logos (contrarian, advocacy, none) are not significantly different in audiences of different ages. Values from our analyses are provided in table \ref{tab:rq2-age}.

\subsubsubsection{RQ2b: Does the ad delivery ratio, $D^R$ differ based on logos present in a solar cell image?} 
\paragraph{U.S. states} In 6 states (\textit{California}, \textit{Michigan}, \textit{Nevada}, \textit{New York}, \textit{Montana}, \textit{South Dakota}), there is a significant difference (N=260, p < 0.05, k=3) in the $D^R$ samples of solar cells containing different types of logos (contrarian, advocacy, and none). Upon investigating the pairwise differences, we find that:
Upon investigating the pairwise differences, we find that:
\begin{itemize}
    \item In \textit{California} and \textit{New York}, $D^R$ samples of solar cells with an advocacy logo are significantly different(N=130, p < 0.05) from those with a contrarian logo.
    \item In \textit{Michigan} and \textit{South Dakota}, $D^R$ samples of solar cells with an advocacy logo are significantly different (N=195, p < 0.05) from those with no logo. In these states, we further find significant difference (N=195, p < 0.05) between $D^R$ samples of solar cells with a contrarian logo and those with no logo.
\end{itemize}

\paragraph{Gender} $D^R$ samples of solar cells do not significantly differ based on logo, in audiences of different genders.

\paragraph{Age} $D^R$ samples of solar cells do not significantly differ based on logo in nearly all age-based audience destinations. However, in audiences in the age groups, \textit{25-44} and \textit{65+}, we find a significant difference in the $D^R$ samples of solar cells with a contrarian logo and those with an advocacy logo. In audiences in the age-groups, \textit{18-34} and \textit{65+}, the $D^R$ samples of solar cells with no logos are significantly different from those with an advocacy logo.

%% file: part2-3-6-RQ3.tex
\subsubsection{RQ3: Can observed ad delivery be consistently attributed to the ad image?}
We test if the observed ad delivery can be consistently attributed to the ad image, by duplicating the ads featuring images without logos. This includes images of solar cells, oil rigs, and controls without logos. Since statistical tests that compare distributions of categorical variables rely on count values, we compare the `Reach' of an ad and its copy among various ad destinations (U.S. states, gender, age). 
\paragraph{U.S states} Audiences in many U.S. states receive 0 views of some ads, and several states receive < 5 views. To satisfy the assumptions of the Fisher's Test, we first group states based on Facebook's population estimates (See table \ref{tab:fb-estimated-state}). States that are expected to receive close to 0\%, 1\%, or 2\% of the ad are grouped together (and their reach counts are summed), while states expected to receive greater than 2\% of the ad are retained as is. This gives us 14 possible state destinations where an ad can be distributed. We use Fisher's Test to compare these observed delivery samples of an ad with its copy. In 89.7\% of the ads, we find that the observed delivery sample of an ad is not significantly different from that of its duplicate (N = 195, p < 0.05). The exact values from our analysis are present in \ref{tab:rq3-state-consistency}.

\paragraph{Gender} Among gender-based destinations, we find that in 100\% of the ads, the observed delivery sample of an ad and its duplicate are not significantly different (N = 195, p < 0.05). The exact values from our analysis are present in \ref{tab:rq3-gender-consistency}.

\paragraph{Age} Among age-based destinations, we find that in 99\% of the ads, the observed delivery sample of an ad and its duplicate are not significantly different (N = 195, p < 0.05). The exact values from our analysis are present in \ref{tab:rq3-age-consistency}.

%% file: part2-3-7-RQ4.tex
\subsubsection{RQ4: Is the observed ad delivery proportional to Facebook’s population estimates within U.S. state, age, and gender-based ad destinations?}
\paragraph{U.S. states} In 47\% of all ads, the observed delivery matches Facebook's population estimates(N=650, p < 0.05; See table \ref{tab:fb-estimated-state} for population estimates by state.). The observed delivery of 64\% of controls (N=130, p < 0.05), and 42.5\% (N=520, p < 0.05) of non-control images (solar cells or oil \rig{s}) matches Facebook's population estimates. The exact H-statistics and p-values are provided in tables \ref{tab:rq4-state-Batch1}, \ref{tab:rq4-state-Batch2}, and \ref{tab:rq4-state-Batch3}

\paragraph{Gender} The observed delivery of 28\% of non-control images (oil \rig{s} or solar cells, N = 520, p < 0.05) and 54\% of control images (N=130, p < 0.05) matches Facebook's population estimates (See table \ref{tab:fb-estimated-gender} for population estimates by gender). The exact H-statistics and p-values are provided in tables \ref{tab:rq4-gender-Batch1}, \ref{tab:rq4-gender-Batch2}, and \ref{tab:rq4-gender-Batch3}

\paragraph{Age} The observed delivery of none of the non-control images (N=520, p < 0.05) and none of the control images (N=130, p < 0.05) matches Facebook's population estimates (See table \ref{tab:fb-estimated-age} for population estimates by age). The exact H-statistics and p-values are provided in tables \ref{tab:rq4-age-Batch1}, \ref{tab:rq4-age-Batch2}, and \ref{tab:rq4-age-Batch3}

%% file: discussion.tex
\section{Discussion}
We show experimentally that within gender, age, and location based ad destinations,  climate ads featuring different content are delivered differently. This suggests that climate advertising is vulnerable to algorithmic decision-making. Further, ad content consistently influences ad delivery, with nearly 90\% of ad pairs featuring the same image having statistically similar delivery. Further, delivery decisions made by Facebook's Algorithmic Decision System (ADS) are not proportional to Facebook's ad audience estimates for gender, age, and U.S. state based ad destinations and could misguide advertisers, suggesting that delivery skew is a feature built into the algorithmic system. While, we do not verify this in our experiment, we note that past research has determined that delivery decisions are largely driven by automated classifications considered by the ADS and not due to interactions of the ad audiences with the ad. Startlingly, ads that appear invisible to a human (but visible to an automatic image classifier system) are delivered similarly to ads that are fully visible to humans, by Facebook \cite{ali2019discrimination}.\\
Our ad experiments control for the spend made on an ad by setting a budget of \$1 on each ad. Our results indicating preferential delivery, therefore, also indicate preferential pricing. It is `cheaper' to advertise images of oil \rig{s} to males and older audiences and images of solar cells to females and younger audiences. Advocacy advertising is decentralized among 482 advertisers associated with 45 advocacy organizations. These organizations are also cash strapped, with 15\% of advocacy ads requesting for donations or subscriptions. Preferential pricing could therefore adversely impact the advertising strategy employed by advocacy organizations. Ad budget has also been shown to have an impact on algorithmic decision making \cite{ali2019discrimination}, with lower budget ads being preferentially delivered to men. Based on our pilot experiments, where the ad budget was set to \$5 instead of \$1, we speculate that increasing the ad budget may further increase the divide between the delivery ratio of contrarian and advocacy ads. However, the role of ad budgets on ad delivery is difficult to test experimentally given the number of logistical constraints on the platform side, and the budgetary constraints when pursuing audits.\\
Our results also highlight how advertising algorithms may impact the consumption of climate communications by audiences that have different psychological, cultural, and political reasons for their response to the climate crisis. The Six Americas Report\cite{leiserowitz2011global} segments the U.S. population into six groups based on their response -- Alarmed, Concerned, Cautious, Disengaged, Doubtful and Dismissive. This order is decreasing in the magnitude of their concern towards the climate crisis. Communication studies have noted that these groups require different persuasion strategies, and information channels, for climate engagement \cite{roser2015engaging}. For example, audiences in the Doubtful and Dismissive category are best engaged by adopting non-confrontational approaches, and by framing messages in ways that are consistent with their values. Past research has shown that directly challenging the beliefs of these groups is likely to trigger counter argumentation rather than persuasion, suggesting that pro-attitudinal messaging is a better advertising strategy than counter-attitudinal messaging \cite{roser2015engaging}. Audiences in these groups are also more likely to be older individuals and male and located in the interior regions of the U.S. \cite{leiserowitz2023global, marlon2020yale} -- demographics and regions where the algorithm preferentially delivers ads featuring pictures of an oil rig, and likely enables contrarian content as found in the analysis of non-targeted ads in our observational study. Advocacy ads featuring images of oil rigs are more likely to use these images to dissuade audiences from fossil fuels. Algorithms direct these ads at male and older audiences who are more likely to be in segments that are Doubtful or Dismissive about the climate crisis, and may thus feel more alienated from the climate action cause. Similarly, contrarian ads featuring renewable energy sources such as solar cells may sometimes be used to show the advertiser as being sustainable, a practice called greenwashing. Research is divided on whether these ads lead individuals to view actors as being more \cite{sengupta2022nytgreenwashing} or less sustainable \cite{menno2018makinggreen}. Therefore the implications of the algorithm's recommendation of ads featuring solar cells to female and younger audiences (or to the 3 states (California, Kentucky and New York) where solar cell ads were delivered preferentially) is unclear. However, our experiment shows that these audiences may be more vulnerable to greenwashing ads, and therefore need to be inoculated more frequently against the practice. \\
The Six Americas report also highlights that liberal audiences are more likely to be in the Alarmed or Concerned segment than conservative audiences. Our finding that non-targeted  contrarian ads are delivered to states that are likely to vote Republican, and non-targeted advocacy ads are delivered to states likely to vote Democrat is concerning, and suggests that the algorithm may be complicating the process of convincing individuals about the harms of climate change. The Six Americas report provides granular information on the six segments that future work can explore to further understand the racial and political implications of algorithmic bias. Overall, we note that social media platforms are a set of new and constantly evolving actors who play an influential role in the climate discourse ecosystem. It is important for social media and disinformation scholars to not just study the proliferation of information on these platforms but to also account for the delivery patterns of the information on these platforms, in order to engage diverse audiences towards climate action.

%% file: consent.tex
\section{Consent and Ethics}
Our research was conducted with an exemption from MIT COUHES, with application ID E-4191. Ad experiments were run using images of objects that had already been used in ads by both the contrarian and advocacy actor whose logos are featured. Ads were launched in full accordance with Facebook's Ad Review policies. Further, our experiment launched ads featuring only images and no text to abstain from propagating climate disinformation.

%% file: acknowledgement.tex
\section{Acknowledgements}
The authors would like to acknowledge funding and support from grant XXXXX. We thank Ben Lipkin, Matt Groh, Hidefuse Okabe, and especially Katya Arquilla for their guidance on statistical inference. We thank Piotr Sapiezynski for guidance on the experimental study. We thank Geoffrey Supran and Cameron Hickey for aggregating and sharing the list of contrarian and advocate actors from peer-reviewed research, from which we retrieved the set of actors considered in this work. We thank XXX for feedback on initial versions of this manuscript. Finally, we thank members of the Climate Disinformation Research Coalition for insightful discussions. We are also grateful to Shreyas Satish, Ajith Ranka, and Surya Narreddi at \href{https://www.ownpath.com/}{OwnPath} for helping us build an interface to explore the climate discourse on Facebook.

%% file: appendix.tex
\clearpage
\section{Appendix}
\subsection{Dataset}\label{appendix:dataset}
The full observational and experimental datasets are available on \href{https://www.dropbox.com/home/climate-data}{Dropbox}. 
\subsection{Various Impression Classes on the Ad Archive} \label{appendix:impClassesAll}
0\_999, 1000\_1999, 2000\_2999, 3000\_3999, 4000\_4999, 5000\_5999, 6000\_6999, 7000\_7999, 8000\_8999, 9000\_9999, 10000\_14999, 15000\_19999, 20000\_24999, 25000\_29999, 30000\_34999, 35000\_39999, 40000\_44999, 45000\_49999, 50000\_59999, 60000\_69999, 70000\_79999, 80000\_89999, 90000\_99999, 100000\_124999, 125000\_149999, 150000\_174999, 175000\_199999, 200000\_249999, 250000\_299999, 300000\_349999, 350000\_399999, 400000\_449999, 450000\_499999, 500000\_599999, 600000\_699999, 700000\_799999, 800000\_899999, 900000\_999999, 1000000\_1000000
\subsection{Metadata}\label{appendix:metadata}
The metadata that is most relevant to our analysis and work are the following:
\begin{itemize}[noitemsep]
    \item \texttt{ad\_reached\_countries} - Facebook delivered the ads in these countries. We use this attribute to filter advertisements that were only shown in the United States.
    \item \texttt{delivery\_by\_region} - A state-wise breakdown of the ad delivery percentage.
    \item \texttt{demographic\_distribution} - A gender and age wise breakdown of the ad delivery percentage.
    \item \texttt{impressions} - A range representing the minimum/maximum number of non-unique Facebook accounts that were shown an ad. The smallest bin represents ads that were shown to between 0 - 999 Facebook accounts and the largest bin contains ads that were shown to $>$ 1M Facebook accounts.
    \item \texttt{spend} - A range representing the minimum/maximum amount that was spent on an ad. The smallest bin represents ads whose expenditure was between \$0 - \$100 and the largest bin represents ads whose expenditure was $>$\$1M.
\end{itemize}

\subsection{Ad Attributes}\label{appendix:adAttributes}
The ad archive associates the following attributes with each ad:
\begin{enumerate}
    \item A unique identifier for the ad
    \item The time that the ad was created
    \item The time at which the ad began running
    \item The time at which the ad stopped running
    \item A unique URL that points to the exact ad
    \item The currency used to pay for the ad
    \item The estimated size of the Facebook account population from which user accounts were sampled to be shown the ad
    \item The budget class for the ad, platforms on which the ad was shown
    \item The impressions class\footnote{An ad impression is a non-unique view that was received by an ad. For example, if an ad was shown to 10 unique Facebook accounts, such that 2 unique accounts were shown the ad 4 times, the total number of impressions received by the ad would be 16 (8 + 2 * 4). Data from the Facebook ad library provides the impressions \textit{class} attribute for each ad, i.e the lower and upper bound of the impressions that would have been received by the ad.} for the ad.
    \item Funding information for the ad
    \item Delivery information for the ad across multiple demographics
\end{enumerate}

\subsection{Methods: Observational Study}\label{appendix:part1-methods}
We analyze the dataset to provide insights into the destinations reached by climate contrarian and advocacy advertising across various location, gender, and age demographics in the United States. 
Studying the delivery information for various ads alongside the associated targeting intent would enable accurate disentanglement of the algorithm's role from the advertiser's role in causing delivery. However, as stated previously, the Facebook Ad Library only provides the delivery information associated with each ad. The data itself does suggest certain proxies for targeting, which are used to draw insights about the entanglement between targeting and delivery. If the delivery volumes are significantly different, we further analyze if contrarian or advocacy advertisements dominate in different destinations. To do this:
\begin{outline}
    \1 For each possible location, gender, or age category, and for each possible category of impression volumes defined above, it is verified if the distribution of advocacy and contrarian ads consists of at least 30 samples.
        \2 If the distribution contains $>=$ 30 samples, it is first verified if the sample distributions are homoscedastic using Levene's Test. A 2 sample Welch's T-test is then used to investigate if the distribution of contrarian ads is significantly different from the distribution of advocacy ads, for each location, gender, and age category. For example, when analysing the difference in delivery volumes for various location destinations, this analysis would be conducted for each U.S. state, and reported at the state level.
        \2 If the distribution contains $<$ 30 samples, a 2 sample Mann-Whitney U Test is used to investigate if the distribution of contrarian ads is significantly different from the distribution of advocacy ads, for each location, gender, and age category, after establishing the homoscedasticity assumption of the sample distributions using a Levene's test.
    \1 If the delivery volume distributions of contrarian and advocacy ads are found to be significantly different (p < 0.05) using the Welch's T-test or the Mann-Whitney U Test, the means or medians of the sample distributions (depending on whether the data was normal or not) are used to further determine if contrarian or advocacy ads dominated in the location, age, or gender category being considered.
    \1 For advertisements delivered to a single location, or gender or age type, normalized advertisement counts are compared for contrarians and advocates.
\end{outline}

\subsection{Tabular Results: Observational Study}
\input{appendix-obs-region-tables}
\input{appendix-obs-gender-tables}
\input{appendix-obs-age-tables}

\subsection{Ad Campaign Attributes}\label{appendix:ad-camp-attr}
We briefly describe the attributes that were used for our ad campaigns.
\begin{itemize}[noitemsep]
    \item \textit{Duration} - The 652 ad campaigns were run in 3 batches, such that each batch was run for a period of 24h in order to reach all timezones of the U.S.\footnote{The 652 ads were run in 3 batches since Facebook has an upper limit of 250 concurrent ads that can be run by an advertiser whose advertising budget is less than \$1,000,000/month. Batch 1 (22 ads per campaign) was run from X to Y on Z. Batch 2 (22 ads per campaign) was run from X to Y on Z. Batch 3 (21 ads per campaign) was run from X to Y on Z. Since the ads are run simultaneously and run for a time period that spans all the timezones in the U.S, we minimize any market effects  to the extent possible.} 
    \item \textit{Ad media} We use images of oil rigs, solar cells or controls (Fig. \ref{fig:exp-design-images}). Each image was modified with the logo of a contrarian or advocacy organization, depending on the treatment group it was assigned to.
    \item \textit{Ad text}  For each ad, we included the text, ``Use our website to tell us what you think about this picture." 
    \item \textit{Desired audience attributes} - The ads were scheduled to be delivered to anyone in the United States who belonged to the default age criteria on Facebook, irrespective of gender and location. We did not use any additional micro-targeting features.
    \item \textit{Ad placement} - We specified that the ads could only be shown on the Facebook platform, and could only be situated on a user account's Facebook feed.
    \item \textit{Ad budget} - We specified a daily ad budget of \$1/day.
    \item \textit{Campaign Objective} - We specified that the ads'  objective was to maximize audience traffic to the \href{htps://imagesweseepage.wordpress.com/}{website}. This website collected opinions about the ad images, when shared by a visitor. It did not contain content that revealed the intentions of our experiment, or a stance on climate action or climate change.
    \item \textit{Ad type} - We ran the ads under the `Social issues, elections or political issues' category, in accordance with Facebook's advertising guidelines.
\end{itemize}

\subsection{Experiment: Sampling control images} \label{appendix:sampling-control}
To sample control images, we utilize the ImageNet-21k dataset \cite{ridnik2021imagenet21k} and the WordNet \cite{fellbaum2010wordnet} hierarchy. The ImageNet-21k dataset contains images grouped under 21,841 classes; WordNet is a large lexical database of English. In WordNet, nouns, verbs, adjectives and adverbs are grouped into sets of cognitive synonyms (synsets)\cite{fellbaum2010wordnet}, each expressing a distinct concept. The 21,841 labels in the ImageNet-21k dataset are a direct mapping to the noun synsets in Wordnet. We devise a methodology to randomly sample diverse ImageNet categories, such that a sampled category contains at least one image of width and height greater than 600px, which is a criteria required by Facebook's Ad Platform.
        
\paragraph{ImageNet Labels Tree} We begin by constructing the WordNet tree for all the labels (synsets) in the ImageNet-21K dataset. The root of this tree is the synset, ``entity"\cite{fellbaum2010wordnet}, level 1 of this tree contains nodes that are descendants of the ``entity" node, level 2 contains descendants of nodes in level 1 and so on. \\
        
We then devise a methodology to randomly sample 300 different terminal nodes of this tree, such that these nodes are not related to each other, and the ImageNet category associated with the node contains images of width and height greater than 600px. We found, empirically, that it was necessary to sample roughly 4x the number of images we needed, in order to gather images that satisfied the Facebook Ad Platform's size criteria. To select 65 control ad images, we therefore sampled 300 categories. \\
        
In order to gather diverse images, we started at a tree level that has $>$ 300 nodes. Level 6 of the tree is the highest level to have > 300 nodes at 1188. We begin by sampling a random category on level 6 of the WordNet tree. For each category sampled on level 6, we sample a random sub category on the subsequent level, repeating this process until we sample a category that has no descendants. We repeat this process 300 times, to sample 300 unique categories from the 21,841 synsets. From each selected category, we sample a random image having at least 600px width and 600px height to satisfy Facebook criteria for ad images; only 103 categories satisfy this condition. We randomly sample 65 categories from this filtered set to get 65 control images. \\
        
We use the \textit{random} library on python for all our sampling needs.

\subsection{Additional Analyses: RQ1, Region axis} \label{appendix:additionalAnalysisRQ1Region}
When we exclude the images containing logos from the groups considered above, the Kruskal-Wallis test finds that in 35 states there is a statistically significant difference (p < 0.05) between the $D^R$ of ads featuring solar cells, oil fields, and controls. Within each state, we then investigate the pairwise differences between the 3 image groups using a Mann Whitney U Test with a Bonferroni correction, and find that in 17 states, there's a statistically significant difference (p < 0.05) between the delivery of images showing solar cells and oil fields. In 24 states, there's a statistically significant difference (p < 0.05) between the $D^R$ of solar cell images and the control images and in 11 states there's a statistically significant difference (p < 0.05) between the $D^R$ of oil field images and the control images. 

When we exclude the control images from our omnibus test, and directly investigate if there's a difference between the solar cell images and the oil field images in different states in the U.S, we actually see that in 29 states, where there is a significant difference between the $D^R$ of solar cells and oil fields.

\input{appendix-rq1}
\input{appendix-rq2}
\input{appendix-rq3}
\input{appendix-rq4}

\subsection{Facebook Estimated Audience Size Estimates}
\begin{table*}[]
    \centering
    \begin{tabular}{|c|c|c|}
    \hline
State & Estimated Audience Size (Lower Bound) & Estimated Audience Size (Upper Bound) \\
\hline
\hline
Alabama & 3400000 & 4000000 \\
Alaska & 530500 & 624100 \\
Arizona & 4900000 & 5800000 \\
Arkansas & 2000000 & 2400000 \\
California & 27200000 & 32000000 \\
Colorado & 3800000 & 4400000 \\
Connecticut & 2400000 & 2800000 \\
Delaware & 644900 & 758700 \\
Florida & 16500000 & 19400000 \\
Georgia & 7500000 & 8900000 \\
Hawaii & 978900 & 1200000 \\
Idaho & 1200000 & 1400000 \\
Illinois & 8200000 & 9600000 \\
Indiana & 4400000 & 5200000 \\
Iowa & 2000000 & 2300000 \\
Kansas & 1900000 & 2300000 \\
Kentucky & 3000000 & 3500000 \\
Louisiana & 3200000 & 3700000 \\
Maine & 889400 & 1000000 \\
Maryland & 4100000 & 4800000 \\
Massachusetts & 4700000 & 5600000 \\
Michigan & 6300000 & 7500000 \\
Minnesota & 3400000 & 4100000 \\
Mississippi & 1900000 & 2300000 \\
Missouri & 3900000 & 4600000 \\
Montana & 656300 & 772100 \\
Nebraska & 1200000 & 1500000 \\
Nevada & 2300000 & 2700000 \\
New Hampshire & 895800 & 1100000 \\
New Jersey & 6200000 & 7300000 \\
New Mexico & 1200000 & 1400000 \\
New York & 13800000 & 16300000 \\
North Carolina & 7300000 & 8600000 \\
North Dakota & 493100 & 580100 \\
Ohio & 7500000 & 8800000 \\
Oklahoma & 2700000 & 3200000 \\
Oregon & 2700000 & 3200000 \\
Pennsylvania & 8000000 & 9400000 \\
Rhode Island & 761100 & 895400 \\
South Carolina & 3500000 & 4200000 \\
South Dakota & 548100 & 644900 \\
Tennessee & 4800000 & 5600000 \\
Texas & 21300000 & 25100000 \\
Utah & 2200000 & 2500000 \\
Vermont & 392800 & 462100 \\
Virginia & 5800000 & 6900000 \\
Washington D. C. & 694200 & 816700 \\
Washington & 4800000 & 5700000 \\
West Virginia & 1100000 & 1300000 \\
Wisconsin & 3700000 & 4300000 \\
Wyoming & 350300 & 412100 \\
\hline
    \end{tabular}
    \caption{Estimated Facebook ad audience size estimates for different U.S. states}
    \label{tab:fb-estimated-state}
\end{table*}

\begin{table*}[]
    \centering
    \begin{tabular}{|c|c|c|}
    \hline
    Gender & Estimated Audience Size (Lower Bound) & Estimated Audience Size (Upper Bound) \\
    \hline
    \hline
    male & 100725600 & 118361400 \\
    female & 121112000 & 142313900 \\
    unknown & 1997800 & 3190900 \\
    \hline
    \end{tabular}
    \caption{Estimated Facebook ad audience size estimates for different genders}
    \label{tab:fb-estimated-gender}
\end{table*}

\begin{table*}[b]
    \centering
    \begin{tabular}{|c|c|c|}
    \hline
    Ages & Estimated Audience Size (Lower Bound) & Estimated Audience Size (Upper Bound) \\
    \hline
    \hline
    18-24 & 42538900 & 50135000 \\
25-34 & 55250800 & 65054000 \\
35-44 & 42453300 & 50032600 \\
45-54 & 31922400 & 37540100 \\
55-64 & 26205100 & 30632600 \\
65+ & 26166800 & 30603100 \\
\hline
    \end{tabular}
    \caption{Estimated Facebook ad audience size estimates for different age groups}
    \label{tab:fb-estimated-age}
\end{table*}

%% file: appendix-obs-region-tables.tex
\subsubsection{All Ads: Region}
T values from statistical analyses for all ads in the dataset are presented in Tables \ref{tab:obs_table_region_1K_all}, \ref{tab:obs_table_region_10K_all}, \ref{tab:obs_table_region_100K_all},  \ref{tab:obs_table_region_1M_all}, and \ref{tab:obs_table_region_1M+_all}.
\begin{table*}[b]
    \centering
\begin{tabular}{|c|c|c|c|c|c|}
    \hline
     U.S. State & t Value & p Value & $N_{\text{contrarians}}$ & $N_{\text{advocates}}$ \\
     \hline
         \hline
Alabama & 13.2836346869786 & \hl{3.739807955784749e-39} & 2952 & 17488 \\
Alaska & 17.240934066892773 & \hl{5.068944745356406e-63} & 2423 & 14916 \\
Arizona & 16.851408808758244 & \hl{7.3456807716259915e-62} & 4230 & 23737 \\
Arkansas & 4.4920164462999645 & \hl{7.2325244104419225e-06} & 2780 & 17127 \\
California & -1.9379163824345829 & 0.05269482670250833 & 4007 & 25750 \\
Colorado & 19.86519910871058 & \hl{1.496703520596214e-84} & 4466 & 24025 \\
Connecticut & 3.958532532578084 & \hl{7.73765850425023e-05} & 2603 & 22342 \\
\hline
    \end{tabular}
    \caption{Observational Study, All Ads: t Values and associated p values for each state destination for ads receiving < 1000 impressions.}
    \label{tab:obs_table_region_1K_all}
\end{table*}
\begin{table*}[b]
    \centering
    \begin{tabular}{|c|c|c|c|c|c|}
    \hline
     U.S. State & t Value & p Value & $N_{\text{contrarians}}$ & $N_{\text{advocates}}$ \\
     \hline
         \hline
\hline
Alabama & 16.149688886981693 & 
\hl{3.2869252201135587e-57} & 4898 & 25557 \\
Alaska & 20.244076369200148 & \hl{8.787456224905471e-88} & 4992 & 25946 \\
Arizona & 7.653594202514507 & \hl{2.193838227984813e-14} & 6165 & 28543 \\
Arkansas & 8.590132248046663 & \hl{1.097462654713791e-17} & 4704 & 26187 \\
California & 4.054436799000379 & \hl{5.073221406324565e-05} & 6717 & 29802 \\
Colorado & 18.576033047586158 & \hl{1.911575665052344e-75} & 6814 & 27954 \\
Connecticut & -1.7339992448694264 & 0.08297898521951685 & 4551 & 27126 \\
\hline
    \end{tabular}
    \caption{Observational Study, All Ads:: t Values and associated p values for each state destination for ads receiving 1K - 10K impressions.}
    \label{tab:obs_table_region_10K_all}
\end{table*}
\begin{table*}[b]
    \centering
    \begin{tabular}{|c|c|c|c|c|c|}
    \hline
     U.S. State & t Value & p Value & $N_{\text{contrarians}}$ & $N_{\text{advocates}}$ \\
     \hline
         \hline
\hline
Alabama & 13.631356101805594 & \hl{2.347104423550024e-41} & 3740 & 15752 \\
Alaska & 18.125965860626245 & \hl{7.627383569721037e-71} & 4064 & 16171 \\
Arizona & -2.2439924895066565 & \hl{0.024862845327724682} & 4476 & 17576 \\
Arkansas & 9.460396888257277 & \hl{4.772452906054855e-21} & 3656 & 16082 \\
California & 3.1536715238791686 & \hl{0.001620360076553684} & 4819 & 18148 \\
Colorado & 17.687716562334458 & \hl{2.5111479870299496e-68} & 5085 & 17078 \\
Connecticut & -2.35916908380124 & \hl{0.01835959494765869} & 3549 & 16612 \\
\hline
    \end{tabular}
    \caption{Observational Study, All Ads: t Values and associated p values for each state destination for ads receiving 10K - 100K impressions.}
    \label{tab:obs_table_region_100K_all}
\end{table*}
\begin{table*}[b]
    \centering
    \begin{tabular}{|c|c|c|c|c|c|}
    \hline
     U.S. State & t Value & p Value & $N_{\text{contrarians}}$ & $N_{\text{advocates}}$ \\
     \hline
         \hline
\hline
Alabama & 8.178159375268372 & \hl{5.585897941052643e-16} & 1544 & 4164 \\
Alaska & 10.662073608617675 & \hl{9.514027133296105e-26} & 1646 & 4246 \\
Arizona & -6.846803505249799 & \hl{8.455106947322627e-12} & 1796 & 4711 \\
Arkansas & 7.684197652096507 & \hl{2.6285203870840744e-14} & 1553 & 4269 \\
California & -7.984971760742617 & \hl{2.0719962022089683e-15} & 1803 & 4683 \\
Colorado & 12.61434735726191 & \hl{2.2639615207653574e-35} & 1964 & 4443 \\
Connecticut & -3.4002155243892975 & \hl{0.0006833492969778226} & 1543 & 4379 \\
\hline
    \end{tabular}
    \caption{Observational Study, All Ads: t Values and associated p values for each state destination for ads receiving 100K - 1M impressions.}
    \label{tab:obs_table_region_1M_all}
\end{table*}
\begin{table*}[b]
    \centering
    \begin{tabular}{|c|c|c|c|c|c|}
    \hline
     U.S. State & t Value & p Value & $N_{\text{contrarians}}$ & $N_{\text{advocates}}$ \\
     \hline
        \hline
Alabama & 3.3313525467526834 & \hl{0.00098231397505672} & 271 & 277 \\
Alaska & -0.01986037893413233 & 0.9841624410937129 & 252 & 272 \\
Arizona & -3.7628159404199204 & \hl{0.00019621104167365276} & 275 & 299 \\
Arkansas & 11.504004748737968 & \hl{4.418094663022708e-27} & 265 & 278 \\
California & -10.159202442023634 & \hl{3.0941416832997245e-22} & 283 & 296 \\
Colorado & 3.4419886079299213 & \hl{0.0006425979913303564} & 291 & 285 \\
Connecticut & -13.56397839515139 & \hl{4.6737615575806076e-36} & 262 & 287 \\
\hline
    \end{tabular}
    \caption{Observational Study, All Ads: t Values and associated p values for each state destination for ads receiving 1M+ impressions.}
    \label{tab:obs_table_region_1M+_all}
\end{table*}

\subsubsection{Targeted Ads: Region}
Mean delivery percentage values from statistical analyses for targeted ads in the dataset are presented in Tables \ref{tab:obs_table_region_1K_t}, \ref{tab:obs_table_region_10K_t}, \ref{tab:obs_table_region_100K_t},  \ref{tab:obs_table_region_1M_t}, and \ref{tab:obs_table_region_1M+_t}.
\begin{table*}[b]
    \centering
\begin{tabular}{|c|c|c|c|c|c|}
\hline
U.S. State & Mean Delivery Percentage (Contrarians) & Mean Delivery Percentage (Advocates) & $N_{\text{contrarians}}$ & $N_{\text{advocates}}$ \\
\hline
\hline
Alabama & 0.0010833935218623257 & 0.0 & 39 & 0 \\
Alaska & 0.007722651258403245 & 0.002880097013794149 & 278 & 133 \\
Arizona & 0.009944996944274682 & 0.009831308603477772 & 358 & 454 \\
Arkansas & 0.0 & 0.0023820351241906493 & 0 & 110 \\
California & 0.013528529362742375 & 0.025682669611728274 & 487 & 1186 \\
Colorado & 0.018862159008833822 & 0.018146776673379675 & 679 & 838 \\
Connecticut & 0.0014167453747430414 & 0.0007795751315533034 & 51 & 36 \\
\hline
\end{tabular}
    \caption{Observational Study, Targeted Ads: Mean delivery percentages in each state destination for ads receiving < 1K impressions.}
    \label{tab:obs_table_region_1K_t}
\end{table*}
\begin{table*}[b]
    \centering
\begin{tabular}{|c|c|c|c|c|c|}
\hline
U.S. State & Mean Delivery Percentage (Contrarians) & Mean Delivery Percentage (Advocates) & $N_{\text{contrarians}}$ & $N_{\text{advocates}}$ \\
\hline
\hline
Alabama & 0.0012500694483026835 & 8.661945906147817e-05 & 45 & 4 \\
Alaska & 0.011917328740485582 & 0.004634141059789081 & 429 & 214 \\
Arizona & 0.005944774709706095 & 0.015634812360596807 & 214 & 722 \\
Arkansas & 0.00019445524751375077 & 0.002035557287944737 & 7 & 94 \\
California & 0.02497360964498028 & 0.036618376318239894 & 899 & 1691 \\
Colorado & 0.022612367353741874 & 0.021113493146235302 & 814 & 975 \\
Connecticut & 0.0008333796322017889 & 0.0011693626973299552 & 30 & 54 \\
\hline
\end{tabular}
    \caption{Observational Study, Targeted Ads: Mean delivery percentages in each state destination for ads receiving 1K - 10K impressions.}
    \label{tab:obs_table_region_10K_t}
\end{table*}
\begin{table*}[b]
    \centering
\begin{tabular}{|c|c|c|c|c|c|}
\hline
U.S. State & Mean Delivery Percentage (Contrarians) & Mean Delivery Percentage (Advocates) & $N_{\text{contrarians}}$ & $N_{\text{advocates}}$ \\
\hline
\hline
Alabama & 0.0003055725318073226 & 0.0002165486476536954 & 11 & 10 \\
Alaska & 0.010056114228568254 & 0.003421468632928387 & 362 & 158 \\
Arizona & 0.002694594144119118 & 0.012148379133372312 & 97 & 561 \\
Arkansas & 0.00013889660536696484 & 0.0006496459429610862 & 5 & 30 \\
California & 0.015778654369687204 & 0.020420537473743477 & 568 & 943 \\
Colorado & 0.01713984110228346 & 0.011931830485718617 & 617 & 551 \\
Connecticut & 0.0006944830268348242 & 0.001039433508737738 & 25 & 48 \\
\hline
\end{tabular}
    \caption{Observational Study, Targeted Ads: Mean delivery percentages in each state destination for ads receiving 10K - 100K impressions.}
    \label{tab:obs_table_region_100K_t}
\end{table*}
\begin{table*}[b]
    \centering
\begin{tabular}{|c|c|c|c|c|c|}
\hline
U.S. State & Mean Delivery Percentage (Contrarians) & Mean Delivery Percentage (Advocates) & $N_{\text{contrarians}}$ & $N_{\text{advocates}}$ \\
\hline
\hline
Alabama & 0.0 & 0.0001082743238268477 & 0 & 5 \\
Alaska & 0.003305739207733763 & 0.00034647783624591267 & 119 & 16 \\
Arizona & 0.0005555864214678594 & 0.0037029818748781913 & 20 & 171 \\
Arkansas & 2.7779321073392966e-05 & 2.1654864765369542e-05 & 1 & 1 \\
California & 0.0022223456858714375 & 0.002923406743324888 & 80 & 135 \\
Colorado & 0.004083560197788766 & 0.002273760800363802 & 147 & 105 \\
Connecticut & 0.00013889660536696484 & 0.00015158405335758679 & 5 & 7 \\
\hline
\end{tabular}
    \caption{Observational Study, Targeted Ads: Mean delivery percentages in each state destination for ads receiving 100K - 1M impressions.}
    \label{tab:obs_table_region_1M_t}
\end{table*}
\begin{table*}[b]
    \centering
\begin{tabular}{|c|c|c|c|c|c|}
\hline
U.S. State & Mean Delivery Percentage (Contrarians) & Mean Delivery Percentage (Advocates) & $N_{\text{contrarians}}$ & $N_{\text{advocates}}$ \\
\hline
\hline
Alabama & 0.0 & 0.0 & 0 & 0 \\
Alaska & 2.7779321073392966e-05 & 0.0 & 1 & 0 \\
Arizona & 2.7779321073392966e-05 & 0.00019489378288832586 & 1 & 9 \\
Arkansas & 0.0 & 0.0 & 0 & 0 \\
California & 0.0 & 4.3309729530739084e-05 & 0 & 2 \\
Colorado & 0.00038891049502750154 & 4.3309729530739084e-05 & 14 & 2 \\
Connecticut & 0.0 & 0.0 & 0 & 0 \\
\hline
\end{tabular}
    \caption{Observational Study, Targeted Ads: Mean delivery percentages in each state destination for ads receiving 1M+ impressions.}
    \label{tab:obs_table_region_1M+_t}
\end{table*}

\subsubsection{Non-Targeted Ads: Region}
T values from statistical analyses for non-targeted ads in the dataset are presented in Tables \ref{tab:obs_table_region_1K_nt}, \ref{tab:obs_table_region_10K_nt}, \ref{tab:obs_table_region_100K_nt},  \ref{tab:obs_table_region_1M_nt}, and \ref{tab:obs_table_region_1M+_nt}.
\begin{table*}[b]
    \centering
\begin{tabular}{|c|c|c|c|c|c|}
\hline
U.S. State & t Value & p Value & $N_{\text{contrarians}}$ & $N_{\text{advocates}}$ \\
\hline
\hline
Alabama & 35.03893920911009 & \hl{2.4353810598518363e-151} & 577 & 3647 \\
Alaska & -1.379772159152534 & 0.16812285685575717 & 522 & 3403 \\
Arizona & -3.1424494668971303 & \hl{0.0017454002486166193} & 577 & 3713 \\
Arkansas & 29.35027253340735 & \hl{2.1632833514226475e-121} & 577 & 3618 \\
California & -38.81734957638038 & \hl{2.871806550854955e-190} & 533 & 3706 \\
Colorado & -1.8398447349627245 & 0.06628593470855605 & 574 & 3713 \\
Connecticut & -28.28811163695321 & \hl{1.884255515998055e-123} & 565 & 3709 \\
\hline
\end{tabular}
    \caption{Observational Study, Non-Targeted Ads: t Values and associated p values for each state destination for ads receiving < 1000 impressions.}
    \label{tab:obs_table_region_1K_nt}
\end{table*}
\begin{table*}[b]
    \centering
\begin{tabular}{|c|c|c|c|c|c|}
\hline
 U.S. State & t Value & p Value & $N_{\text{contrarians}}$ & $N_{\text{advocates}}$ \\
 \hline
 \hline
 Alabama & 56.40480876947256 & \hl{0.0} & 2864 & 15021 \\
Alaska & -6.702428681923706 & \hl{2.1131832296872146e-11} & 2787 & 14964 \\
Arizona & -6.867574291879497 & \hl{6.976119308152953e-12} & 2865 & 15166 \\
Arkansas & 67.60831541962507 & \hl{0.0} & 2864 & 15090 \\
California & -54.290808751783615 & \hl{0.0} & 2779 & 15140 \\
Colorado & -5.341363173639038 & \hl{9.826906877583027e-08} & 2855 & 15148 \\
Connecticut & -83.84132818688015 & \hl{0.0} & 2829 & 15165 \\
\hline
\end{tabular}
    \caption{Observational Study, Non-Targeted Ads: t Values and associated p values for each state destination for ads receiving 1K - 10K impressions.}
    \label{tab:obs_table_region_10K_nt}
\end{table*}
\begin{table*}[b]
    \centering
\begin{tabular}{|c|c|c|c|c|c|}
\hline
 U.S. State & t Value & p Value & $N_{\text{contrarians}}$ & $N_{\text{advocates}}$ \\
 \hline
 \hline
 Alabama & 29.337242287432815 & \hl{2.1657971941601314e-165} & 2746 & 9927 \\
Alaska & -5.42790809789668 & \hl{5.807387231283471e-08} & 2685 & 9927 \\
Arizona & -9.625441130612389 & \hl{7.505900836060769e-22} & 2746 & 9962 \\
Arkansas & 57.586758996324825 & \hl{0.0} & 2746 & 9954 \\
California & -50.036238993682986 & \hl{0.0} & 2744 & 9947 \\
Colorado & 1.6801365404771518 & 0.09303455161933309 & 2730 & 9931 \\
Connecticut & -83.83337069621916 & \hl{0.0} & 2744 & 9963 \\
\hline
\end{tabular}
    \caption{Observational Study, Non-Targeted Ads: t Values and associated p values for each state destination for ads receiving 10K - 100K impressions.}
    \label{tab:obs_table_region_100K_nt}
\end{table*}
\begin{table*}[b]
    \centering
\begin{tabular}{|c|c|c|c|c|c|}
\hline
 U.S. State & t Value & p Value & $N_{\text{contrarians}}$ & $N_{\text{advocates}}$ \\
 \hline
 \hline
 Alabama & 36.6355152596689 & \hl{9.465273742117823e-211} & 1174 & 2759 \\
Alaska & -13.63175370114572 & \hl{5.470423540390248e-41} & 1149 & 2750 \\
Arizona & -5.9053936956825135 & \hl{3.817779948155841e-09} & 1174 & 2757 \\
Arkansas & 33.61570023146241 & \hl{9.002299876604879e-185} & 1174 & 2760 \\
California & -31.596846124572686 & \hl{3.5426906860853187e-177} & 1174 & 2761 \\
Colorado & 9.274439714327457 & \hl{8.318981943736975e-20} & 1172 & 2746 \\
Connecticut & -49.86321948306464 & \hl{0.0} & 1174 & 2758 \\
\hline
\end{tabular}
    \caption{Observational Study, Non-Targeted Ads: t Values and associated p values for each state destination for ads receiving 100K - 1M impressions.}
    \label{tab:obs_table_region_1M_nt}
\end{table*}
\begin{table*}[b]
    \centering
\begin{tabular}{|c|c|c|c|c|c|}
\hline
 U.S. State & t Value & p Value & $N_{\text{contrarians}}$ & $N_{\text{advocates}}$ \\
 \hline
 \hline
 Alabama & 12.147375988438649 & \hl{1.4167945330679985e-27} & 166 & 187 \\
Alaska & -3.4890194108552675 & \hl{0.0005479199848896388} & 164 & 183 \\
Arizona & -4.244923825966461 & \hl{2.9833834498326144e-05} & 166 & 187 \\
Arkansas & 12.569147189027678 & \hl{1.878946774200173e-28} & 166 & 187 \\
California & -15.944919193482804 & \hl{9.804233845260503e-42} & 166 & 187 \\
Colorado & 2.3595100517698775 & \hl{0.019463836836519015} & 166 & 187 \\
Connecticut & -17.249010472423567 & \hl{5.0127050407154034e-48} & 166 & 187 \\
\hline
\end{tabular}
    \caption{Observational Study, Non-Targeted Ads: t Values and associated p values for each state destination for ads receiving 1M+ impressions.}
    \label{tab:obs_table_region_1M+_nt}
\end{table*}

%% file: appendix-obs-gender-tables.tex
\subsubsection{All Ads: Gender}
T values from statistical analyses for all ads in the dataset are presented in Table \ref{tab:obs_table_gender_all}
\begin{table*}[b]
    \centering
    \begin{tabular}{|c|c|c|c|c|c|}
    \hline
Gender & t Value & p Value & Impression sub-division & $N_{\text{contrarians}}$ & $N_{\text{advocates}}$ \\
\hline
\hline
unknown & -59.60088925907019 & \hl{0.0} & < 1K & 28010 & 111087 \\
female & -97.10111437343016 & \hl{0.0} & < 1K & 106506 & 236915 \\
male & 102.44394249827462 & \hl{0.0} & < 1K & 108962 & 232187 \\
unknown & -168.77328336281914 & \hl{0.0} & < 10K & 95883 & 253225 \\
female & -124.58471487192271 & \hl{0.0} & < 10K & 141922 & 292969 \\
male & 115.7166994377231 & \hl{0.0} & < 10K & 143406 & 289853 \\
unknown & -121.16766460719253 & \hl{0.0} & < 100K & 77940 & 173670 \\
female & -100.70275663539485 & \hl{0.0} & < 100K & 84305 & 178618 \\
male & 89.31311023365433 & \hl{0.0} & < 100K & 84922 & 177609 \\
unknown & -32.957934721755535 & \hl{7.962730352268168e-236} & < 1M & 22021 & 43209 \\
female & -45.784473656143255 & \hl{0.0} & < 1M & 22450 & 43395 \\
male & 39.67812445269567 & \hl{0.0} & < 1M & 22549 & 43066 \\
unknown & -13.046496788671366 & \hl{8.95595785159375e-38} & 1M+ & 2191 & 2393 \\
female & -11.544436256615667 & \hl{2.1485467554794404e-30} & 1M+ & 2183 & 2341 \\
male & 12.265838192270873 & \hl{5.229702111203309e-34} & 1M+ & 2212 & 2308 \\
\hline
    \end{tabular}
    \caption{Observational Study: t Values and associated p values for each gender destination for ads across all impression classes.}
    \label{tab:obs_table_gender_all}
\end{table*}

\subsubsection{Targeted Ads: Gender}
Mean delivery percentages for males and females from targeted contrarian ads in the dataset are presented in Table \ref{tab:obs_table_gender_targeted}. Note that there were no advocacy ads that were targeted by gender, hence no analyses were conducted for advocacy ads by gender. 
\begin{table*}[b]
    \centering
    \begin{tabular}{|c|c|c|c|c|}
    \hline
Mean Delivery Percentage (Males) & Mean Delivery Percentage (Females)  & Impression sub-division & $N_\text{ads targeted at males}$ & $N_\text{ads targeted at females}$ \\
\hline
\hline
355.999989 & 363.000008 & < 1K & 1511 & 1441 \\
710.9999829999999 & 435.000001 & 1K - 10K & 3150 & 1839 \\
310.000006 & 183.999995 & 10K - 100K & 1329 & 841 \\
99.000006 & 88.000002 & 100K - 1M & 373 & 325 \\
30.0 & 14.999998000000001 & 1M+ & 70 & 46 \\
\hline
    \end{tabular}
    \caption{Observational Study: t Values and associated p values for each gender destination for ads across all impression classes.}
    \label{tab:obs_table_gender_targeted}
\end{table*}

\subsubsection{Non-Targeted Ads: Gender}
T values from statistical analyses for non-targeted ads in the dataset are presented in Table \ref{tab:obs_table_gender_non_targeted}

\begin{table*}[b]
    \centering
    \begin{tabular}{|c|c|c|c|c|c|}
    \hline
Gender & t Value & p Value & Impression sub-division & $N_{\text{contrarians}}$ & $N_{\text{advocates}}$ \\
\hline
\hline
unknown & -30.28802255890723 & \hl{3.1754600502463557e-177} & < 1K & 1483 & 15499 \\
male & 32.61562753682927 & \hl{1.990943082909287e-205} & < 1K & 3428 & 22077 \\
female & -34.572363564160426 & \hl{7.406451724015186e-234} & < 1K & 3359 & 22190 \\
unknown & -22.213467618583373 & \hl{4.2084778862737495e-98} & < 10K & 738 & 3982 \\
male & 13.264519919956074 & \hl{4.767025239339843e-37} & < 10K & 826 & 4124 \\
female & -12.343077251302159 & \hl{2.9166788230990206e-33} & < 10K & 825 & 4125 \\
unknown & -19.3228838763063 & \hl{1.795806652367272e-76} & < 100K & 497 & 1526 \\
male & 11.924245970453422 & \hl{7.060401628376165e-30} & < 100K & 511 & 1529 \\
female & -11.237579420828638 & \hl{6.123916371010068e-28} & < 100K & 509 & 1531 \\
unknown & -4.777145065199948 & \hl{3.1367517374290717e-06} & < 1M & 80 & 161 \\
male & 4.557210833796518 & \hl{1.3254789205093724e-05} & < 1M & 81 & 152 \\
female & -4.2583601247444935 & \hl{3.1002777419026356e-05} & < 1M & 79 & 152 \\
unknown & -11.750517442129782 & \hl{1.8095059573890114e-30} & 1M+ & 1005 & 1157 \\
male & 13.894464689721922 & \hl{9.075404859213381e-42} & 1M+ & 974 & 1111 \\
female & -10.504585088444019 & \hl{3.823310689178778e-25} & 1M+ & 974 & 1110 \\
\hline
    \end{tabular}
    \caption{Observational Study: t Values and associated p values for each gender destination for non-targeted ads across all impression classes.}
    \label{tab:obs_table_gender_non_targeted}
\end{table*}

%% file: appendix-obs-age-tables.tex
\subsubsection{All Ads: Age}
T values from statistical analyses for all ads in the dataset are presented in Table \ref{tab:obs_table_age_all}
\begin{table*}[b]
    \centering
    \begin{tabular}{|c|c|c|c|c|c|}
\hline
Age & t Value & p Value & Impression Sub-Division & 
$N_{\text{contrarians}}$ & $N_{\text{advocates}}$ \\
\hline
\hline
65+ & 1.637078292124389 & 0.10161800665557924 &  < 1K &  42124 & 99609 \\
35 & 0.8237781751756732 & 0.4100680857744594 &  < 1K &  41180 & 100674 \\
55 & 17.593452567321243 & \hl{3.8033242249070307e-69} &  < 1K &  41265 & 97510 \\
45 & 13.339167570988117 & \hl{1.527758309195544e-40} &  < 1K &  41164 & 96888 \\
25 & 5.400874757979404 & \hl{6.652028202123496e-08} &  < 1K &  40572 & 100522 \\
18 & 21.748839575685555 & \hl{1.696347907025595e-104} &  < 1K &  36038 & 80995 \\
13 & 6.516764902706327 & \hl{9.705057981521989e-11} &  < 1K &  1130 & 3933 \\
\hline
    \end{tabular}
    \caption{Observational Study: t Values and associated p values for each age destination for ads across all impression classes.}
    \label{tab:obs_table_age_all}
\end{table*}

\subsubsection{Targeted Ads: Age}
Mean delivery percentages for different age groups from targeted contrarian ads in the dataset are presented in Table \ref{tab:obs_table_age_targeted}. Note that there were no advocacy ads that were targeted by age, hence no analyses were conducted for advocacy ads by age. 
\begin{table*}[b]
    \centering
    \begin{tabular}{|c|c|c|c|}
    \hline
Age & Mean Delivery Percentage & Impression sub-division & $N_{\text{contrarians}}$ \\
\hline
\hline
18 & 12.0 & 1K - 10K & 18 \\
65+ & 45.999998000000005 & 1K - 10K & 135 \\
35 & 22.0 & 1K - 10K & 24 \\
55 & 8.0 & 1K - 10K & 8 \\
45 & 21.0 & 1K - 10K & 21 \\
25 & 18.0 & 1K - 10K & 18 \\
13 & 0.0 & 1K - 10K & 0 \\
18 & 12.000001000000001 & 10K - 100K & 14 \\
65+ & 7.000001 & 10K - 100K & 21 \\
35 & 8.0 & 10K - 100K & 14 \\
55 & 1.9999999999999998 & 10K - 100K & 6 \\
45 & 1.0 & 10K - 100K & 3 \\
25 & 11.0 & 10K - 100K & 15 \\
13 & 0.0 & 10K - 100K & 0 \\
18 & 0.999999 & 100K - 1M & 3 \\
65+ & 0.0 & 100K - 1M & 0 \\
35 & 0.0 & 100K - 1M & 0 \\
55 & 0.0 & 100K - 1M & 0 \\
45 & 0.0 & 100K - 1M & 0 \\
25 & 0.0 & 100K - 1M & 0 \\
13 & 0.0 & 100K - 1M & 0 \\
\hline
    \end{tabular}
    \caption{Observational Study, Targeted Ads: Mean delivery percentages for each age destination for ads across all impression classes.}
    \label{tab:obs_table_age_targeted}
\end{table*}

\subsubsection{Non-Targeted Ads: Age}
T values from statistical analyses for non-targeted ads in the dataset are presented in Table \ref{tab:obs_table_age_non_targeted}
\begin{table*}[b]
    \centering
    \begin{tabular}{|c|c|c|c|c|c|}
\hline
Age & t Value & p Value & Impression Sub-Division & 
$N_{\text{contrarians}}$ & $N_{\text{advocates}}$ \\
\hline
\hline
55 & 10.13253509968088 & \hl{1.619963675143212e-23} & < 1K & 1510 & 9933 \\
65+ & 6.115165452338112 & \hl{1.1616996418389305e-09} & < 1K & 1548 & 10314 \\
35 & -2.0167290698189344 & \hl{0.04388236127624781} & < 1K & 1350 & 9987 \\
18 & -2.212795992772528 & \hl{0.027064477797172033} & < 1K & 1189 & 9551 \\
25 & -8.598329728749883 & \hl{1.7270045316697664e-17} & < 1K & 1323 & 10282 \\
45 & 8.075492254596814 & \hl{1.2935354585329297e-15} & < 1K & 1393 & 9769 \\
13 & 303.0 & \hl{0.019840951141236008} & < 1K & 7 & 56 \\
\hline
    \end{tabular}
    \caption{Observational Study, Non-Targeted Ads: t Values and associated p values for each age destination for non-targeted ads across all impression classes.}
    \label{tab:obs_table_age_non_targeted}
\end{table*}

%% file: appendix-rq1.tex
\subsection{Ad Delivery and Objects in an Ad Image}
H-statistic and associated p-values from the Kruskal-Wallis test investigating if the population medians of ads featuring solar cells and oil rigs with no logo, logo of an advocacy organization and logo of a contrarian organization are significantly different. Results for U.S. State based ad destinations, gender based ad destinations and age based ad destinations are available in tables

\begin{table*}[b]
    \centering
    \begin{tabular}{|c|c|}
    \hline
    U.S. State & H-statistic, p-value \\
    \hline
    \hline
    Alabama & \hl{(10.38, 0.01)} \\
Alaska & \hl{(20.63, 0.0)} \\
Arizona & \hl{(41.19, 0.0)} \\
Arkansas & (0.7, 0.71) \\
California & \hl{(13.54, 0.0)} \\
Colorado & \hl{(41.2, 0.0)} \\
Connecticut & \hl{(31.24, 0.0)} \\
Delaware & \hl{(21.66, 0.0)} \\
Florida & (3.23, 0.2) \\
Georgia & (1.01, 0.6) \\
Hawaii & \hl{(13.36, 0.0)} \\
Idaho & \hl{(10.49, 0.01)} \\
Illinois & \hl{(9.74, 0.01)} \\
Indiana & (4.66, 0.1) \\
Iowa & \hl{(10.03, 0.01)} \\
Kansas & \hl{(8.49, 0.01)} \\
Kentucky & \hl{(14.16, 0.0)} \\
Louisiana & \hl{(8.67, 0.01)} \\
Maine & \hl{(27.79, 0.0)} \\
Maryland & \hl{(31.25, 0.0)} \\
Massachusetts & \hl{(43.94, 0.0)} \\
Michigan & (4.73, 0.09) \\
Minnesota & \hl{(13.15, 0.0)} \\
Mississippi & (4.1, 0.13) \\
Missouri & (1.42, 0.49) \\
Montana & \hl{(18.5, 0.0)} \\
Nebraska & \hl{(7.93, 0.02)} \\
Nevada & \hl{(29.7, 0.0)} \\
New Hampshire & \hl{(29.08, 0.0)} \\
New Jersey & \hl{(13.19, 0.0)} \\
New Mexico & \hl{(13.23, 0.0)} \\
New York & \hl{(20.47, 0.0)} \\
North Carolina & (2.52, 0.28) \\
North Dakota & \hl{(26.96, 0.0)} \\
Ohio & \hl{(12.55, 0.0)} \\
Oklahoma & \hl{(32.05, 0.0)} \\
Oregon & \hl{(22.74, 0.0)} \\
Pennsylvania & \hl{(11.89, 0.0)} \\
Rhode Island & \hl{(11.39, 0.0)} \\
South Carolina & (2.54, 0.28) \\
South Dakota & \hl{(10.34, 0.01)} \\
Tennessee & (5.81, 0.05) \\
Texas & \hl{(42.88, 0.0)} \\
Utah & \hl{(20.56, 0.0)} \\
Vermont & \hl{(21.09, 0.0)} \\
Virginia & (5.12, 0.08) \\
Washington & \hl{(15.18, 0.0)} \\
West Virginia & (1.08, 0.58) \\
Wisconsin & \hl{(7.83, 0.02)} \\
Wyoming & \hl{(16.69, 0.0)} \\
\hline
    \end{tabular}
    \caption{ H-statistic and p-values from the Kruskal-Wallis H-tests testing the null hypothesis that the population median of ads featuring solar
cells, oil rigs, and controls are different.}
    \label{tab:rq1-region}
\end{table*}

\begin{table*}[b]
    \centering
    \begin{tabular}{|c|c|}
    \hline
    Gender & (H-statistic, p-value) \\
    \hline
    \hline
Female & (324.71, 0.0) \\
Male & (321.33, 0.0) \\
Unknown & (36.96, 0.0) \\
\hline
    \end{tabular}
    \caption{ H-statistic and p-values from the Kruskal-Wallis H-tests testing the null hypothesis that the population median of ads featuring solar cells, oil rigs, and controls are different.}
    \label{tab:rq1-gender}
\end{table*}

\begin{table*}[b]
    \centering
    \begin{tabular}{|c|c|}
    \hline
    Age & (H-statistic, p-value) \\
    \hline
    \hline
18-24 & (113.96, 0.0) \\
25-34 & (73.85, 0.0) \\
35-44 & (7.59, 0.02) \\
45-54 & (6.29, 0.04) \\
55-64 & (19.34, 0.0) \\
65+ & (48.68, 0.0) \\
\hline
    \end{tabular}
    \caption{ H-statistic and p-values from the Kruskal-Wallis H-tests testing the null hypothesis that the population median of ads featuring solar cells, oil rigs, and controls are different.}
    \label{tab:rq1-age}
\end{table*}

%% file: appendix-rq2.tex
\subsection{Ad Delivery and Logos in an Ad Image}
H-statistic and associated p-values from the Kruskal-Wallis test investigating if the population medians of ads featuring solar cells and oil rigs are significantly different. Results for U.S. State based ad destinations, gender based ad destinations and age based ad destinations are available in tables \ref{tab:rq2-region}, \ref{tab:rq2-gender}, and \ref{tab:rq2-age}.

\begin{table*}[b]
    \centering
    \begin{tabular}{|c|c|c|}
    \hline
    Region & H-statistic and p-value (Oil \rig{s}) & H-statistic and p-value (Solar Cells) \\ 
    \hline
    \hline
    Alabama & (1.13, 0.57) & (0.52, 0.77) \\
    Alaska & (3.91, 0.14) & (0.05, 0.97) \\
    Arizona & (1.55, 0.46) & (2.51, 0.29) \\
    Arkansas & (1.74, 0.42) & (3.9, 0.14) \\
    California & (1.01, 0.6) & \hl{(7.14, 0.03)} \\
    Colorado & (1.59, 0.45) & (1.21, 0.55) \\
    Connecticut & \hl{(7.5, 0.02)} & (1.1, 0.58) \\
    Delaware & (0.88, 0.65) & (0.86, 0.65) \\
    Florida & (2.93, 0.23) & (0.84, 0.66) \\
    Georgia & (4.06, 0.13) & (2.8, 0.25) \\
    Hawaii & (4.2, 0.12) & (2.15, 0.34) \\
    Idaho & \hl{(14.07, 0.0)} & (0.24, 0.89) \\
    Illinois & (0.41, 0.82) & (1.85, 0.4) \\
    Indiana & (0.42, 0.81) & (1.05, 0.59) \\
    Iowa & (0.1, 0.95) & (0.47, 0.79) \\
    Kansas & (0.65, 0.72) & (0.6, 0.74) \\
    Kentucky & (1.03, 0.6) & (1.79, 0.41) \\
    Louisiana & (0.02, 0.99) & (0.43, 0.81) \\
    Maine & (0.37, 0.83) & (0.69, 0.71) \\
    Maryland & (1.78, 0.41) & (0.48, 0.79) \\
    Massachusetts & (4.06, 0.13) & \hl{(5.94, 0.05)} \\
    Michigan & (0.58, 0.75) & \hl{(6.57, 0.04)} \\
    Minnesota & (3.07, 0.22) & (0.37, 0.83) \\
    Mississippi & (5.19, 0.07) & (0.05, 0.98) \\
    Missouri & (3.87, 0.14) & (4.28, 0.12) \\
    Montana & (3.35, 0.19) & \hl{(7.67, 0.02)} \\
    Nebraska & (1.95, 0.38) & (0.46, 0.8) \\
    Nevada & (0.83, 0.66) & \hl{(7.74, 0.02)} \\
    New Hampshire & (0.35, 0.84) & (2.41, 0.3) \\
    New Jersey & (0.45, 0.8) & (1.2, 0.55) \\
    New Mexico & \hl{(6.79, 0.03)} & (0.89, 0.64) \\
    New York & (0.78, 0.68) & \hl{(6.66, 0.04)} \\
    North Carolina & (1.58, 0.45) & (1.04, 0.6) \\
    North Dakota & (0.39, 0.82) & (2.03, 0.36) \\
    Ohio & (0.67, 0.72) & (3.16, 0.21) \\
    Oklahoma & (0.07, 0.97) & (1.0, 0.61) \\
    Oregon & (0.61, 0.74) & (0.25, 0.88) \\
    Pennsylvania & (0.68, 0.71) & (0.29, 0.86) \\
    Rhode Island & (0.0, 1.0) & (2.09, 0.35) \\
    South Carolina & (4.54, 0.1) & (0.14, 0.93) \\
    South Dakota & (0.64, 0.73) & \hl{(8.78, 0.01)} \\
    Tennessee & (2.65, 0.27) & (1.63, 0.44) \\
    Texas & (1.0, 0.61) & (1.04, 0.6) \\
    Utah & (4.75, 0.09) & (1.1, 0.58) \\
    Vermont & (3.61, 0.16) & (0.02, 0.99) \\
    Virginia & (5.35, 0.07) & (1.46, 0.48) \\
    Washington & (0.96, 0.62) & (4.16, 0.12) \\
    West Virginia & (4.35, 0.11) & (1.78, 0.41) \\
    Wisconsin & (0.33, 0.85) & (1.75, 0.42) \\
    Wyoming & (4.22, 0.12) & (0.17, 0.92) \\
    \hline
    \end{tabular}
    \caption{H-statistic and p-values from the Kruskal-Wallis H-tests testing the null hypothesis that the population medians of ads featuring solar cells and oil \rig{s} with no logo, with the logo of an advocacy organization and the logo of a contrarian organization are different.}
    \label{tab:rq2-region}
\end{table*}

\begin{table*}[b]
    \centering
    \begin{tabular}{|c|c|c|}
    \hline
    Gender & H Statistic and p-value (Oil \rig{s}) & H Statistic and p-value (Solar cells) \\
    \hline
    \hline
    female & (0.09,0.96) & (5.7, 0.06) \\
    male & (0.18, 0.91) & (5.36, 0.07) \\
    unknown & \hl{(6.68, 0.04)} & (0.71, 0.7) \\
    \hline
    \end{tabular}
    \caption{H-statistic and p-values from the Kruskal-Wallis H-tests testing the null hypothesis that the population medians of ads featuring solar cells and oil \rig{s} with no logo, with the logo of an advocacy organization and the logo of a contrarian organization are different.}
    \label{tab:rq2-gender}
\end{table*}

\begin{table*}[b]
    \centering
    \begin{tabular}{|c|c|c|}
    \hline
    Age & H Statistic and p-value (Oil \rig{s}) & H Statistic and p-value (Solar cells) \\
    \hline
    \hline
    18-24 & (2.74 0.25) & \hl{(6.88, 0.03)} \\
    25-34 & (0.05 0.98) & \hl{(23.63, 0.0)} \\
    35-44 & (1.84 0.4) & \hl{(11.59, 0.0)} \\
    45-54 & (3.11 0.21) & (1.31, 0.52) \\
    55-64 & (1.07 0.59) & (3.46, 0.18) \\
    65+ & (0.21 0.9) & \hl{(15.82, 0.0)} \\
    \hline
    \end{tabular}
    \caption{H-statistic and p-values from the Kruskal-Wallis H-tests testing the null hypothesis that the population medians of ads featuring solar cells and oil \rig{s} with no logo, with the logo of an advocacy organization and the logo of a contrarian organization are different.}
    \label{tab:rq2-age}
\end{table*}

%% file: appendix-rq3.tex
\subsection{Ad Delivery Attribution to Ad Image}
p-values from Fisher's exact test, comparing the delivery of two ads featuring the same image and running at the same time, for U.S. State based ad destinations, gender based ad destinations and age based ad destinations are available in tables \ref{tab:rq3-state-consistency}, \ref{tab:rq3-gender-consistency} and \ref{tab:rq3-age-consistency}

\begin{table*}[b]
    \centering
    \begin{tabular}{|c|c|c|c|c|c|c|c|c|c|c|c|c|c|c|c|c|c|}
\hline
& \multicolumn{3}{c|}{Solar Cells} & \multicolumn{3}{c|}{Oil \rig{s}} & \multicolumn{3}{c|}{Controls} \\
\hline
Ad ID & Batch 1 & Batch 2 & Batch 3 & Batch 1 & Batch 2 & Batch 3 & Batch 1 & Batch 2 & Batch 3 \\
\hline
\hline
1  & 0.12 & 0.56 & 0.53 & \hl{0.01} & 1.0 & 0.34 & 1.0 & 0.73 & 0.55 \\
2  & 1.0 & 0.06 & 1.0 & 0.49 & \hl{0.01} & 1.0 & 0.53 & 1.0 & 0.15 \\
3  & 1.0 & 1.0 & 0.54 & \hl{0.01} & 0.77 & 0.37 & 0.56 & 0.19 & 0.28 \\
4  & 0.5 & 1.0 & 0.23 & 0.46 & 0.36 & 0.69 & 0.11 & 0.63 & 0.26 \\
5  & 1.0 & 0.67 & 0.52 & 0.63 & 0.56 & 1.0 & 0.65 & \hl{0.02} & 0.23 \\
6  & 1.0 & 1.0 & 0.44 & 0.69 & 0.13 & 0.48 & 0.41 & \hl{0.02} & 0.76 \\
7  & 0.6 & 1.0 & 0.31 & 1.0 & 1.0 & 0.14 & 0.39 & 0.76 & 0.34 \\
8  & 1.0 & 0.17 & 0.42 & 1.0 & 1.0 & 0.9 & 1.0 & 0.27 & 1.0 \\
9  & 1.0 & 0.36 & 0.53 & \hl{0.03} & 0.73 & 1.0 & 0.14 & 0.07 & 0.2 \\
10 & 1.0 & 0.92 & 0.22 & 1.0 & 0.34 & 0.04 & 0.06 & 1.0 & 1.0 \\
11 & 0.59 & 1.0 & 0.09 & 1.0 & 0.77 & 0.06 & 0.63 & 0.58 & 0.02 \\
12 & 0.25 & 0.17 & 1.0 & 0.53 & \hl{0.0} & 0.2 & 0.54 & 0.74 & 1.0 \\
13 & 0.1 & 1.0 & 0.26 & 0.64 & 1.0 & 0.85 & 0.42 & 0.71 & 0.75 \\
14 & 0.3 & \hl{0.03} & \hl{0.03} & 0.54 & 0.33 & 0.89 & 0.54 & 0.06 & 1.0 \\
15 & 0.38 & 1.0 & 0.73 & 0.25 & 0.14 & 0.9 & 0.12 & 0.11 & 0.12 \\
16 & 0.08 & 0.48 & \hl{0.0} & 0.73 & \hl{0.0} & 1.0 & 0.48 & 0.44 & 0.91 \\
17 & 1.0 & 1.0 & \hl{0.04} & 0.35 & 1.0 & 0.1 & 0.06 & 0.57 & 0.35 \\
18 & \hl{0.04} & 1.0 & 0.09 & 0.04 & 0.69 & 1.0 & 0.76 & 0.02 & 1.0 \\
19 & 0.49 & 1.0 & \hl{0.03} & 0.27 & 0.45 & 0.08 & 0.55 & 0.51 & 0.5 \\
20 & 0.12 & 0.64 & 1.0 & 1.0 & \hl{0.0} & 0.48 & 0.81 & 1.0 & 0.71 \\
21 & 1.0 & 0.55 & 1.0 & 0.51 & \hl{0.0} & 0.5 & 1.0 & 0.59 & 0.32 \\
22 & 0.46 & 0.06 & & 0.51 & 0.08 & & 0.14 & 1.0 & \\
\hline
    \end{tabular}
    \caption{Table showing p-values for the Two Sided Fisher's Exact Test. The test measures if observed ad delivery in different U.S. state based ad destinations was consistent between 2 ads  featuring the same image and run at the same time. p-values were calculated using the exact test, without using Monte-carlo simulations. Confidence intervals, and an estimate of the odds ratio are not available since this is a 14x2 dataset. Delivery of 65 associated solar cell, oil \rig{} and control ad pairs split into 3 batches were compared.}    
    \label{tab:rq3-state-consistency}
\end{table*}

\begin{table*}[b]
    \centering
    \begin{tabular}{|c|c|c|c|c|c|c|c|c|c|c|c|c|c|c|c|c|c|}
\hline
& \multicolumn{3}{c|}{Solar Cells} & \multicolumn{3}{c|}{Oil \rig{s}} & \multicolumn{3}{c|}{Controls} \\
\hline
Ad ID & Batch 1 & Batch 2 & Batch 3 & Batch 1 & Batch 2 & Batch 3 & Batch 1 & Batch 2 & Batch 3 \\
\hline
\hline
1 & 1.0 & 1.0 & 1.0 & 1.0 & 1.0 & 1.0 &  1.0 & 1.0 & 1.0 \\
2 & 1.0 & 1.0 & 1.0 & 1.0 & 1.0 & 1.0 &  1.0 & 1.0 & 1.0 \\
3 & 1.0 & 1.0 & 1.0 & 1.0 & 1.0 & 1.0 &  1.0 & 1.0 & 1.0 \\
4 & 1.0 & 1.0 & 1.0 & 1.0 & 1.0 & 1.0 &  1.0 & 1.0 & 1.0 \\
5 & 1.0 & 1.0 & 1.0 & 1.0 & 1.0 & 1.0 &  1.0 & 1.0 & 1.0 \\
6 & 1.0 & 1.0 & 1.0 & 1.0 & 1.0 & 1.0 &  1.0 & 1.0 & 1.0 \\
7 & 1.0 & 1.0 & 1.0 & 1.0 & 1.0 & 1.0 &  1.0 & 1.0 & 1.0 \\
8 & 1.0 & 1.0 & 1.0 & 1.0 & 1.0 & 1.0 &  1.0 & 1.0 & 1.0 \\
9 & 1.0 & 1.0 & 1.0 & 1.0 & 1.0 & 1.0 &  1.0 & 1.0 & 1.0 \\
10 & 1.0 & 1.0 & 1.0 & 1.0 & 1.0 & 1.0 &  1.0 & 1.0 & 1.0 \\
11 & 1.0 & 1.0 & 1.0 & 1.0 & 1.0 & 1.0 &  1.0 & 1.0 & 1.0 \\
12 & 1.0 & 1.0 & 1.0 & 1.0 & 1.0 & 1.0 &  1.0 & 1.0 & 1.0 \\
13 & 1.0 & 1.0 & 1.0 & 1.0 & 1.0 & 1.0 &  1.0 & 1.0 & 1.0 \\
14 & 1.0 & 1.0 & 1.0 & 1.0 & 1.0 & 1.0 &  1.0 & 1.0 & 1.0 \\
15 & 1.0 & 1.0 & 1.0 & 1.0 & 1.0 & 1.0 &  1.0 & 1.0 & 1.0 \\
16 & 1.0 & 1.0 & 1.0 & 1.0 & 1.0 & 1.0 &  1.0 & 1.0 & 1.0 \\
17 & 1.0 & 1.0 & 1.0 & 1.0 & 1.0 & 1.0 &  1.0 & 1.0 & 1.0 \\
18 & 1.0 & 1.0 & 1.0 & 1.0 & 1.0 & 1.0 &  1.0 & 1.0 & 1.0 \\
19 & 1.0 & 1.0 & 1.0 & 1.0 & 1.0 & 1.0 &  1.0 & 1.0 & 1.0 \\
20 & 1.0 & 1.0 & 1.0 & 1.0 & 1.0 & 1.0 &  1.0 & 1.0 & 1.0 \\
21 & 1.0 & 1.0 & 1.0 & 1.0 & 1.0 & 1.0 &  1.0 & 1.0 & 1.0 \\
22 & 1.0 & 1.0 & &1.0 & 1.0 & & 1.0 & 1.0 & \\
\hline
    \end{tabular}
    \caption{Table showing p-values for the Two Sided Fisher's Exact Test. The test measures if observed ad delivery in different gender based ad destinations was consistent between 2 ads  featuring the same image and run at the same time. p-values were calculated using the exact test, without using Monte-carlo simulations. Confidence intervals, and an estimate of the odds ratio are not available since this is a 3x2 dataset. Delivery of 65 associated solar cell, oil \rig{} and control ad pairs split into 3 batches were compared.}    
    \label{tab:rq3-gender-consistency}
\end{table*}

\begin{table*}[b]
    \centering
    \begin{tabular}{|c|c|c|c|c|c|c|c|c|c|c|c|c|c|c|c|c|c|}
\hline
& \multicolumn{3}{c|}{Solar Cells} & \multicolumn{3}{c|}{Oil \rig{s}} & \multicolumn{3}{c|}{Controls} \\
\hline
Ad ID & Batch 1 & Batch 2 & Batch 3 & Batch 1 & Batch 2 & Batch 3 & Batch 1 & Batch 2 & Batch 3 \\
\hline
\hline
1 & 1.0 & 1.0 & 1.0 &  1.0 & 1.0 & 1.0 &  0.2 & 1.0 & 1.0 \\
2 & 0.07 & 1.0 & 1.0 &  1.0 & 1.0 & 0.5 &  0.07 & 1.0 & 0.5 \\
3 & 1.0 & 1.0 & 1.0 &  1.0 & 0.4 & 0.2 &  1.0 & 1.0 & 0.2 \\
4 & 1.0 & 1.0 & 1.0 &  1.0 & 1.0 & 1.0 &  1.0 & 0.5 & 0.07 \\
5 & 1.0 & 1.0 & 1.0 &  1.0 & 1.0 & 1.0 &  1.0 & 0.07 & \hl{0.05} \\
6 & 1.0 & 0.2 & 1.0 &  0.2 & 1.0 & 1.0 &  1.0 & 1.0 & 0.13 \\
7 & 0.2 & 1.0 & 1.0 &  1.0 & 1.0 & 1.0 &  1.0 & 0.4 & 1.0 \\
8 & 1.0 & 1.0 & 1.0 &  1.0 & 1.0 & 0.5 &  1.0 & 0.07 & 1.0 \\
9 & 1.0 & 1.0 & 1.0 &  1.0 & 1.0 & 1.0 &  1.0 & 1.0 & 1.0 \\
10 & 1.0 & 0.4 & 1.0 &  0.5 & 1.0 & 1.0 &  0.13 & 1.0 & 1.0 \\
11 & 1.0 & 1.0 & 1.0 &  1.0 & 1.0 & 0.07 &  1.0 & 1.0 & 0.4 \\
12 & 1.0 & 1.0 & 1.0 &  1.0 & 1.0 & 0.4 &  1.0 & 0.07 & 1.0 \\
13 & 1.0 & 0.2 & 0.2 &  0.2 & 0.4 & 1.0 &  1.0 & 0.2 & 1.0 \\
14 & 1.0 & 1.0 & 1.0 &  0.07 & 0.5 & 1.0 &  1.0 & 0.4 & 1.0 \\
15 & 1.0 & 1.0 & 1.0 &  1.0 & 0.2 & 1.0 &  1.0 & 1.0 & 0.2 \\
16 & 1.0 & 1.0 & 1.0 &  1.0 & 1.0 & 1.0 &  0.07 & 0.2 & 1.0 \\
17 & 1.0 & 0.4 & 1.0 &  1.0 & 1.0 & 1.0 &  1.0 & 0.2 & 1.0 \\
18 & 1.0 & 1.0 & 1.0 &  1.0 & 1.0 & 0.02 &  0.4 & 0.2 & 0.07 \\
19 & 0.2 & 1.0 & 1.0 &  1.0 & 0.13 & 1.0 &  1.0 & 1.0 & 1.0 \\
20 & 1.0 & 1.0 & 1.0 &  0.07 & 1.0 & 1.0 &  1.0 & 1.0 & 1.0 \\
21 & 1.0 & 1.0 & 1.0 &  1.0 & 1.0 & 1.0 &  1.0 & 1.0 & 1.0 \\
22 & 1.0 & 1.0 & & 1.0 & 1.0 & & 1.0 & 1.0 &    \\
\hline
    \end{tabular}
    \caption{Table showing p-values for the Two Sided Fisher's Exact Test. The test measures if observed ad delivery in different age based ad destinations was consistent between 2 ads featuring the same image and run at the same time. p-values were calculated using the exact test, without using Monte-carlo simulations. Confidence intervals, and an estimate of the odds ratio are not available since this is a 6x2 dataset. Delivery of 65 associated solar cell, oil \rig{} and control ad pairs split into 3 batches were compared.}    
    \label{tab:rq3-age-consistency}
\end{table*}

%% file: appendix-rq4.tex
\subsection{Ad Delivery vs Facebook's Population Estimates}

\subsubsection{U.S. States}
$\chi^2$ statistics and associated p-values for the 3 batches of ads in our experiment show whether observed ad delivery was proportional to Facebook's population estimates. We find that in a majority of cases, the values were not proportional, as shown in tables \ref{tab:rq4-state-Batch1}, \ref{tab:rq4-state-Batch2} and \ref{tab:rq4-state-Batch3}
\begin{table*}[b]
\resizebox{\textwidth}{!}{%
 \centering
    \begin{tabular}{|c|c|c|c|c|c|c|c|c|c|c|c|}
    \hline
    Ad ID & Solar Cells & Solar Cells (Copy) & Solar Cells + Contrarian Logo & Solar Cells + Advocacy Logo & Oil \rig{s} & Oil \rig{s} (Copy) & Oil \rig{s} + Contrarian Logo & Oil \rig{s} + Advocacy Logo & Controls & Controls (Copy) & df \\
    \hline
    \hline
1 & \hl{(104.95, 0.0)} & \hl{(106.24, 0.0)} & \hl{(116.97, 0.0)} & \hl{(76.82, 0.01)} & \hl{(89.09, 0.0)} & \hl{(62.19, 0.12)} & \hl{(93.37, 0.0)} & \hl{(118.92, 0.0)} & \hl{(118.72, 0.0)} & \hl{(77.22, 0.01)} & 50.0 \\
2 & \hl{(109.51, 0.0)} & \hl{(86.17, 0.0)} & \hl{(78.15, 0.01)} & \hl{(79.53, 0.0)} & \hl{(76.79, 0.01)} & \hl{(90.42, 0.0)} & \hl{(108.25, 0.0)} & \hl{(92.04, 0.0)} & (66.72, 0.06) & \hl{(73.0, 0.02)} & 50.0 \\
3 & \hl{(88.01, 0.0)} & \hl{(112.27, 0.0)} & \hl{(82.11, 0.0)} & \hl{(108.85, 0.0)} & \hl{(75.29, 0.01)} & \hl{(128.11, 0.0)} & \hl{(72.49, 0.02)} & \hl{(94.48, 0.0)} & \hl{(109.75, 0.0)} & \hl{(76.67, 0.01)} & 50.0 \\
4 & \hl{(163.81, 0.0)} & \hl{(208.42, 0.0)} & \hl{(157.19, 0.0)} & \hl{(91.36, 0.0)} & \hl{(120.11, 0.0)} & (59.43, 0.17) & (59.77, 0.16) & \hl{(88.37, 0.0)} & (44.52, 0.69) & (50.24, 0.46) & 50.0 \\
5 & \hl{(143.41, 0.0)} & (52.59, 0.37) & \hl{(51.76, 0.41)} & \hl{(223.41, 0.0)} & \hl{(78.73, 0.01)} & \hl{(111.96, 0.0)} & \hl{(96.15, 0.0)} & \hl{(97.94, 0.0)} & \hl{(112.85, 0.0)} & (48.15, 0.55) & 50.0 \\
6 & \hl{(73.44, 0.02)} & \hl{(106.57, 0.0)} & \hl{(78.25, 0.01)} & \hl{(69.79, 0.03)} & (56.11, 0.26) & \hl{(117.85, 0.0)} & \hl{(98.64, 0.0)} & \hl{(63.06, 0.1)} & \hl{(63.02, 0.1)} & \hl{(83.72, 0.0)} & 50.0 \\
7 & \hl{(127.21, 0.0)} & \hl{(78.81, 0.01)} & \hl{(91.82, 0.0)} & \hl{(126.46, 0.0)} & \hl{(70.61, 0.03)} & \hl{(129.3, 0.0)} & (63.54, 0.09) & \hl{(51.12, 0.43)} & \hl{(88.53, 0.0)} & \hl{(69.61, 0.03)} & 50.0 \\
8 & (59.21, 0.17) & \hl{(102.69, 0.0)} & \hl{(92.73, 0.0)} & \hl{(85.69, 0.0)} & \hl{(72.79, 0.02)} & \hl{(129.47, 0.0)} & (37.93, 0.89) & \hl{(92.84, 0.0)} & \hl{(100.01, 0.0)} & (49.32, 0.5) & 50.0 \\
9 & \hl{(51.3, 0.42)} & \hl{(95.26, 0.0)} & \hl{(140.48, 0.0)} & (39.67, 0.85) & \hl{(80.14, 0.0)} & \hl{(77.47, 0.01)} & \hl{(74.27, 0.01)} & \hl{(69.22, 0.04)} & (55.8, 0.27) & (65.28, 0.07) & 50.0 \\
10 & \hl{(76.54, 0.01)} & \hl{(71.17, 0.03)} & \hl{(84.64, 0.0)} & \hl{(76.53, 0.01)} & \hl{(84.02, 0.0)} & \hl{(80.62, 0.0)} & (64.81, 0.08) & \hl{(153.8, 0.0)} & (64.28, 0.08) & (66.76, 0.06) & 50.0 \\
11 & \hl{(110.5, 0.0)} & \hl{(76.3, 0.01)} & \hl{(77.94, 0.01)} & \hl{(85.99, 0.0)} & \hl{(91.35, 0.0)} & \hl{(75.54, 0.01)} & \hl{(88.27, 0.0)} & \hl{(78.17, 0.01)} & \hl{(88.79, 0.0)} & (67.95, 0.05) & 50.0 \\
12 & (60.0, 0.16) & \hl{(91.5, 0.0)} & (48.01, 0.55) & (58.38, 0.19) & (59.82, 0.16) & (66.35, 0.06) & \hl{(111.38, 0.0)} & \hl{(100.16, 0.0)} & (52.23, 0.39) & \hl{(57.04, 0.23)} & 50.0 \\
13 & \hl{(54.32, 0.31)} & \hl{(91.8, 0.0)} & \hl{(103.74, 0.0)} & \hl{(111.11, 0.0)} & \hl{(73.33, 0.02)} & \hl{(104.07, 0.0)} & \hl{(62.15, 0.12)} & \hl{(80.67, 0.0)} & \hl{(79.55, 0.0)} & \hl{(126.3, 0.0)} & 50.0 \\
14 & (56.35, 0.25) & \hl{(86.02, 0.0)} & \hl{(113.53, 0.0)} & \hl{(71.24, 0.03)} & \hl{(79.7, 0.0)} & \hl{(106.38, 0.0)} & (59.57, 0.17) & (67.45, 0.05) & (55.78, 0.27) & \hl{(76.21, 0.01)} & 50.0 \\
15 & \hl{(108.33, 0.0)} & \hl{(157.93, 0.0)} & \hl{(89.92, 0.0)} & \hl{(131.43, 0.0)} & \hl{(87.01, 0.0)} & \hl{(75.94, 0.01)} & \hl{(54.7, 0.3)} & (50.23, 0.46) & \hl{(69.54, 0.04)} & \hl{(51.05, 0.43)} & 50.0 \\
16 & \hl{(91.1, 0.0)} & (63.91, 0.09) & \hl{(69.22, 0.04)} & \hl{(144.7, 0.0)} & \hl{(95.62, 0.0)} & \hl{(98.78, 0.0)} & \hl{(98.1, 0.0)} & (55.75, 0.27) & (58.98, 0.18) & (64.39, 0.08) & 50.0 \\
17 & \hl{(88.82, 0.0)} & \hl{(73.08, 0.02)} & \hl{(107.32, 0.0)} & \hl{(88.96, 0.0)} & \hl{(99.38, 0.0)} & \hl{(100.44, 0.0)} & \hl{(96.55, 0.0)} & (68.06, 0.05) & \hl{(70.22, 0.03)} & \hl{(83.82, 0.0)} & 50.0 \\
18 & \hl{(84.16, 0.0)} & \hl{(57.33, 0.22)} & (59.56, 0.17) & \hl{(86.1, 0.0)} & \hl{(63.16, 0.1)} & (66.81, 0.06) & \hl{(84.33, 0.0)} & \hl{(71.63, 0.02)} & \hl{(119.09, 0.0)} & \hl{(91.66, 0.0)} & 50.0 \\
19 & \hl{(128.59, 0.0)} & \hl{(99.51, 0.0)} & \hl{(87.01, 0.0)} & \hl{(86.74, 0.0)} & \hl{(82.97, 0.0)} & \hl{(86.93, 0.0)} & \hl{(94.82, 0.0)} & \hl{(264.14, 0.0)} & (38.93, 0.87) & \hl{(69.34, 0.04)} & 50.0 \\
20 & \hl{(106.34, 0.0)} & \hl{(93.06, 0.0)} & \hl{(76.36, 0.01)} & \hl{(105.89, 0.0)} & \hl{(71.71, 0.02)} & \hl{(56.53, 0.24)} & \hl{(54.42, 0.31)} & \hl{(81.16, 0.0)} & (59.0, 0.18) & \hl{(51.02, 0.43)} & 50.0 \\
21 & \hl{(95.32, 0.0)} & \hl{(83.27, 0.0)} & \hl{(71.39, 0.03)} & \hl{(150.18, 0.0)} & \hl{(117.22, 0.0)} & \hl{(92.15, 0.0)} & \hl{(146.39, 0.0)} & \hl{(63.03, 0.1)} & \hl{(71.51, 0.02)} & \hl{(54.35, 0.31)} & 50.0 \\
22 & \hl{(188.79, 0.0)} & \hl{(117.99, 0.0)} & \hl{(68.71, 0.04)} & (64.27, 0.08) & \hl{(80.6, 0.0)} & \hl{(126.53, 0.0)} & \hl{(100.78, 0.0)} & \hl{(128.06, 0.0)} & (48.36, 0.54) & \hl{(77.07, 0.01)} & 50.0 \\
\hline
    \end{tabular}}
    \caption{Batch1: $\chi^2$ statistic and p values to test the hypothesis that the observed ad delivery in different U.S. states is proportional to the population estimates provided by Facebook. In a majority of ads, the observed ad delivery is not proportional to the population estimates from Facebook. The Chi Square test was not able to be performed for the highlighted values.}
    \label{tab:rq4-state-Batch1}
\end{table*}

\begin{table*}[b]
\resizebox{\textwidth}{!}{%
 \centering
    \begin{tabular}{|c|c|c|c|c|c|c|c|c|c|c|c|}
    \hline
    Ad ID & Solar Cells & Solar Cells (Copy) & Solar Cells + Contrarian Logo & Solar Cells + Advocacy Logo & Oil \rig{s} & Oil \rig{s} (Copy) & Oil \rig{s} + Contrarian Logo & Oil \rig{s} + Advocacy Logo & Controls & Controls (Copy) & df \\
    \hline
    \hline
    1 & \hl{(98.56, 0.0)} & \hl{(74.01, 0.02)} & (50.34, 0.46) & (55.08, 0.29) & \hl{(76.01, 0.01)} & \hl{(172.15, 0.0)} & \hl{(85.09, 0.0)} & \hl{(78.18, 0.01)} & \hl{(71.18, 0.03)} & \hl{(82.85, 0.0)} & 50.0 \\
2 & \hl{(102.4, 0.0)} & \hl{(80.59, 0.0)} & \hl{(99.13, 0.0)} & \hl{(86.78, 0.0)} & \hl{(121.49, 0.0)} & \hl{(80.14, 0.0)} & \hl{(77.16, 0.01)} & \hl{(83.76, 0.0)} & (45.42, 0.66) & \hl{(103.1, 0.0)} & 50.0 \\
3 & \hl{(80.91, 0.0)} & \hl{(90.0, 0.0)} & \hl{(129.87, 0.0)} & (43.34, 0.74) & \hl{(124.5, 0.0)} & \hl{(72.14, 0.02)} & \hl{(95.83, 0.0)} & \hl{(116.18, 0.0)} & \hl{(87.31, 0.0)} & \hl{(89.58, 0.0)} & 50.0 \\
4 & \hl{(93.65, 0.0)} & \hl{(137.15, 0.0)} & (60.01, 0.16) & \hl{(107.83, 0.0)} & \hl{(78.88, 0.01)} & (53.07, 0.36) & \hl{(84.38, 0.0)} & \hl{(98.94, 0.0)} & \hl{(61.23, 0.13)} & (52.87, 0.36) & 50.0 \\
5 & \hl{(86.26, 0.0)} & \hl{(80.23, 0.0)} & \hl{(85.87, 0.0)} & \hl{(101.13, 0.0)} & \hl{(99.32, 0.0)} & \hl{(101.56, 0.0)} & \hl{(99.09, 0.0)} & \hl{(122.46, 0.0)} & \hl{(87.79, 0.0)} & \hl{(69.13, 0.04)} & 50.0 \\
6 & \hl{(112.19, 0.0)} & \hl{(110.53, 0.0)} & \hl{(90.26, 0.0)} & \hl{(109.3, 0.0)} & \hl{(84.24, 0.0)} & \hl{(69.09, 0.04)} & \hl{(84.98, 0.0)} & \hl{(77.28, 0.01)} & (56.45, 0.25) & (59.43, 0.17) & 50.0 \\
7 & \hl{(106.44, 0.0)} & \hl{(132.26, 0.0)} & \hl{(95.13, 0.0)} & \hl{(122.91, 0.0)} & \hl{(72.3, 0.02)} & \hl{(74.6, 0.01)} & \hl{(79.02, 0.01)} & \hl{(61.13, 0.13)} & \hl{(90.89, 0.0)} & (64.1, 0.09) & 50.0 \\
8 & (60.02, 0.16) & (60.08, 0.16) & \hl{(88.67, 0.0)} & \hl{(122.13, 0.0)} & (67.03, 0.05) & \hl{(89.8, 0.0)} & \hl{(167.53, 0.0)} & \hl{(93.99, 0.0)} & \hl{(84.59, 0.0)} & \hl{(80.59, 0.0)} & 50.0 \\
9 & \hl{(74.45, 0.01)} & \hl{(104.81, 0.0)} & \hl{(58.04, 0.2)} & \hl{(82.17, 0.0)} & \hl{(62.61, 0.11)} & \hl{(90.72, 0.0)} & \hl{(54.09, 0.32)} & \hl{(105.85, 0.0)} & \hl{(75.41, 0.01)} & \hl{(79.02, 0.01)} & 50.0 \\
10 & \hl{(114.68, 0.0)} & \hl{(199.3, 0.0)} & \hl{(179.95, 0.0)} & \hl{(73.14, 0.02)} & \hl{(92.24, 0.0)} & \hl{(112.91, 0.0)} & \hl{(88.13, 0.0)} & \hl{(63.02, 0.1)} & \hl{(75.46, 0.01)} & \hl{(89.4, 0.0)} & 50.0 \\
11 & \hl{(139.0, 0.0)} & \hl{(99.67, 0.0)} & \hl{(131.25, 0.0)} & \hl{(92.24, 0.0)} & \hl{(91.44, 0.0)} & \hl{(98.07, 0.0)} & \hl{(133.64, 0.0)} & \hl{(81.82, 0.0)} & (46.02, 0.63) & (55.47, 0.28) & 50.0 \\
12 & \hl{(136.44, 0.0)} & (63.68, 0.09) & \hl{(93.33, 0.0)} & \hl{(92.44, 0.0)} & \hl{(84.12, 0.0)} & \hl{(129.34, 0.0)} & \hl{(92.0, 0.0)} & \hl{(83.06, 0.0)} & (65.37, 0.07) & \hl{(51.25, 0.42)} & 50.0 \\
13 & (40.51, 0.83) & \hl{(172.55, 0.0)} & \hl{(95.83, 0.0)} & \hl{(96.4, 0.0)} & \hl{(85.92, 0.0)} & \hl{(85.23, 0.0)} & \hl{(71.49, 0.02)} & \hl{(57.52, 0.22)} & \hl{(76.78, 0.01)} & \hl{(68.3, 0.04)} & 50.0 \\
14 & \hl{(71.67, 0.02)} & \hl{(130.78, 0.0)} & \hl{(164.91, 0.0)} & \hl{(87.88, 0.0)} & \hl{(72.98, 0.02)} & \hl{(118.84, 0.0)} & (47.32, 0.58) & \hl{(74.04, 0.02)} & \hl{(101.62, 0.0)} & \hl{(84.82, 0.0)} & 50.0 \\
15 & \hl{(96.68, 0.0)} & \hl{(98.11, 0.0)} & \hl{(82.54, 0.0)} & \hl{(69.58, 0.03)} & \hl{(97.61, 0.0)} & \hl{(85.45, 0.0)} & \hl{(116.35, 0.0)} & \hl{(79.89, 0.0)} & \hl{(104.76, 0.0)} & (31.7, 0.98) & 50.0 \\
16 & \hl{(74.01, 0.02)} & \hl{(138.75, 0.0)} & \hl{(98.81, 0.0)} & \hl{(158.47, 0.0)} & \hl{(126.15, 0.0)} & \hl{(152.81, 0.0)} & \hl{(96.34, 0.0)} & (67.31, 0.05) & \hl{(103.95, 0.0)} & \hl{(106.62, 0.0)} & 50.0 \\
17 & \hl{(82.5, 0.0)} & \hl{(128.22, 0.0)} & \hl{(106.67, 0.0)} & \hl{(77.7, 0.01)} & \hl{(109.42, 0.0)} & \hl{(88.15, 0.0)} & \hl{(99.85, 0.0)} & \hl{(114.56, 0.0)} & \hl{(69.81, 0.03)} & \hl{(107.25, 0.0)} & 50.0 \\
18 & \hl{(76.04, 0.01)} & \hl{(74.81, 0.01)} & \hl{(109.8, 0.0)} & \hl{(111.11, 0.0)} & \hl{(81.8, 0.0)} & \hl{(75.17, 0.01)} & \hl{(108.48, 0.0)} & \hl{(99.75, 0.0)} & \hl{(76.01, 0.01)} & (49.49, 0.49) & 50.0 \\
19 & \hl{(94.75, 0.0)} & \hl{(90.15, 0.0)} & \hl{(144.54, 0.0)} & \hl{(62.77, 0.11)} & (49.68, 0.49) & \hl{(96.47, 0.0)} & \hl{(88.63, 0.0)} & \hl{(125.28, 0.0)} & \hl{(82.67, 0.0)} & (59.48, 0.17) & 50.0 \\
20 & \hl{(124.37, 0.0)} & \hl{(63.24, 0.1)} & \hl{(50.85, 0.44)} & \hl{(106.72, 0.0)} & \hl{(80.8, 0.0)} & \hl{(79.84, 0.0)} & \hl{(76.14, 0.01)} & (66.45, 0.06) & \hl{(76.0, 0.01)} & \hl{(143.05, 0.0)} & 50.0 \\
21 & \hl{(109.56, 0.0)} & \hl{(133.78, 0.0)} & \hl{(87.25, 0.0)} & \hl{(136.85, 0.0)} & \hl{(84.68, 0.0)} & (66.16, 0.06) & (60.3, 0.15) & \hl{(72.58, 0.02)} & \hl{(50.99, 0.43)} & (58.4, 0.19) & 50.0 \\
22 & \hl{(83.9, 0.0)} & (47.85, 0.56) & (60.39, 0.15) & \hl{(99.85, 0.0)} & \hl{(70.22, 0.03)} & \hl{(90.22, 0.0)} & \hl{(89.96, 0.0)} & \hl{(147.39, 0.0)} & \hl{(91.55, 0.0)} & \hl{(76.6, 0.01)} & 50.0 \\
\hline
    \end{tabular}}
    \caption{Batch2: $\chi^2$ statistic and p values to test the hypothesis that the observed ad delivery in different U.S. states is proportional to the population estimates provided by Facebook. In a majority of ads, the observed ad delivery is not proportional to the population estimates from Facebook. The Chi Square test was not able to be performed for the highlighted values.}
    \label{tab:rq4-state-Batch2}
\end{table*}

\begin{table*}[b]
\resizebox{\textwidth}{!}{%
 \centering
    \begin{tabular}{|c|c|c|c|c|c|c|c|c|c|c|c|}
    \hline
    Ad ID & Solar Cells & Solar Cells (Copy) & Solar Cells + Contrarian Logo & Solar Cells + Advocacy Logo & Oil \rig{s} & Oil \rig{s} (Copy) & Oil \rig{s} + Contrarian Logo & Oil \rig{s} + Advocacy Logo & Controls & Controls (Copy) & df \\
    \hline
    \hline
    1 & (56.15, 0.26) & \hl{(76.29, 0.01)} & \hl{(120.82, 0.0)} & \hl{(129.39, 0.0)} & \hl{(82.96, 0.0)} & (67.05, 0.05) & \hl{(111.55, 0.0)} & \hl{(144.61, 0.0)} & \hl{(82.41, 0.0)} & \hl{(75.7, 0.01)} & 50.0 \\ 
2 & \hl{(103.46, 0.0)} & (45.43, 0.66) & \hl{(69.82, 0.03)} & \hl{(91.97, 0.0)} & \hl{(88.38, 0.0)} & \hl{(112.53, 0.0)} & (56.23, 0.25) & \hl{(81.43, 0.0)} & \hl{(83.0, 0.0)} & \hl{(51.76, 0.41)} & 50.0 \\ 
3 & \hl{(100.08, 0.0)} & \hl{(82.08, 0.0)} & \hl{(94.15, 0.0)} & \hl{(106.96, 0.0)} & (35.4, 0.94) & \hl{(99.9, 0.0)} & \hl{(87.68, 0.0)} & \hl{(74.96, 0.01)} & \hl{(93.73, 0.0)} & \hl{(96.57, 0.0)} & 50.0 \\ 
4 & \hl{(87.17, 0.0)} & \hl{(75.94, 0.01)} & \hl{(72.36, 0.02)} & (59.21, 0.17) & \hl{(68.65, 0.04)} & \hl{(173.72, 0.0)} & \hl{(63.46, 0.1)} & \hl{(104.6, 0.0)} & \hl{(98.66, 0.0)} & \hl{(56.61, 0.24)} & 50.0 \\ 
5 & \hl{(175.35, 0.0)} & \hl{(87.64, 0.0)} & \hl{(106.15, 0.0)} & \hl{(56.92, 0.23)} & \hl{(108.4, 0.0)} & \hl{(103.01, 0.0)} & \hl{(81.05, 0.0)} & \hl{(124.17, 0.0)} & \hl{(73.78, 0.02)} & \hl{(82.22, 0.0)} & 50.0 \\ 
6 & \hl{(85.33, 0.0)} & \hl{(78.0, 0.01)} & \hl{(62.29, 0.11)} & \hl{(73.52, 0.02)} & \hl{(90.17, 0.0)} & \hl{(95.55, 0.0)} & (65.63, 0.07) & (58.39, 0.19) & \hl{(87.8, 0.0)} & \hl{(107.88, 0.0)} & 50.0 \\ 
7 & \hl{(114.37, 0.0)} & (65.28, 0.07) & \hl{(80.3, 0.0)} & (37.1, 0.91) & \hl{(80.11, 0.0)} & \hl{(91.93, 0.0)} & \hl{(148.37, 0.0)} & (55.74, 0.27) & (46.31, 0.62) & (42.23, 0.77) & 50.0 \\ 
8 & \hl{(119.44, 0.0)} & \hl{(83.23, 0.0)} & \hl{(96.1, 0.0)} & \hl{(162.83, 0.0)} & (58.64, 0.19) & \hl{(117.64, 0.0)} & \hl{(140.48, 0.0)} & \hl{(82.33, 0.0)} & \hl{(137.94, 0.0)} & \hl{(72.56, 0.02)} & 50.0 \\ 
9 & \hl{(80.43, 0.0)} & \hl{(167.22, 0.0)} & \hl{(105.45, 0.0)} & \hl{(78.67, 0.01)} & \hl{(85.61, 0.0)} & (65.75, 0.07) & \hl{(88.78, 0.0)} & \hl{(72.07, 0.02)} & \hl{(100.58, 0.0)} & \hl{(178.92, 0.0)} & 50.0 \\ 
10 & \hl{(98.4, 0.0)} & \hl{(114.45, 0.0)} & (58.7, 0.19) & \hl{(56.52, 0.24)} & \hl{(123.32, 0.0)} & \hl{(82.62, 0.0)} & \hl{(83.53, 0.0)} & \hl{(105.16, 0.0)} & \hl{(115.62, 0.0)} & \hl{(99.46, 0.0)} & 50.0 \\ 
11 & \hl{(145.62, 0.0)} & \hl{(132.34, 0.0)} & \hl{(84.09, 0.0)} & (67.4, 0.05) & \hl{(98.36, 0.0)} & \hl{(78.12, 0.01)} & \hl{(81.24, 0.0)} & \hl{(76.28, 0.01)} & \hl{(107.2, 0.0)} & \hl{(71.97, 0.02)} & 50.0 \\ 
12 & \hl{(108.4, 0.0)} & \hl{(82.18, 0.0)} & \hl{(124.75, 0.0)} & \hl{(111.42, 0.0)} & \hl{(99.05, 0.0)} & \hl{(77.79, 0.01)} & (60.49, 0.15) & \hl{(79.04, 0.01)} & \hl{(121.28, 0.0)} & \hl{(68.45, 0.04)} & 50.0 \\ 
13 & \hl{(76.28, 0.01)} & \hl{(139.4, 0.0)} & \hl{(84.2, 0.0)} & \hl{(109.85, 0.0)} & \hl{(84.53, 0.0)} & \hl{(106.61, 0.0)} & \hl{(89.0, 0.0)} & \hl{(99.93, 0.0)} & \hl{(70.95, 0.03)} & \hl{(83.73, 0.0)} & 50.0 \\ 
14 & (45.35, 0.66) & \hl{(101.0, 0.0)} & \hl{(79.78, 0.0)} & \hl{(51.63, 0.41)} & (60.56, 0.15) & (50.03, 0.47) & \hl{(102.07, 0.0)} & \hl{(71.49, 0.02)} & \hl{(99.34, 0.0)} & \hl{(120.22, 0.0)} & 50.0 \\ 
15 & \hl{(74.58, 0.01)} & \hl{(88.26, 0.0)} & \hl{(100.78, 0.0)} & \hl{(126.71, 0.0)} & (66.83, 0.06) & \hl{(79.3, 0.01)} & \hl{(88.17, 0.0)} & \hl{(78.04, 0.01)} & \hl{(137.73, 0.0)} & (64.26, 0.08) & 50.0 \\ 
16 & \hl{(68.78, 0.04)} & \hl{(62.84, 0.1)} & (59.53, 0.17) & \hl{(75.44, 0.01)} & \hl{(94.0, 0.0)} & \hl{(63.42, 0.1)} & \hl{(114.76, 0.0)} & \hl{(62.78, 0.11)} & (64.39, 0.08) & \hl{(57.58, 0.22)} & 50.0 \\ 
17 & \hl{(79.97, 0.0)} & \hl{(134.0, 0.0)} & \hl{(72.11, 0.02)} & \hl{(71.85, 0.02)} & \hl{(97.8, 0.0)} & \hl{(73.4, 0.02)} & (43.85, 0.72) & \hl{(119.27, 0.0)} & (47.49, 0.57) & \hl{(61.72, 0.12)} & 50.0 \\ 
18 & (50.12, 0.47) & \hl{(125.92, 0.0)} & \hl{(94.28, 0.0)} & \hl{(79.73, 0.0)} & \hl{(88.83, 0.0)} & \hl{(87.96, 0.0)} & \hl{(73.34, 0.02)} & \hl{(74.46, 0.01)} & \hl{(68.28, 0.04)} & \hl{(76.84, 0.01)} & 50.0 \\ 
19 & \hl{(62.31, 0.11)} & (52.73, 0.37) & \hl{(93.47, 0.0)} & (46.61, 0.61) & \hl{(117.6, 0.0)} & \hl{(60.83, 0.14)} & \hl{(136.66, 0.0)} & \hl{(75.28, 0.01)} & \hl{(88.84, 0.0)} & \hl{(96.01, 0.0)} & 50.0 \\ 
20 & \hl{(91.08, 0.0)} & \hl{(126.32, 0.0)} & \hl{(118.75, 0.0)} & \hl{(90.96, 0.0)} & \hl{(85.47, 0.0)} & \hl{(91.01, 0.0)} & \hl{(86.94, 0.0)} & \hl{(93.59, 0.0)} & \hl{(80.74, 0.0)} & \hl{(79.47, 0.01)} & 50.0 \\ 
21 & \hl{(108.18, 0.0)} & \hl{(130.64, 0.0)} & \hl{(94.78, 0.0)} & \hl{(80.13, 0.0)} & \hl{(85.01, 0.0)} & (66.4, 0.06) & \hl{(84.7, 0.0)} & \hl{(105.86, 0.0)} & \hl{(95.75, 0.0)} & \hl{(99.19, 0.0)} & 50.0 \\
    \hline
    \end{tabular}}
    \caption{Batch3: $\chi^2$ statistic and p values to test the hypothesis that the observed ad delivery in different U.S. states is proportional to the population estimates provided by Facebook. In a majority of ads, the observed ad delivery is not proportional to the population estimates from Facebook. The Chi Square test was not able to be performed for the highlighted values.}
    \label{tab:rq4-state-Batch3}
\end{table*}

\subsubsection{Gender}
$\chi^2$ statistics and associated p-values for the 3 batches of ads in our experiment show whether observed ad delivery was proportional to Facebook's population estimates. We find that in a majority of cases, the values were not proportional, as shown in tables \ref{tab:rq4-gender-Batch1}, \ref{tab:rq4-gender-Batch2} and \ref{tab:rq4-gender-Batch3}

\begin{table*}[]
\resizebox{\textwidth}{!}{%
    \centering
    \begin{tabular}{|c|c|c|c|c|c|c|c|c|c|c|c|}
    \hline
    Ad ID & Solar Cells & Solar Cells (Copy) & Solar Cells + Contrarian Logo & Solar Cells + Advocacy Logo & Oil \rig{s} & Oil \rig{s} (Copy) & Oil \rig{s} + Contrarian Logo & Oil \rig{s} + Advocacy Logo & Controls & Controls (Copy) & df \\
    \hline
    \hline
    1 & \hl{(16.12, 0.0)} & \hl{(9.14, 0.01)} & \hl{(25.14, 0.0)} & \hl{(23.77, 0.0)} & \hl{(50.83, 0.0)} & \hl{(88.08, 0.0)} & \hl{(40.5, 0.0)} & \hl{(39.07, 0.0)} & \hl{(75.92, 0.0)} & \hl{(87.77, 0.0)} & 2.0 \\
2 & \hl{(16.26, 0.0)} & (3.78, 0.15) & \hl{(23.31, 0.0)} & \hl{(38.66, 0.0)} & \hl{(51.56, 0.0)} & \hl{(58.89, 0.0)} & \hl{(36.93, 0.0)} & \hl{(44.47, 0.0)} & \hl{(41.48, 0.0)} & \hl{(41.88, 0.0)} & 2.0 \\
3 & \hl{(52.66, 0.0)} & \hl{(30.63, 0.0)} & \hl{(22.8, 0.0)} & \hl{(2.86, 0.24)} & \hl{(47.2, 0.0)} & \hl{(42.7, 0.0)} & \hl{(45.58, 0.0)} & \hl{(3.24, 0.2)} & \hl{(65.37, 0.0)} & \hl{(65.76, 0.0)} & 2.0 \\
4 & \hl{(10.44, 0.01)} & \hl{(2.9, 0.23)} & \hl{(30.92, 0.0)} & \hl{(16.03, 0.0)} & \hl{(50.83, 0.0)} & \hl{(42.58, 0.0)} & \hl{(45.95, 0.0)} & \hl{(74.85, 0.0)} & \hl{(53.74, 0.0)} & \hl{(79.24, 0.0)} & 2.0 \\
5 & (1.19, 0.55) & \hl{(49.32, 0.0)} & \hl{(42.23, 0.0)} & \hl{(53.52, 0.0)} & \hl{(32.22, 0.0)} & \hl{(43.06, 0.0)} & \hl{(65.23, 0.0)} & \hl{(53.24, 0.0)} & \hl{(30.53, 0.0)} & \hl{(56.01, 0.0)} & 2.0 \\
6 & \hl{(4.16, 0.12)} & \hl{(12.13, 0.0)} & \hl{(28.86, 0.0)} & \hl{(30.81, 0.0)} & \hl{(37.3, 0.0)} & \hl{(62.94, 0.0)} & \hl{(40.67, 0.0)} & \hl{(60.41, 0.0)} & (1.53, 0.47) & (3.64, 0.16) & 2.0 \\
7 & \hl{(10.62, 0.0)} & \hl{(6.41, 0.04)} & (0.72, 0.7) & (0.74, 0.69) & \hl{(54.86, 0.0)} & \hl{(67.38, 0.0)} & \hl{(41.88, 0.0)} & \hl{(88.93, 0.0)} & \hl{(7.36, 0.03)} & \hl{(10.93, 0.0)} & 2.0 \\
8 & \hl{(7.4, 0.02)} & (3.78, 0.15) & \hl{(18.41, 0.0)} & \hl{(2.33, 0.31)} & \hl{(49.35, 0.0)} & \hl{(49.09, 0.0)} & \hl{(33.42, 0.0)} & \hl{(56.85, 0.0)} & \hl{(3.99, 0.14)} & \hl{(6.35, 0.04)} & 2.0 \\
9 & \hl{(15.09, 0.0)} & \hl{(7.28, 0.03)} & \hl{(39.72, 0.0)} & \hl{(9.25, 0.01)} & \hl{(39.34, 0.0)} & \hl{(30.29, 0.0)} & \hl{(52.75, 0.0)} & \hl{(56.22, 0.0)} & \hl{(15.46, 0.0)} & \hl{(16.45, 0.0)} & 2.0 \\
10 & (1.11, 0.57) & \hl{(6.9, 0.03)} & \hl{(14.66, 0.0)} & \hl{(23.27, 0.0)} & \hl{(45.81, 0.0)} & \hl{(43.08, 0.0)} & \hl{(64.21, 0.0)} & \hl{(22.19, 0.0)} & (4.92, 0.09) & (0.73, 0.69) & 2.0 \\
11 & \hl{(2.87, 0.24)} & \hl{(2.33, 0.31)} & \hl{(44.44, 0.0)} & \hl{(8.53, 0.01)} & \hl{(76.58, 0.0)} & \hl{(71.44, 0.0)} & \hl{(56.23, 0.0)} & \hl{(72.08, 0.0)} & (0.87, 0.65) & (1.39, 0.5) & 2.0 \\
12 & \hl{(28.08, 0.0)} & \hl{(25.09, 0.0)} & \hl{(19.99, 0.0)} & (5.91, 0.05) & \hl{(55.93, 0.0)} & \hl{(20.08, 0.0)} & \hl{(36.31, 0.0)} & \hl{(74.86, 0.0)} & \hl{(82.86, 0.0)} & \hl{(58.39, 0.0)} & 2.0 \\
13 & \hl{(14.94, 0.0)} & \hl{(24.7, 0.0)} & \hl{(26.21, 0.0)} & \hl{(35.05, 0.0)} & \hl{(46.47, 0.0)} & \hl{(51.3, 0.0)} & \hl{(77.6, 0.0)} & \hl{(81.78, 0.0)} & (2.65, 0.27) & \hl{(2.21, 0.33)} & 2.0 \\
14 & \hl{(22.26, 0.0)} & (1.62, 0.45) & \hl{(26.61, 0.0)} & \hl{(3.17, 0.21)} & \hl{(16.75, 0.0)} & \hl{(28.12, 0.0)} & \hl{(25.21, 0.0)} & \hl{(21.38, 0.0)} & \hl{(6.75, 0.03)} & (5.92, 0.05) & 2.0 \\
15 & \hl{(8.83, 0.01)} & \hl{(11.26, 0.0)} & \hl{(11.46, 0.0)} & \hl{(4.67, 0.1)} & \hl{(35.61, 0.0)} & \hl{(42.25, 0.0)} & \hl{(29.25, 0.0)} & \hl{(48.21, 0.0)} & (2.74, 0.25) & \hl{(7.94, 0.02)} & 2.0 \\
16 & \hl{(15.65, 0.0)} & \hl{(15.55, 0.0)} & \hl{(22.97, 0.0)} & \hl{(17.2, 0.0)} & \hl{(21.59, 0.0)} & \hl{(10.91, 0.0)} & \hl{(28.34, 0.0)} & \hl{(72.67, 0.0)} & \hl{(8.19, 0.02)} & \hl{(2.16, 0.34)} & 2.0 \\
17 & \hl{(22.58, 0.0)} & \hl{(48.87, 0.0)} & \hl{(32.43, 0.0)} & \hl{(23.99, 0.0)} & \hl{(53.51, 0.0)} & \hl{(46.78, 0.0)} & \hl{(59.3, 0.0)} & \hl{(56.97, 0.0)} & \hl{(3.21, 0.2)} & (0.75, 0.69) & 2.0 \\
18 & \hl{(12.19, 0.0)} & \hl{(6.63, 0.04)} & \hl{(21.24, 0.0)} & \hl{(15.05, 0.0)} & \hl{(39.62, 0.0)} & \hl{(60.54, 0.0)} & \hl{(40.06, 0.0)} & \hl{(66.24, 0.0)} & \hl{(32.08, 0.0)} & \hl{(28.26, 0.0)} & 2.0 \\
19 & \hl{(2.34, 0.31)} & \hl{(7.41, 0.02)} & (1.5, 0.47) & \hl{(1.68, 0.43)} & \hl{(50.19, 0.0)} & \hl{(39.69, 0.0)} & \hl{(46.12, 0.0)} & \hl{(86.43, 0.0)} & \hl{(11.26, 0.0)} & \hl{(4.6, 0.1)} & 2.0 \\
20 & (5.01, 0.08) & (0.16, 0.92) & \hl{(15.55, 0.0)} & \hl{(9.31, 0.01)} & \hl{(71.32, 0.0)} & \hl{(65.09, 0.0)} & \hl{(58.39, 0.0)} & \hl{(69.39, 0.0)} & \hl{(70.2, 0.0)} & \hl{(78.02, 0.0)} & 2.0 \\
21 & \hl{(29.55, 0.0)} & \hl{(16.22, 0.0)} & \hl{(35.49, 0.0)} & \hl{(8.66, 0.01)} & \hl{(89.27, 0.0)} & \hl{(63.42, 0.0)} & \hl{(31.57, 0.0)} & \hl{(61.9, 0.0)} & (0.75, 0.69) & (1.08, 0.58) & 2.0 \\
22 & \hl{(1.76, 0.41)} & (2.67, 0.26) & (6.02, 0.05) & (0.81, 0.67) & \hl{(53.2, 0.0)} & \hl{(64.18, 0.0)} & \hl{(63.98, 0.0)} & \hl{(93.0, 0.0)} & \hl{(23.03, 0.0)} & \hl{(30.34, 0.0)} & 2.0 \\
\hline
    \end{tabular}}
    \caption{Batch1: $\chi^2$ statistic and p values to test the hypothesis that the observed ad delivery in different gender based ad destinations is proportional to the
population estimates provided by Facebook. In a majority of ads, the observed ad delivery is not proportional to the population estimates from
Facebook.}
    \label{tab:rq4-gender-Batch1}
\end{table*}

\begin{table*}[]
\resizebox{\textwidth}{!}{%
    \centering
    \begin{tabular}{|c|c|c|c|c|c|c|c|c|c|c|c|}
    \hline
    Ad ID & Solar Cells & Solar Cells (Copy) & Solar Cells + Contrarian Logo & Solar Cells + Advocacy Logo & Oil \rig{s} & Oil \rig{s} (Copy) & Oil \rig{s} + Contrarian Logo & Oil \rig{s} + Advocacy Logo & Controls & Controls (Copy) & df \\
    \hline
    \hline
    1 & \hl{(13.15, 0.0)} & \hl{(13.54, 0.0)} & \hl{(19.51, 0.0)} & \hl{(19.35, 0.0)} & \hl{(43.08, 0.0)} & \hl{(48.36, 0.0)} & \hl{(36.02, 0.0)} & \hl{(38.27, 0.0)} & \hl{(23.12, 0.0)} & \hl{(16.98, 0.0)} & 2.0 \\
2 & \hl{(6.36, 0.04)} & \hl{(8.66, 0.01)} & \hl{(13.02, 0.0)} & \hl{(27.35, 0.0)} & \hl{(41.96, 0.0)} & \hl{(50.33, 0.0)} & \hl{(20.44, 0.0)} & \hl{(44.13, 0.0)} & \hl{(43.32, 0.0)} & \hl{(48.34, 0.0)} & 2.0 \\
3 & (0.1, 0.95) & (1.88, 0.39) & \hl{(12.98, 0.0)} & \hl{(6.58, 0.04)} & \hl{(28.94, 0.0)} & \hl{(44.72, 0.0)} & \hl{(34.86, 0.0)} & \hl{(31.57, 0.0)} & \hl{(4.15, 0.13)} & (1.11, 0.58) & 2.0 \\
4 & \hl{(4.15, 0.13)} & (0.93, 0.63) & \hl{(7.71, 0.02)} & (5.04, 0.08) & \hl{(35.49, 0.0)} & \hl{(29.24, 0.0)} & \hl{(39.93, 0.0)} & \hl{(42.25, 0.0)} & (3.67, 0.16) & (2.52, 0.28) & 2.0 \\
5 & \hl{(1.8, 0.41)} & (0.8, 0.67) & (5.29, 0.07) & \hl{(12.08, 0.0)} & \hl{(45.68, 0.0)} & \hl{(30.06, 0.0)} & \hl{(29.74, 0.0)} & \hl{(43.96, 0.0)} & \hl{(13.22, 0.0)} & (0.7, 0.71) & 2.0 \\
6 & (5.02, 0.08) & \hl{(27.25, 0.0)} & \hl{(15.24, 0.0)} & \hl{(32.66, 0.0)} & \hl{(46.2, 0.0)} & \hl{(42.89, 0.0)} & \hl{(29.24, 0.0)} & \hl{(36.31, 0.0)} & \hl{(15.63, 0.0)} & \hl{(25.89, 0.0)} & 2.0 \\
7 & (0.62, 0.73) & (1.45, 0.48) & (1.24, 0.54) & (1.21, 0.55) & \hl{(46.9, 0.0)} & \hl{(37.57, 0.0)} & \hl{(32.17, 0.0)} & \hl{(44.45, 0.0)} & (5.46, 0.07) & \hl{(12.39, 0.0)} & 2.0 \\
8 & \hl{(11.68, 0.0)} & \hl{(15.83, 0.0)} & \hl{(21.01, 0.0)} & \hl{(21.34, 0.0)} & \hl{(19.35, 0.0)} & \hl{(22.34, 0.0)} & \hl{(29.8, 0.0)} & \hl{(26.61, 0.0)} & \hl{(21.14, 0.0)} & \hl{(10.0, 0.01)} & 2.0 \\
9 & \hl{(7.75, 0.02)} & \hl{(17.76, 0.0)} & \hl{(16.68, 0.0)} & \hl{(10.8, 0.0)} & \hl{(32.24, 0.0)} & \hl{(19.59, 0.0)} & \hl{(42.23, 0.0)} & \hl{(25.35, 0.0)} & (0.62, 0.73) & (1.14, 0.57) & 2.0 \\
10 & \hl{(8.57, 0.01)} & \hl{(9.78, 0.01)} & \hl{(6.82, 0.03)} & \hl{(13.52, 0.0)} & \hl{(44.89, 0.0)} & \hl{(46.08, 0.0)} & \hl{(45.04, 0.0)} & \hl{(21.24, 0.0)} & \hl{(22.64, 0.0)} & \hl{(21.59, 0.0)} & 2.0 \\
11 & \hl{(23.07, 0.0)} & \hl{(7.83, 0.02)} & \hl{(25.72, 0.0)} & \hl{(28.07, 0.0)} & \hl{(32.81, 0.0)} & \hl{(50.33, 0.0)} & \hl{(31.83, 0.0)} & \hl{(52.46, 0.0)} & \hl{(42.25, 0.0)} & \hl{(38.69, 0.0)} & 2.0 \\
12 & (0.8, 0.67) & (3.67, 0.16) & \hl{(4.64, 0.1)} & \hl{(12.75, 0.0)} & \hl{(68.25, 0.0)} & \hl{(37.73, 0.0)} & \hl{(61.95, 0.0)} & \hl{(33.42, 0.0)} & (1.07, 0.59) & (0.87, 0.65) & 2.0 \\
13 & \hl{(23.55, 0.0)} & \hl{(24.22, 0.0)} & \hl{(19.3, 0.0)} & (5.92, 0.05) & \hl{(52.15, 0.0)} & \hl{(61.95, 0.0)} & \hl{(51.31, 0.0)} & \hl{(36.31, 0.0)} & (0.71, 0.7) & (2.53, 0.28) & 2.0 \\
14 & \hl{(20.27, 0.0)} & \hl{(32.3, 0.0)} & \hl{(30.63, 0.0)} & \hl{(20.02, 0.0)} & \hl{(40.1, 0.0)} & \hl{(21.37, 0.0)} & \hl{(6.7, 0.04)} & \hl{(37.37, 0.0)} & \hl{(12.74, 0.0)} & \hl{(20.28, 0.0)} & 2.0 \\
15 & (3.43, 0.18) & \hl{(6.47, 0.04)} & \hl{(4.18, 0.12)} & \hl{(17.1, 0.0)} & \hl{(45.04, 0.0)} & \hl{(66.64, 0.0)} & \hl{(55.17, 0.0)} & \hl{(56.27, 0.0)} & \hl{(62.28, 0.0)} & \hl{(30.07, 0.0)} & 2.0 \\
16 & \hl{(6.52, 0.04)} & (5.03, 0.08) & \hl{(8.97, 0.01)} & \hl{(2.23, 0.33)} & \hl{(53.74, 0.0)} & \hl{(16.32, 0.0)} & \hl{(35.9, 0.0)} & \hl{(38.07, 0.0)} & \hl{(11.8, 0.0)} & (4.98, 0.08) & 2.0 \\
17 & \hl{(15.63, 0.0)} & \hl{(2.39, 0.3)} & \hl{(20.27, 0.0)} & \hl{(22.73, 0.0)} & \hl{(69.74, 0.0)} & \hl{(32.17, 0.0)} & \hl{(33.42, 0.0)} & \hl{(53.08, 0.0)} & \hl{(2.95, 0.23)} & (1.91, 0.39) & 2.0 \\
18 & \hl{(20.76, 0.0)} & \hl{(28.73, 0.0)} & (1.18, 0.55) & (5.63, 0.06) & \hl{(51.64, 0.0)} & \hl{(35.57, 0.0)} & \hl{(46.63, 0.0)} & \hl{(54.0, 0.0)} & \hl{(17.65, 0.0)} & \hl{(23.36, 0.0)} & 2.0 \\
19 & \hl{(15.91, 0.0)} & \hl{(21.0, 0.0)} & \hl{(15.65, 0.0)} & \hl{(14.85, 0.0)} & \hl{(18.43, 0.0)} & \hl{(14.12, 0.0)} & \hl{(20.72, 0.0)} & \hl{(43.73, 0.0)} & (5.02, 0.08) & \hl{(16.43, 0.0)} & 2.0 \\
20 & \hl{(7.69, 0.02)} & (5.59, 0.06) & \hl{(4.6, 0.1)} & (4.81, 0.09) & \hl{(27.4, 0.0)} & \hl{(62.71, 0.0)} & \hl{(22.96, 0.0)} & \hl{(25.22, 0.0)} & \hl{(37.73, 0.0)} & \hl{(57.77, 0.0)} & 2.0 \\
21 & (5.0, 0.08) & \hl{(13.87, 0.0)} & (2.66, 0.26) & \hl{(16.76, 0.0)} & \hl{(72.81, 0.0)} & \hl{(44.36, 0.0)} & \hl{(50.15, 0.0)} & \hl{(40.06, 0.0)} & \hl{(10.89, 0.0)} & \hl{(12.07, 0.0)} & 2.0 \\
22 & \hl{(13.27, 0.0)} & \hl{(4.12, 0.13)} & (2.05, 0.36) & \hl{(11.01, 0.0)} & \hl{(31.11, 0.0)} & \hl{(32.02, 0.0)} & \hl{(17.05, 0.0)} & \hl{(50.76, 0.0)} & \hl{(25.42, 0.0)} & \hl{(31.7, 0.0)} & 2.0 \\
\hline
    \end{tabular}}
    \caption{Batch2: $\chi^2$ statistic and p values to test the hypothesis that the observed ad delivery in different gender based ad destinations is proportional to the
population estimates provided by Facebook. In a majority of ads, the observed ad delivery is not proportional to the population estimates from
Facebook.}
    \label{tab:rq4-gender-Batch2}
\end{table*}
\begin{table*}[]
\resizebox{\textwidth}{!}{%
    \centering
    \begin{tabular}{|c|c|c|c|c|c|c|c|c|c|c|c|}
    \hline
    Ad ID & Solar Cells & Solar Cells (Copy) & Solar Cells + Contrarian Logo & Solar Cells + Advocacy Logo & Oil \rig{s} & Oil \rig{s} (Copy) & Oil \rig{s} + Contrarian Logo & Oil \rig{s} + Advocacy Logo & Controls & Controls (Copy) & df \\
    \hline
    \hline
    1 & \hl{(15.11, 0.0)} & \hl{(8.34, 0.02)} & (5.68, 0.06) & \hl{(4.1, 0.13)} & \hl{(29.38, 0.0)} & \hl{(41.0, 0.0)} & \hl{(42.93, 0.0)} & \hl{(18.44, 0.0)} & \hl{(2.41, 0.3)} & (0.92, 0.63) & 2.0 \\
2 & (2.01, 0.37) & (2.02, 0.36) & \hl{(21.01, 0.0)} & \hl{(18.57, 0.0)} & \hl{(26.88, 0.0)} & \hl{(15.92, 0.0)} & \hl{(27.84, 0.0)} & \hl{(50.68, 0.0)} & \hl{(26.17, 0.0)} & \hl{(18.83, 0.0)} & 2.0 \\
3 & (5.55, 0.06) & \hl{(12.94, 0.0)} & \hl{(13.95, 0.0)} & \hl{(7.4, 0.02)} & \hl{(20.01, 0.0)} & \hl{(18.17, 0.0)} & \hl{(52.4, 0.0)} & \hl{(39.28, 0.0)} & \hl{(2.21, 0.33)} & \hl{(8.07, 0.02)} & 2.0 \\
4 & \hl{(15.85, 0.0)} & \hl{(18.57, 0.0)} & (0.65, 0.72) & \hl{(10.15, 0.01)} & \hl{(38.34, 0.0)} & \hl{(12.01, 0.0)} & \hl{(20.09, 0.0)} & \hl{(47.73, 0.0)} & \hl{(2.43, 0.3)} & (0.65, 0.72) & 2.0 \\
5 & \hl{(17.18, 0.0)} & \hl{(13.51, 0.0)} & \hl{(10.62, 0.0)} & \hl{(2.85, 0.24)} & \hl{(39.81, 0.0)} & \hl{(37.08, 0.0)} & \hl{(34.45, 0.0)} & \hl{(27.29, 0.0)} & \hl{(4.12, 0.13)} & \hl{(10.49, 0.01)} & 2.0 \\
6 & \hl{(11.92, 0.0)} & (5.46, 0.07) & (1.18, 0.55) & (2.02, 0.36) & \hl{(23.58, 0.0)} & \hl{(19.89, 0.0)} & \hl{(37.07, 0.0)} & \hl{(47.22, 0.0)} & (0.57, 0.75) & (0.56, 0.76) & 2.0 \\
7 & \hl{(23.16, 0.0)} & \hl{(13.4, 0.0)} & \hl{(10.06, 0.01)} & \hl{(15.85, 0.0)} & \hl{(47.08, 0.0)} & \hl{(40.41, 0.0)} & \hl{(37.42, 0.0)} & \hl{(29.76, 0.0)} & \hl{(33.42, 0.0)} & \hl{(76.46, 0.0)} & 2.0 \\
8 & \hl{(22.49, 0.0)} & \hl{(18.74, 0.0)} & \hl{(16.61, 0.0)} & \hl{(9.19, 0.01)} & \hl{(20.42, 0.0)} & \hl{(27.88, 0.0)} & \hl{(32.22, 0.0)} & \hl{(28.91, 0.0)} & \hl{(19.93, 0.0)} & \hl{(7.18, 0.03)} & 2.0 \\
9 & \hl{(16.6, 0.0)} & \hl{(25.17, 0.0)} & (0.47, 0.79) & \hl{(2.94, 0.23)} & \hl{(19.87, 0.0)} & \hl{(53.24, 0.0)} & \hl{(26.0, 0.0)} & \hl{(25.1, 0.0)} & \hl{(96.59, 0.0)} & \hl{(81.96, 0.0)} & 2.0 \\
10 & \hl{(13.96, 0.0)} & \hl{(3.0, 0.22)} & \hl{(20.16, 0.0)} & \hl{(7.28, 0.03)} & \hl{(40.41, 0.0)} & \hl{(37.58, 0.0)} & \hl{(37.48, 0.0)} & \hl{(42.85, 0.0)} & \hl{(32.39, 0.0)} & \hl{(31.47, 0.0)} & 2.0 \\
11 & \hl{(18.34, 0.0)} & (0.11, 0.95) & \hl{(15.45, 0.0)} & \hl{(12.82, 0.0)} & \hl{(12.98, 0.0)} & \hl{(47.29, 0.0)} & \hl{(24.14, 0.0)} & \hl{(33.38, 0.0)} & \hl{(2.13, 0.34)} & (0.53, 0.77) & 2.0 \\
12 & \hl{(12.82, 0.0)} & \hl{(17.18, 0.0)} & \hl{(3.22, 0.2)} & \hl{(8.05, 0.02)} & \hl{(13.96, 0.0)} & \hl{(17.61, 0.0)} & \hl{(31.9, 0.0)} & \hl{(36.31, 0.0)} & \hl{(8.49, 0.01)} & (3.65, 0.16) & 2.0 \\
13 & (0.24, 0.89) & (1.25, 0.53) & \hl{(8.18, 0.02)} & \hl{(2.96, 0.23)} & \hl{(32.48, 0.0)} & \hl{(38.63, 0.0)} & \hl{(52.15, 0.0)} & \hl{(42.56, 0.0)} & (2.66, 0.26) & (1.07, 0.59) & 2.0 \\
14 & \hl{(7.8, 0.02)} & \hl{(6.87, 0.03)} & \hl{(1.63, 0.44)} & (3.64, 0.16) & \hl{(41.26, 0.0)} & \hl{(36.21, 0.0)} & \hl{(35.72, 0.0)} & \hl{(20.82, 0.0)} & (2.05, 0.36) & (0.73, 0.69) & 2.0 \\
15 & \hl{(10.7, 0.0)} & (5.47, 0.06) & \hl{(28.86, 0.0)} & \hl{(6.97, 0.03)} & \hl{(23.88, 0.0)} & \hl{(23.17, 0.0)} & \hl{(39.34, 0.0)} & \hl{(15.25, 0.0)} & \hl{(8.76, 0.01)} & \hl{(16.17, 0.0)} & 2.0 \\
16 & \hl{(10.62, 0.0)} & \hl{(11.17, 0.0)} & \hl{(24.14, 0.0)} & \hl{(15.55, 0.0)} & \hl{(49.94, 0.0)} & \hl{(30.71, 0.0)} & \hl{(26.25, 0.0)} & \hl{(23.72, 0.0)} & \hl{(1.79, 0.41)} & (2.11, 0.35) & 2.0 \\
17 & (3.77, 0.15) & \hl{(3.02, 0.22)} & \hl{(18.22, 0.0)} & (2.45, 0.29) & \hl{(29.88, 0.0)} & \hl{(27.83, 0.0)} & \hl{(37.54, 0.0)} & \hl{(30.83, 0.0)} & (3.54, 0.17) & \hl{(9.17, 0.01)} & 2.0 \\
18 & \hl{(11.64, 0.0)} & \hl{(7.98, 0.02)} & \hl{(18.26, 0.0)} & \hl{(4.55, 0.1)} & \hl{(31.57, 0.0)} & \hl{(45.95, 0.0)} & \hl{(28.75, 0.0)} & \hl{(17.71, 0.0)} & \hl{(42.25, 0.0)} & \hl{(37.21, 0.0)} & 2.0 \\
19 & \hl{(12.48, 0.0)} & \hl{(8.18, 0.02)} & \hl{(1.8, 0.41)} & \hl{(11.73, 0.0)} & \hl{(30.6, 0.0)} & \hl{(43.29, 0.0)} & \hl{(39.0, 0.0)} & \hl{(18.53, 0.0)} & \hl{(1.63, 0.44)} & (1.08, 0.58) & 2.0 \\
20 & (3.42, 0.18) & (5.09, 0.08) & \hl{(3.25, 0.2)} & \hl{(8.02, 0.02)} & \hl{(59.64, 0.0)} & \hl{(40.24, 0.0)} & \hl{(51.62, 0.0)} & \hl{(32.48, 0.0)} & \hl{(42.12, 0.0)} & \hl{(52.65, 0.0)} & 2.0 \\
21 & \hl{(7.76, 0.02)} & \hl{(15.83, 0.0)} & \hl{(8.25, 0.02)} & \hl{(20.84, 0.0)} & \hl{(26.87, 0.0)} & \hl{(42.25, 0.0)} & \hl{(33.14, 0.0)} & \hl{(62.08, 0.0)} & \hl{(32.65, 0.0)} & \hl{(44.65, 0.0)} & 2.0 \\
\hline
    \end{tabular}}
    \caption{Batch3: $\chi^2$ statistic and p values to test the hypothesis that the observed ad delivery in different gender based ad destinations is proportional to the
population estimates provided by Facebook. In a majority of ads, the observed ad delivery is not proportional to the population estimates from
Facebook.}
    \label{tab:rq4-gender-Batch3}
\end{table*}

\subsubsection{Age}
$\chi^2$ statistics and associated p-values for the 3 batches of ads in our experiment show whether observed ad delivery was proportional to Facebook's population estimates. We find that in a majority of cases, the values were not proportional, as shown in tables \ref{tab:rq4-age-Batch1}, \ref{tab:rq4-age-Batch2} and \ref{tab:rq4-age-Batch3}
\begin{table*}[]
\resizebox{\textwidth}{!}{%
    \centering
    \begin{tabular}{|c|c|c|c|c|c|c|c|c|c|c|c|}
    \hline
    Ad ID & Solar Cells & Solar Cells (Copy) & Solar Cells + Contrarian Logo & Solar Cells + Advocacy Logo & Oil \rig{s} & Oil \rig{s} (Copy) & Oil \rig{s} + Contrarian Logo & Oil \rig{s} + Advocacy Logo & Controls & Controls (Copy) & df \\
    \hline
    \hline
    1 & \hl{(302.51, 0.0)} & \hl{(385.41, 0.0)} & \hl{(369.35, 0.0)} & \hl{(294.82, 0.0)} & \hl{(205.49, 0.0)} & \hl{(233.23, 0.0)} & \hl{(234.73, 0.0)} & \hl{(289.0, 0.0)} & \hl{(463.34, 0.0)} & \hl{(417.05, 0.0)} & 5.0 \\
2 & \hl{(301.61, 0.0)} & \hl{(282.14, 0.0)} & \hl{(277.93, 0.0)} & \hl{(381.29, 0.0)} & \hl{(339.91, 0.0)} & \hl{(341.34, 0.0)} & \hl{(421.03, 0.0)} & \hl{(362.77, 0.0)} & \hl{(187.0, 0.0)} & \hl{(312.64, 0.0)} & 5.0 \\
3 & \hl{(166.25, 0.0)} & \hl{(255.77, 0.0)} & \hl{(296.22, 0.0)} & \hl{(296.22, 0.0)} & \hl{(264.97, 0.0)} & \hl{(225.35, 0.0)} & \hl{(270.68, 0.0)} & \hl{(353.26, 0.0)} & \hl{(355.09, 0.0)} & \hl{(381.51, 0.0)} & 5.0 \\
4 & \hl{(364.75, 0.0)} & \hl{(426.58, 0.0)} & \hl{(394.32, 0.0)} & \hl{(321.63, 0.0)} & \hl{(412.74, 0.0)} & \hl{(290.78, 0.0)} & \hl{(357.23, 0.0)} & \hl{(388.61, 0.0)} & \hl{(120.63, 0.0)} & \hl{(135.6, 0.0)} & 5.0 \\
5 & \hl{(425.98, 0.0)} & \hl{(243.8, 0.0)} & \hl{(161.96, 0.0)} & \hl{(332.67, 0.0)} & \hl{(375.14, 0.0)} & \hl{(403.02, 0.0)} & \hl{(398.8, 0.0)} & \hl{(558.47, 0.0)} & \hl{(318.63, 0.0)} & \hl{(307.82, 0.0)} & 5.0 \\
6 & \hl{(304.33, 0.0)} & \hl{(246.75, 0.0)} & \hl{(205.39, 0.0)} & \hl{(286.74, 0.0)} & \hl{(288.23, 0.0)} & \hl{(432.48, 0.0)} & \hl{(431.28, 0.0)} & \hl{(362.94, 0.0)} & \hl{(333.75, 0.0)} & \hl{(322.45, 0.0)} & 5.0 \\
7 & \hl{(442.47, 0.0)} & \hl{(378.23, 0.0)} & \hl{(333.58, 0.0)} & \hl{(599.85, 0.0)} & \hl{(206.75, 0.0)} & \hl{(265.03, 0.0)} & \hl{(396.46, 0.0)} & \hl{(227.15, 0.0)} & \hl{(350.31, 0.0)} & \hl{(369.44, 0.0)} & 5.0 \\
8 & \hl{(430.26, 0.0)} & \hl{(360.6, 0.0)} & \hl{(421.0, 0.0)} & \hl{(379.51, 0.0)} & \hl{(319.55, 0.0)} & \hl{(287.94, 0.0)} & \hl{(212.48, 0.0)} & \hl{(247.41, 0.0)} & \hl{(385.89, 0.0)} & \hl{(404.44, 0.0)} & 5.0 \\
9 & \hl{(245.63, 0.0)} & \hl{(267.54, 0.0)} & \hl{(212.99, 0.0)} & \hl{(449.16, 0.0)} & \hl{(284.62, 0.0)} & \hl{(343.23, 0.0)} & \hl{(370.28, 0.0)} & \hl{(336.25, 0.0)} & \hl{(231.4, 0.0)} & \hl{(276.03, 0.0)} & 5.0 \\
10 & \hl{(363.57, 0.0)} & \hl{(357.38, 0.0)} & \hl{(279.11, 0.0)} & \hl{(357.89, 0.0)} & \hl{(463.32, 0.0)} & \hl{(482.42, 0.0)} & \hl{(336.6, 0.0)} & \hl{(489.62, 0.0)} & \hl{(293.96, 0.0)} & \hl{(359.55, 0.0)} & 5.0 \\
11 & \hl{(341.33, 0.0)} & \hl{(315.76, 0.0)} & \hl{(236.36, 0.0)} & \hl{(375.15, 0.0)} & \hl{(475.94, 0.0)} & \hl{(335.32, 0.0)} & \hl{(404.95, 0.0)} & \hl{(475.71, 0.0)} & \hl{(201.3, 0.0)} & \hl{(240.13, 0.0)} & 5.0 \\
12 & \hl{(144.08, 0.0)} & \hl{(147.04, 0.0)} & \hl{(55.71, 0.0)} & \hl{(268.9, 0.0)} & \hl{(335.24, 0.0)} & \hl{(302.92, 0.0)} & \hl{(231.59, 0.0)} & \hl{(419.68, 0.0)} & \hl{(277.81, 0.0)} & \hl{(207.62, 0.0)} & 5.0 \\
13 & \hl{(241.98, 0.0)} & \hl{(282.95, 0.0)} & \hl{(201.61, 0.0)} & \hl{(478.99, 0.0)} & \hl{(345.43, 0.0)} & \hl{(392.46, 0.0)} & \hl{(242.41, 0.0)} & \hl{(355.85, 0.0)} & \hl{(318.22, 0.0)} & \hl{(310.22, 0.0)} & 5.0 \\
14 & \hl{(354.13, 0.0)} & \hl{(399.64, 0.0)} & \hl{(278.52, 0.0)} & \hl{(453.76, 0.0)} & \hl{(280.75, 0.0)} & \hl{(302.37, 0.0)} & \hl{(330.82, 0.0)} & \hl{(328.58, 0.0)} & \hl{(355.09, 0.0)} & \hl{(286.91, 0.0)} & 5.0 \\
15 & \hl{(404.76, 0.0)} & \hl{(474.57, 0.0)} & \hl{(275.91, 0.0)} & \hl{(460.24, 0.0)} & \hl{(348.76, 0.0)} & \hl{(349.85, 0.0)} & \hl{(387.33, 0.0)} & \hl{(275.88, 0.0)} & \hl{(329.08, 0.0)} & \hl{(308.53, 0.0)} & 5.0 \\
16 & \hl{(370.94, 0.0)} & \hl{(282.51, 0.0)} & \hl{(269.82, 0.0)} & \hl{(452.03, 0.0)} & \hl{(374.04, 0.0)} & \hl{(339.74, 0.0)} & \hl{(321.55, 0.0)} & \hl{(289.84, 0.0)} & \hl{(420.82, 0.0)} & \hl{(326.29, 0.0)} & 5.0 \\
17 & \hl{(412.24, 0.0)} & \hl{(292.65, 0.0)} & \hl{(227.68, 0.0)} & \hl{(283.6, 0.0)} & \hl{(363.34, 0.0)} & \hl{(365.32, 0.0)} & \hl{(425.82, 0.0)} & \hl{(418.92, 0.0)} & \hl{(159.56, 0.0)} & \hl{(133.34, 0.0)} & 5.0 \\
18 & \hl{(357.67, 0.0)} & \hl{(251.11, 0.0)} & \hl{(181.95, 0.0)} & \hl{(288.47, 0.0)} & \hl{(189.9, 0.0)} & \hl{(281.19, 0.0)} & \hl{(161.1, 0.0)} & \hl{(294.71, 0.0)} & \hl{(417.72, 0.0)} & \hl{(488.35, 0.0)} & 5.0 \\
19 & \hl{(519.54, 0.0)} & \hl{(506.76, 0.0)} & \hl{(298.38, 0.0)} & \hl{(314.85, 0.0)} & \hl{(223.94, 0.0)} & \hl{(243.3, 0.0)} & \hl{(322.27, 0.0)} & \hl{(417.26, 0.0)} & \hl{(163.83, 0.0)} & \hl{(182.85, 0.0)} & 5.0 \\
20 & \hl{(282.22, 0.0)} & \hl{(332.25, 0.0)} & \hl{(339.77, 0.0)} & \hl{(334.1, 0.0)} & \hl{(294.91, 0.0)} & \hl{(232.2, 0.0)} & \hl{(333.32, 0.0)} & \hl{(409.62, 0.0)} & \hl{(308.75, 0.0)} & \hl{(304.12, 0.0)} & 5.0 \\
21 & \hl{(330.17, 0.0)} & \hl{(394.24, 0.0)} & \hl{(139.97, 0.0)} & \hl{(348.03, 0.0)} & \hl{(349.87, 0.0)} & \hl{(217.6, 0.0)} & \hl{(331.03, 0.0)} & \hl{(181.07, 0.0)} & \hl{(254.78, 0.0)} & \hl{(295.49, 0.0)} & 5.0 \\
22 & \hl{(348.37, 0.0)} & \hl{(191.77, 0.0)} & \hl{(226.59, 0.0)} & \hl{(307.12, 0.0)} & \hl{(353.62, 0.0)} & \hl{(258.88, 0.0)} & \hl{(312.89, 0.0)} & \hl{(385.93, 0.0)} & \hl{(105.25, 0.0)} & \hl{(86.11, 0.0)} & 5.0 \\
\hline
    \end{tabular}}
    \caption{Batch1: $\chi^2$ statistic and p values to test the hypothesis that the observed ad delivery in different age based ad destinations is proportional to the
population estimates provided by Facebook. In a majority of ads, the observed ad delivery is not proportional to the population estimates from
Facebook.}
    \label{tab:rq4-age-Batch1}
\end{table*}

\begin{table*}[]
\resizebox{\textwidth}{!}{%
    \centering
    \begin{tabular}{|c|c|c|c|c|c|c|c|c|c|c|c|}
    \hline
    Ad ID & Solar Cells & Solar Cells (Copy) & Solar Cells + Contrarian Logo & Solar Cells + Advocacy Logo & Oil \rig{s} & Oil \rig{s} (Copy) & Oil \rig{s} + Contrarian Logo & Oil \rig{s} + Advocacy Logo & Controls & Controls (Copy) & df \\
    \hline
    \hline
    1 & \hl{(240.66, 0.0)} & \hl{(264.51, 0.0)} & \hl{(350.49, 0.0)} & \hl{(323.42, 0.0)} & \hl{(337.91, 0.0)} & \hl{(555.35, 0.0)} & \hl{(461.74, 0.0)} & \hl{(261.57, 0.0)} & \hl{(360.69, 0.0)} & \hl{(360.18, 0.0)} & 5.0 \\
2 & \hl{(415.28, 0.0)} & \hl{(400.2, 0.0)} & \hl{(436.58, 0.0)} & \hl{(420.91, 0.0)} & \hl{(369.26, 0.0)} & \hl{(409.56, 0.0)} & \hl{(384.11, 0.0)} & \hl{(400.43, 0.0)} & \hl{(294.99, 0.0)} & \hl{(308.44, 0.0)} & 5.0 \\
3 & \hl{(404.84, 0.0)} & \hl{(315.4, 0.0)} & \hl{(232.59, 0.0)} & \hl{(306.24, 0.0)} & \hl{(426.16, 0.0)} & \hl{(469.0, 0.0)} & \hl{(319.51, 0.0)} & \hl{(434.67, 0.0)} & \hl{(273.88, 0.0)} & \hl{(323.4, 0.0)} & 5.0 \\
4 & \hl{(414.86, 0.0)} & \hl{(547.21, 0.0)} & \hl{(328.6, 0.0)} & \hl{(506.07, 0.0)} & \hl{(356.52, 0.0)} & \hl{(356.09, 0.0)} & \hl{(353.92, 0.0)} & \hl{(372.42, 0.0)} & \hl{(374.36, 0.0)} & \hl{(359.82, 0.0)} & 5.0 \\
5 & \hl{(365.52, 0.0)} & \hl{(419.78, 0.0)} & \hl{(406.02, 0.0)} & \hl{(408.53, 0.0)} & \hl{(488.13, 0.0)} & \hl{(431.04, 0.0)} & \hl{(458.54, 0.0)} & \hl{(403.56, 0.0)} & \hl{(330.86, 0.0)} & \hl{(350.85, 0.0)} & 5.0 \\
6 & \hl{(489.75, 0.0)} & \hl{(332.99, 0.0)} & \hl{(356.26, 0.0)} & \hl{(324.06, 0.0)} & \hl{(470.71, 0.0)} & \hl{(365.23, 0.0)} & \hl{(422.13, 0.0)} & \hl{(385.0, 0.0)} & \hl{(393.81, 0.0)} & \hl{(285.07, 0.0)} & 5.0 \\
7 & \hl{(432.63, 0.0)} & \hl{(384.93, 0.0)} & \hl{(315.52, 0.0)} & \hl{(438.9, 0.0)} & \hl{(260.67, 0.0)} & \hl{(412.63, 0.0)} & \hl{(314.84, 0.0)} & \hl{(474.65, 0.0)} & \hl{(329.19, 0.0)} & \hl{(495.3, 0.0)} & 5.0 \\
8 & \hl{(211.03, 0.0)} & \hl{(332.04, 0.0)} & \hl{(339.74, 0.0)} & \hl{(415.89, 0.0)} & \hl{(354.98, 0.0)} & \hl{(403.07, 0.0)} & \hl{(343.7, 0.0)} & \hl{(487.06, 0.0)} & \hl{(391.13, 0.0)} & \hl{(407.68, 0.0)} & 5.0 \\
9 & \hl{(314.9, 0.0)} & \hl{(201.05, 0.0)} & \hl{(379.61, 0.0)} & \hl{(412.39, 0.0)} & \hl{(366.69, 0.0)} & \hl{(360.67, 0.0)} & \hl{(312.42, 0.0)} & \hl{(451.65, 0.0)} & \hl{(318.41, 0.0)} & \hl{(399.02, 0.0)} & 5.0 \\
10 & \hl{(396.9, 0.0)} & \hl{(480.14, 0.0)} & \hl{(552.1, 0.0)} & \hl{(365.92, 0.0)} & \hl{(418.11, 0.0)} & \hl{(407.0, 0.0)} & \hl{(396.09, 0.0)} & \hl{(445.16, 0.0)} & \hl{(364.76, 0.0)} & \hl{(395.97, 0.0)} & 5.0 \\
11 & \hl{(368.89, 0.0)} & \hl{(471.84, 0.0)} & \hl{(409.81, 0.0)} & \hl{(402.1, 0.0)} & \hl{(377.42, 0.0)} & \hl{(465.99, 0.0)} & \hl{(427.14, 0.0)} & \hl{(442.5, 0.0)} & \hl{(376.81, 0.0)} & \hl{(398.4, 0.0)} & 5.0 \\
12 & \hl{(416.23, 0.0)} & \hl{(404.96, 0.0)} & \hl{(437.94, 0.0)} & \hl{(519.32, 0.0)} & \hl{(406.97, 0.0)} & \hl{(378.15, 0.0)} & \hl{(434.89, 0.0)} & \hl{(330.76, 0.0)} & \hl{(414.12, 0.0)} & \hl{(557.19, 0.0)} & 5.0 \\
13 & \hl{(336.55, 0.0)} & \hl{(322.36, 0.0)} & \hl{(412.29, 0.0)} & \hl{(415.35, 0.0)} & \hl{(423.83, 0.0)} & \hl{(415.47, 0.0)} & \hl{(285.06, 0.0)} & \hl{(436.22, 0.0)} & \hl{(346.21, 0.0)} & \hl{(430.66, 0.0)} & 5.0 \\
14 & \hl{(359.34, 0.0)} & \hl{(324.33, 0.0)} & \hl{(329.79, 0.0)} & \hl{(346.03, 0.0)} & \hl{(340.7, 0.0)} & \hl{(322.75, 0.0)} & \hl{(282.52, 0.0)} & \hl{(334.25, 0.0)} & \hl{(338.0, 0.0)} & \hl{(420.29, 0.0)} & 5.0 \\
15 & \hl{(295.33, 0.0)} & \hl{(387.94, 0.0)} & \hl{(504.13, 0.0)} & \hl{(394.17, 0.0)} & \hl{(364.65, 0.0)} & \hl{(435.43, 0.0)} & \hl{(309.52, 0.0)} & \hl{(379.58, 0.0)} & \hl{(387.64, 0.0)} & \hl{(390.2, 0.0)} & 5.0 \\
16 & \hl{(327.19, 0.0)} & \hl{(359.88, 0.0)} & \hl{(425.26, 0.0)} & \hl{(433.55, 0.0)} & \hl{(425.44, 0.0)} & \hl{(471.97, 0.0)} & \hl{(427.99, 0.0)} & \hl{(460.84, 0.0)} & \hl{(358.2, 0.0)} & \hl{(359.79, 0.0)} & 5.0 \\
17 & \hl{(374.68, 0.0)} & \hl{(369.65, 0.0)} & \hl{(398.02, 0.0)} & \hl{(501.06, 0.0)} & \hl{(609.47, 0.0)} & \hl{(400.75, 0.0)} & \hl{(494.88, 0.0)} & \hl{(440.04, 0.0)} & \hl{(435.27, 0.0)} & \hl{(353.39, 0.0)} & 5.0 \\
18 & \hl{(319.14, 0.0)} & \hl{(309.92, 0.0)} & \hl{(359.35, 0.0)} & \hl{(519.02, 0.0)} & \hl{(299.78, 0.0)} & \hl{(382.94, 0.0)} & \hl{(391.96, 0.0)} & \hl{(481.99, 0.0)} & \hl{(398.34, 0.0)} & \hl{(384.75, 0.0)} & 5.0 \\
19 & \hl{(359.61, 0.0)} & \hl{(467.86, 0.0)} & \hl{(346.51, 0.0)} & \hl{(397.74, 0.0)} & \hl{(388.51, 0.0)} & \hl{(501.11, 0.0)} & \hl{(525.51, 0.0)} & \hl{(563.1, 0.0)} & \hl{(324.68, 0.0)} & \hl{(328.59, 0.0)} & 5.0 \\
20 & \hl{(476.38, 0.0)} & \hl{(397.95, 0.0)} & \hl{(378.06, 0.0)} & \hl{(400.8, 0.0)} & \hl{(441.96, 0.0)} & \hl{(463.14, 0.0)} & \hl{(489.42, 0.0)} & \hl{(496.54, 0.0)} & \hl{(347.52, 0.0)} & \hl{(357.86, 0.0)} & 5.0 \\
21 & \hl{(340.65, 0.0)} & \hl{(419.51, 0.0)} & \hl{(478.39, 0.0)} & \hl{(444.12, 0.0)} & \hl{(511.78, 0.0)} & \hl{(400.6, 0.0)} & \hl{(341.76, 0.0)} & \hl{(434.09, 0.0)} & \hl{(363.68, 0.0)} & \hl{(314.35, 0.0)} & 5.0 \\
22 & \hl{(463.49, 0.0)} & \hl{(297.57, 0.0)} & \hl{(407.21, 0.0)} & \hl{(442.11, 0.0)} & \hl{(339.03, 0.0)} & \hl{(432.22, 0.0)} & \hl{(359.29, 0.0)} & \hl{(533.67, 0.0)} & \hl{(342.47, 0.0)} & \hl{(467.64, 0.0)} & 5.0 \\
\hline
    \end{tabular}}
    \caption{Batch2: $\chi^2$ statistic and p values to test the hypothesis that the observed ad delivery in different age based ad destinations is proportional to the
population estimates provided by Facebook. In a majority of ads, the observed ad delivery is not proportional to the population estimates from
Facebook.}
    \label{tab:rq4-age-Batch2}
\end{table*}

\begin{table*}[]
\resizebox{\textwidth}{!}{%
    \centering
    \begin{tabular}{|c|c|c|c|c|c|c|c|c|c|c|c|}
    \hline
    Ad ID & Solar Cells & Solar Cells (Copy) & Solar Cells + Contrarian Logo & Solar Cells + Advocacy Logo & Oil \rig{s} & Oil \rig{s} (Copy) & Oil \rig{s} + Contrarian Logo & Oil \rig{s} + Advocacy Logo & Controls & Controls (Copy) & df \\
    \hline
    \hline
    1 & \hl{(253.28, 0.0)} & \hl{(237.93, 0.0)} & \hl{(257.05, 0.0)} & \hl{(410.96, 0.0)} & \hl{(401.91, 0.0)} & \hl{(360.31, 0.0)} & \hl{(296.15, 0.0)} & \hl{(328.25, 0.0)} & \hl{(372.85, 0.0)} & \hl{(392.32, 0.0)} & 5.0 \\
2 & \hl{(441.44, 0.0)} & \hl{(375.16, 0.0)} & \hl{(232.82, 0.0)} & \hl{(314.09, 0.0)} & \hl{(412.43, 0.0)} & \hl{(348.91, 0.0)} & \hl{(330.51, 0.0)} & \hl{(441.48, 0.0)} & \hl{(340.34, 0.0)} & \hl{(390.5, 0.0)} & 5.0 \\
3 & \hl{(251.24, 0.0)} & \hl{(268.49, 0.0)} & \hl{(269.96, 0.0)} & \hl{(298.99, 0.0)} & \hl{(197.26, 0.0)} & \hl{(363.22, 0.0)} & \hl{(239.99, 0.0)} & \hl{(329.79, 0.0)} & \hl{(352.6, 0.0)} & \hl{(379.0, 0.0)} & 5.0 \\
4 & \hl{(259.29, 0.0)} & \hl{(271.41, 0.0)} & \hl{(252.32, 0.0)} & \hl{(342.74, 0.0)} & \hl{(353.05, 0.0)} & \hl{(330.0, 0.0)} & \hl{(308.8, 0.0)} & \hl{(349.5, 0.0)} & \hl{(413.49, 0.0)} & \hl{(406.07, 0.0)} & 5.0 \\
5 & \hl{(386.55, 0.0)} & \hl{(380.63, 0.0)} & \hl{(339.54, 0.0)} & \hl{(395.81, 0.0)} & \hl{(350.5, 0.0)} & \hl{(338.25, 0.0)} & \hl{(357.79, 0.0)} & \hl{(334.68, 0.0)} & \hl{(434.17, 0.0)} & \hl{(365.28, 0.0)} & 5.0 \\
6 & \hl{(203.4, 0.0)} & \hl{(262.52, 0.0)} & \hl{(262.56, 0.0)} & \hl{(273.75, 0.0)} & \hl{(346.04, 0.0)} & \hl{(287.88, 0.0)} & \hl{(369.62, 0.0)} & \hl{(378.96, 0.0)} & \hl{(310.31, 0.0)} & \hl{(339.41, 0.0)} & 5.0 \\
7 & \hl{(304.04, 0.0)} & \hl{(222.57, 0.0)} & \hl{(215.38, 0.0)} & \hl{(278.87, 0.0)} & \hl{(343.1, 0.0)} & \hl{(390.82, 0.0)} & \hl{(240.37, 0.0)} & \hl{(341.34, 0.0)} & \hl{(204.08, 0.0)} & \hl{(252.15, 0.0)} & 5.0 \\
8 & \hl{(252.87, 0.0)} & \hl{(312.8, 0.0)} & \hl{(276.86, 0.0)} & \hl{(369.99, 0.0)} & \hl{(348.62, 0.0)} & \hl{(376.37, 0.0)} & \hl{(233.06, 0.0)} & \hl{(346.53, 0.0)} & \hl{(297.64, 0.0)} & \hl{(304.85, 0.0)} & 5.0 \\
9 & \hl{(277.78, 0.0)} & \hl{(310.07, 0.0)} & \hl{(360.29, 0.0)} & \hl{(362.94, 0.0)} & \hl{(323.56, 0.0)} & \hl{(290.05, 0.0)} & \hl{(291.7, 0.0)} & \hl{(308.02, 0.0)} & \hl{(219.66, 0.0)} & \hl{(247.69, 0.0)} & 5.0 \\
10 & \hl{(313.84, 0.0)} & \hl{(489.51, 0.0)} & \hl{(247.21, 0.0)} & \hl{(383.54, 0.0)} & \hl{(210.4, 0.0)} & \hl{(321.68, 0.0)} & \hl{(201.07, 0.0)} & \hl{(353.41, 0.0)} & \hl{(250.04, 0.0)} & \hl{(259.75, 0.0)} & 5.0 \\
11 & \hl{(295.57, 0.0)} & \hl{(367.21, 0.0)} & \hl{(301.12, 0.0)} & \hl{(346.91, 0.0)} & \hl{(318.6, 0.0)} & \hl{(343.32, 0.0)} & \hl{(194.6, 0.0)} & \hl{(293.48, 0.0)} & \hl{(361.54, 0.0)} & \hl{(366.6, 0.0)} & 5.0 \\
12 & \hl{(282.94, 0.0)} & \hl{(309.55, 0.0)} & \hl{(313.07, 0.0)} & \hl{(401.81, 0.0)} & \hl{(342.97, 0.0)} & \hl{(411.61, 0.0)} & \hl{(252.58, 0.0)} & \hl{(371.43, 0.0)} & \hl{(328.99, 0.0)} & \hl{(294.08, 0.0)} & 5.0 \\
13 & \hl{(387.32, 0.0)} & \hl{(335.41, 0.0)} & \hl{(347.62, 0.0)} & \hl{(332.63, 0.0)} & \hl{(359.76, 0.0)} & \hl{(323.31, 0.0)} & \hl{(339.42, 0.0)} & \hl{(304.9, 0.0)} & \hl{(389.55, 0.0)} & \hl{(267.69, 0.0)} & 5.0 \\
14 & \hl{(285.8, 0.0)} & \hl{(361.84, 0.0)} & \hl{(289.23, 0.0)} & \hl{(373.78, 0.0)} & \hl{(299.52, 0.0)} & \hl{(201.56, 0.0)} & \hl{(387.61, 0.0)} & \hl{(254.25, 0.0)} & \hl{(314.79, 0.0)} & \hl{(300.0, 0.0)} & 5.0 \\
15 & \hl{(343.33, 0.0)} & \hl{(219.31, 0.0)} & \hl{(211.65, 0.0)} & \hl{(333.45, 0.0)} & \hl{(229.25, 0.0)} & \hl{(216.21, 0.0)} & \hl{(221.94, 0.0)} & \hl{(314.36, 0.0)} & \hl{(360.07, 0.0)} & \hl{(263.1, 0.0)} & 5.0 \\
16 & \hl{(238.9, 0.0)} & \hl{(261.31, 0.0)} & \hl{(220.22, 0.0)} & \hl{(232.16, 0.0)} & \hl{(224.09, 0.0)} & \hl{(313.06, 0.0)} & \hl{(285.35, 0.0)} & \hl{(241.99, 0.0)} & \hl{(312.41, 0.0)} & \hl{(244.19, 0.0)} & 5.0 \\
17 & \hl{(363.2, 0.0)} & \hl{(302.95, 0.0)} & \hl{(246.09, 0.0)} & \hl{(316.5, 0.0)} & \hl{(277.05, 0.0)} & \hl{(248.77, 0.0)} & \hl{(230.11, 0.0)} & \hl{(308.77, 0.0)} & \hl{(259.78, 0.0)} & \hl{(411.23, 0.0)} & 5.0 \\
18 & \hl{(452.7, 0.0)} & \hl{(445.95, 0.0)} & \hl{(411.74, 0.0)} & \hl{(397.19, 0.0)} & \hl{(352.02, 0.0)} & \hl{(407.03, 0.0)} & \hl{(298.23, 0.0)} & \hl{(285.33, 0.0)} & \hl{(286.37, 0.0)} & \hl{(299.62, 0.0)} & 5.0 \\
19 & \hl{(261.09, 0.0)} & \hl{(270.92, 0.0)} & \hl{(288.69, 0.0)} & \hl{(262.53, 0.0)} & \hl{(348.25, 0.0)} & \hl{(354.32, 0.0)} & \hl{(310.8, 0.0)} & \hl{(224.69, 0.0)} & \hl{(362.8, 0.0)} & \hl{(334.55, 0.0)} & 5.0 \\
20 & \hl{(348.0, 0.0)} & \hl{(472.12, 0.0)} & \hl{(285.83, 0.0)} & \hl{(288.63, 0.0)} & \hl{(380.67, 0.0)} & \hl{(219.06, 0.0)} & \hl{(220.82, 0.0)} & \hl{(345.33, 0.0)} & \hl{(270.62, 0.0)} & \hl{(384.15, 0.0)} & 5.0 \\
21 & \hl{(352.55, 0.0)} & \hl{(287.36, 0.0)} & \hl{(349.64, 0.0)} & \hl{(350.32, 0.0)} & \hl{(227.33, 0.0)} & \hl{(324.19, 0.0)} & \hl{(336.45, 0.0)} & \hl{(337.68, 0.0)} & \hl{(471.93, 0.0)} & \hl{(318.36, 0.0)} & 5.0 \\
\hline
    \end{tabular}}
    \caption{Batch3: $\chi^2$ statistic and p values to test the hypothesis that the observed ad delivery in different age based ad destinations is proportional to the
population estimates provided by Facebook. In a majority of ads, the observed ad delivery is not proportional to the population estimates from
Facebook.}
    \label{tab:rq4-age-Batch3}
\end{table*}